\documentclass[letterpaper,12pt]{article}

\usepackage{amssymb,latexsym,amsmath}
\usepackage{fullpage,mathrsfs}
\usepackage{subfigure}
\usepackage{color}

\makeatletter \@addtoreset{equation}{section}
\makeatother

\def\be{\begin{equation}}
\def\ee{\end{equation}}
\def\bea{\begin{eqnarray}}
\def\eea{\end{eqnarray}}
\def\de{\partial}
\newcommand{\nn}{\nonumber}

\newcommand{\diff}{\mathrm{d}}
\newcommand{\ii}{\mathrm{i}} 
\newcommand{\sq}{v} 
\newcommand{\onshE}{I} 

\newcommand{\trho}{\rho}
\newcommand{\newrho}{r} 
\newcommand{\cL}{c} 
\newcommand{\irp}{\xi} 
\newcommand{\ttrho}{\tilde\trho} 
\newcommand{\eind}{E_\mathrm{Casimir}}
\newcommand{\Anm}{A^\mathrm{nm}}
\newcommand{\Vnm}{V^\mathrm{nm}}
\newcommand{\vcks}{\mathbb{V}}
\newcommand{\Acks}{A^{\mathrm{cs}}}
\newcommand{\prefac}{{\cal F}}
\newcommand{\aano}{\mathbf{a}}
\newcommand{\cano}{\mathbf{c}}
\newcommand{\Egrav}{E}
\newcommand{\rano}{\mathbf{R}}
\newcommand{\eqft}{H}
\newcommand{\Hopf}{{\cal H}}
\newcommand{\Mfive}{M}
\newcommand{\tE}{t}
\newcommand{\complstr}{{\mathcal{J}}}

\usepackage[pdftex]{graphicx}

\numberwithin{equation}{section}       

\begin{document}

\begin{titlepage}

\begin{center}

\today

\vskip 2.3 cm 

\vskip 5mm

{\Large \bf The gravity dual of supersymmetric gauge theories \\[4mm] 
on a squashed $S^1\times S^3$}

\vskip 15mm

{Davide Cassani and Dario Martelli}

\vskip 1cm

\textit{Department of Mathematics, King's College London, \\ [1mm]
The Strand, London WC2R 2LS,  United Kingdom\\}

\end{center}

\vskip 2 cm

\begin{abstract}

\noindent 
We present a new one-parameter family of supersymmetric solutions deforming AdS$_5$. This is constructed as an asymptotically locally anti de Sitter (AlAdS)
solution of five-dimensional minimal gauged supergravity, with topology $\mathbb{R} \times \mathbb{R}^4$ and a non-trivial graviphoton field, and can be uplifted 
to ten or eleven dimensional supergravities. An analytic continuation of this solution yields the gravity dual to a class of four-dimensional ${\cal N}=1$ 
supersymmetric gauge theories on a curved manifold with topology $S^1\times S^3$, comprising an $SU(2)\times U(1)$-symmetric squashed three-sphere, with a non-trivial 
background gauge field coupling to the R-symmetry current.  We compute the holographically renormalised on-shell action and interpret it in terms of the 
Casimir energy of the dual field theory. We also determine the holographic conserved  charges of the solution and discuss relations between them. 

\end{abstract}

\end{titlepage}

\pagestyle{plain}
\setcounter{page}{1}
\newcounter{bean}
\baselineskip18pt

\tableofcontents

\baselineskip 6 mm

\newpage

\section{Introduction}
\label{intro}

Starting with the work of \cite{Pestun:2007rz}, it has been appreciated that the technique of localisation provides 
a powerful tool for performing exact non-perturbative computations in supersymmetric field theories defined on curved manifolds. 
While \cite{Pestun:2007rz} computed the exact path integral of four-dimensional ${\cal N}=2$  gauge theories on a round four-sphere, later the 
attention shifted to three-dimensional theories. The authors of references \cite{Kapustin:2009kz,Jafferis:2010un,Hama:2010av} computed 
the partition function of general ${\cal N}=2$ superconformal gauge theories, with Chern--Simons terms 
and matter in arbitrary representations, placed on the round three-sphere. 
If the theory has an R-symmetry, then it is possible to define it on more general manifolds, 
while preserving supersymmetry, by turning on a background non-dynamical gauge field coupled to the 
R-symmetry current~\cite{FestucciaSeiberg,KTZ,DFS,Cassani:2012ri,CDFK3d}. 
The first examples of such backgrounds were considered in \cite{Hama:2011ea,Imamura:2011wg}, where the exact partition function of 
supersymmetric gauge theories defined on different squashed  three-spheres was computed. 
It was later shown in \cite{Alday:2013lba} that  the  ``ellipsoid'' partition function of \cite{Hama:2011ea} extends to a very general class 
of supersymmetric three-dimensional geometries. An asymptotically locally AdS (AlAdS) supergravity solution dual to the construction of \cite{Hama:2011ea} was presented in \cite{Martelli:2011fu}, and various extensions were obtained in \cite{Martelli:2011fw,Martelli:2012sz,Martelli:2013aqa}.
 
In this paper we will start exploring the gauge/gravity duality for four-dimensional ${\cal N}=1$ superconformal 
field theories defined on curved manifolds, with the aim of comparing gravity 
calculations with exact field theory results. This problem in four (boundary) dimensions turns out to be
more complicated than in one dimension lower for different reasons.  When the new minimal or conformal supergravity formulations of rigid supersymmetry are used, 
unbroken supersymmetry in Euclidean signature implies that the field theory must be defined on a Hermitian manifold \cite{KTZ,DFS}.
However, exact calculations of the path integral using localisation in field theories on such manifolds have been less developed than their three-dimensional analogues so far (but see the recent \cite{Closset:2013sxa}). Moreover, in four dimensions the path integral generically has logarithmic divergences, 
and extracting physical information from its finite part is more subtle than in three dimensions due to a number of ambiguities.

These issues are under better control if one considers supersymmetric manifolds with topology $S^1 \times S^3$. 
In these cases the path integral  with periodic boundary conditions is expected 
to be proportional to a supersymmetric index~\cite{Aharony:2003sx,Romelsberger:2005eg,Kinney:2005ej},  refined with fugacities related to a choice of complex structure 
on $S^1 \times S^3$ \cite{CDFK}. For superconformal theories, the index may be equivalently defined as the generating function of operators weighted by their fermion number, 
so that the contributions from the long multiplets cancel out, 
and as shown in~\cite{Romelsberger:2005eg} this may be obtained by putting the theory on $\mathbb{R}\times S^3$. In~\cite{Gadde:2010en,Eager:2012hx} the superconformal 
index was successfully compared with the spectrum of Kaluza--Klein modes 
dual to the protected operators being counted by the index. However, from the perspective of the Euclidean path integral, it is more natural to compare the field theory 
partition function with the holographically renormalised supergravity on-shell action \cite{Gubser:1998bc,WittenAdS}.  
In  the large $N$ limit the index scales like ${\cal O}(N^0)$ \cite{Kinney:2005ej},  while the on-shell action scales like ${\cal O}(N^2)$. 
Therefore, it would appear that the latter cannot capture new information about the field theory, 
apart from the Weyl anomaly \cite{HenningsonSkenderis}. When the field theory is placed on 
a round (i.e.\ conformally flat) $S^1\times S^3$, the gravity dual is simply  global AdS$_5$ 
with Euclideanized time, and the renormalised  gravity action has been argued to reproduce the \emph{Casimir energy} of 
the field theory\footnote{At least when the field theory is ${\cal N}=4$ super Yang--Mills, and the Casimir energy can be computed at weak coupling.} on $S^3$ \cite{WittenAdS,BalasubramanianKraus}. More generally, the on-shell action is expected to capture the ratio between the 
path integral and the index,  which has been dubbed ``index Casimir energy'' in references \cite{Kim:2009wb,Kim:2012ava,Kim:2013nva}. 
This quantity is sensitive to the regularisation of the path integral, and will be discussed in more detail towards the end of the paper. 

On the gravity side, constructing five-dimensional supersymmetric AlAdS solutions is technically more difficult than obtaining analogous solutions in one dimension less. 
One reason is that in odd dimensions the asymptotic form of the fields (in a Fefferman--Graham coordinate system \cite{feffe}) contains logarithmic terms, 
suggesting that analytic solutions are much harder to find. Ref.~\cite{GGS} presented an AlAdS solution that is special in this regard, as the logarithmic terms do not arise. 
This solution has a conformal boundary with the topology of $\mathbb{R} \times S^3$, however its isometry group is $\mathbb{R}\times SU(2)$, with the non-compact  time being generated by a null (in the boundary) 
Killing vector, hence it is not possible to analytically continue it to Euclidean signature, while keeping a real metric. 

Although in order to compare with localisation results we are ultimately interested in Euclidean conformal boundaries, we will start working with minimal five-dimensional gauged 
supergravity in Lorentzian signature. This will allow us to construct a  supersymmetric solution utilising  the formalism of \cite{GaGuAllMinimalGauged}, thus avoiding 
complications arising in Euclidean signature, where we would have to deal with a complex graviphoton field from the outset.

The solution that we will present is a one-parameter family of supersymmetric deformations of AdS$_5$, whose conformal boundary includes a biaxially squashed three-sphere with $SU(2)\times U(1)$ symmetry, as well as a non-trivial gauge field. It preserves two of the eight supercharges of minimal gauged supergravity.
Globally, the space-time is equivalent to AdS$_5$, with topology $\mathbb{R}\times \mathbb{R}^4$, where the first factor is a time direction, and the 
$\mathbb{R}^4$ factor arises from the three-sphere of the boundary smoothly shrinking to zero size in the interior. In particular, the space-time does not have a horizon. 
The metric, gauge field, and Killing spinor of the solution restricted to the asymptotic boundary reduce precisely to a background \cite{SuperWeylAnomaly_paper} solving the charged conformal Killing spinor equation~\cite{Cassani:2012ri}, as well as the new minimal version of rigid supersymmetry in curved space~\cite{FestucciaSeiberg}. Therefore we interpret our solution as the gravity dual to an ${\cal N}=1$ superconformal gauge theory on a curved, non conformally-flat, background preserving two supercharges. 
Another feature of this background is that the aforementioned logarithmic divergences do not appear in the partition function~\cite{SuperWeylAnomaly_paper}. Our five-dimensional solution can be uplifted either to type IIB or to eleven-dimensional supergravity, with the specific dual field theory depending on the choice of internal manifold~$Y$. In the uplift to type IIB supergravity \cite{Buchel:2006gb,Gauntlett:2007ma}, $Y$ can be a Sasaki--Einstein five-manifold, as well as a more general manifold 
arising from the classification of~\cite{Gauntlett:2005ww}. In the uplift to eleven-dimensional supergravity~\cite{GauntlettColgainVarela}, $Y$ can be any of the manifolds in the classification of~\cite{AdS5Mtheory}.

A main feature of the solution is that although it has been constructed using a combination of perturbative and numerical analysis, all the quantities of physical interest,
namely the on-shell action and the holographic conserved charges, have been obtained analytically as a function of the boundary squashing parameter.  Upon a simple Wick rotation,  and compactification of the Euclidean time, 
the boundary metric describes a squashed $S^1\times S^3$, and the background gauge field becomes complex, consistently with the results of \cite{KTZ,DFS}.  
While the boundary metric remains real, in the bulk both the gauge field and the metric become \emph{complex}. 
However, the renormalised on-shell action is real and we show that it reduces to a boundary term.  Moreover, it receives a crucial contribution from the Chern--Simons term, 
with the result depending on the gauge choice for the graviphoton. We will explain that demanding the Killing spinors (on the boundary as well as in the bulk) to be invariant 
with respect to a specific choice of Killing vector $\de/\de t$ generating time translations, fixes the gauge uniquely.  This is necessary in order to have well-defined spinors 
after performing the Wick rotation, and on the field theory side it is required for defining the Euclidean  path integral with supersymmetric boundary 
conditions  \cite{FestucciaSeiberg}.

We will show that the holographic charges obey relations consistent with the rigid supersymmetry algebras of the field theories defined on the boundary geometry.
Moreover, in the Euclidean setting, we will see that the renormalised on-shell action reproduces  a suitably defined Casimir energy, arising from the supersymmetric path integral, up to terms that we will propose 
to be removable by a choice of finite local counterterms. In particular, the Casimir energy depends linearly  on the complex structure modulus $\beta$ of 
the boundary manifold and the $\mathbf{a}$ anomaly coefficient,  while 
we will show that there exists a choice of finite local counterterms,
 constructed from the supergravity fields and the complex structure, which cancels the remainder terms. 

The rest of the paper is organised as follows. In  section \ref{susysec} we review the derivation of  the relevant ODE stemming from supersymmetry, following \cite{GutowskiReall}.
In section \ref{newsol_section} we present our new solution, combining analytical methods and a numerical integration of the 
ODE. Section~\ref{HoloRenSection} is devoted to holographic renormalisation and the 
computation of conserved charges. In section~\ref{SummarySol} we summarize the main features of the solution. 
In section \ref{fielddisc} we discuss key properties of the dual field theories, and compare these with the gravity results.
Section~\ref{disc} concludes summarising our findings and outlining directions for future work. Two appendices contain technical 
details on the solution and the holographic quantities.

\section{Supersymmetry equations}
\label{susysec}

We begin by presenting the supersymmetry equations of minimal  five-dimensional gauged supergravity, adopting the formalism of 
\cite{GaGuAllMinimalGauged}. In particular, we will be  interested in supersymmetric solutions of the time-like class, derived from an ansatz 
possessing (locally) $SU(2)\times U(1)\times U(1)$ symmetry. This leads precisely to the same equations 
obeyed by the supersymmetric black hole solutions
found in \cite{GutowskiReall}, and in this section we will follow closely the derivation therein. In the next section, we will present a new solution to these equations. 
The reader familiar with eq.~\eqref{SixthOrderEq} below, or not interested in its derivation, can safely skip to the next section.

In the conventions of \cite{GutowskiReall}, the action for the bosonic sector of minimal five-dimensional gauged supergravity  reads
\be
S_{\rm bulk} \ = \ \frac{1}{16\pi G}\int \left[ \diff^5x \sqrt{g}\left( R - F_{\mu\nu}F^{\mu\nu} +\frac{12}{\ell^2} \right)  - \frac{8}{3\sqrt 3} A \wedge F\wedge F  \right]\,.
\label{sugraction}
\ee
Here $R$ denotes the Ricci scalar of the five-dimensional metric $g_{\mu\nu}$, and $g = |\!\det g_{\mu\nu}|$.\footnote{Our Riemann tensor is defined 
as $R^\mu{}_{\nu\kappa\lambda} = \partial_\kappa \Gamma^\mu_{\nu\lambda} + \Gamma^\mu_{\kappa\sigma}\Gamma^\sigma_{\nu\lambda} - \kappa \leftrightarrow \lambda$,
and the Ricci tensor is $R_{\mu\nu}= R^\lambda{}_{\mu\lambda\nu}$.\label{FootnoteConventionsCurvature}}  The graviphoton $A$ is an Abelian gauge field with field 
strength $F = \diff A$. Moreover, $G$ is the five-dimensional Newton constant and $\ell$  
is a parameter with dimensions of length, related to the cosmological constant.
The equations of motion derived from (\ref{sugraction}) read
\bea
\label{EinsteinEq}
R_{\mu\nu} + 2 F_{\mu\rho}F^\rho{}_\nu + g_{\mu\nu} \left( \frac{4}{\ell^2} + \frac{1}{3}F_{\rho\sigma}F^{\rho\sigma} \right) & = & 0 \, ,\\
\label{MaxwellEq}
\diff *F + \frac{2}{\sqrt 3} F \wedge F & = & 0\,.
\eea
For most of this paper we will work in Lorentzian signature $(-,+,+,+,+)$, although we will also discuss an analytic continuation later. 

A solution is supersymmetric if there is a non-trivial Dirac spinor $\epsilon$ satisfying the equation
\be\label{KillingSpEqDirac}
\left[ \nabla_\mu + \frac{\ii}{4\sqrt 3}\left( \gamma_{\mu}{}^{\nu\lambda} - 4\delta^\nu_\mu\gamma^\lambda\right) F_{\nu\lambda} - \frac{1}{2\ell}  \big( \gamma_\mu - 2\sqrt 3\,\ii A_\mu \big)\right]\epsilon  \,=\, 0\,,
\ee
where  $\gamma_\mu$ generate the Clifford algebra Cliff$(1, 4)$, so $\{\gamma_\mu,\gamma_\nu\} = 2 g_{\mu\nu}$. 
Any such solution uplifts (locally) to a supersymmetric 
solution of type IIB supergravity~\cite{Buchel:2006gb,Gauntlett:2007ma} or of eleven-dimensional supergravity~\cite{GauntlettColgainVarela}. Although global aspects of this uplift can be subtle 
\cite{Martelli:2012sz}, for the solution that we will discuss the gauge field $A$ will be a \emph{globally defined} one-form on the five-dimensional space, and therefore
there are no global obstructions to uplifting it to ten or eleven dimensions. 

Reference \cite{GaGuAllMinimalGauged} showed that solutions to (\ref{KillingSpEqDirac}) possess a Killing vector $V$ (constructed as bilinear in the spinor $\epsilon$), 
and fall into two classes, depending on whether this is everywhere null, or timelike at least in some open set. We will focus on the timelike class, 
which is further specified by a K\"ahler metric $\diff s^2_{B}$ on the four-dimensional ``base'' $B$ transverse to $V$.  Thus, 
in coordinates such that $V=\de /\de y$, the five-dimensional metric takes the form 
\be
\diff s^2 \, =\, - f^2 (\diff y + \omega)^2 + f^{-1}\diff s^2_{B}\,,
\label{generalggmetric}
\ee
where $f$ and $\omega$ are a positive function and a transverse one-form, respectively. 
The base is characterized by a K\"ahler form $X^1$, inducing a complex structure $(X^1)_a{}^b$, 
and a complex two-form $\Omega$, which is $(2,0)$ with respect to the complex structure and satisfies
\hbox{$\diff \Omega +  \ii P\wedge \Omega  =  0$}, where $P$ is the Ricci one-form potential. 
Namely, given the Ricci curvature two-form $\mathcal R_{ab} = \frac{1}{2} R_{abcd}(X^1)^{cd}$, where $R_{abcd}$ is the Riemann tensor of the K\"ahler metric, 
the Ricci potential is defined by ${\cal R}= \diff  P$. Equivalently, splitting $\Omega$ into its real and imaginary parts as \hbox{$\Omega =  X^{2}+ \ii X^{3}$}, we have a triplet of real two-forms $X^{I}$, $I=1,2,3$. Choosing ${\rm vol}_B = -\frac{1}{2} X^1 \wedge X^1$ as orientation on $B$,  
the $X^I$ are anti-self-dual: $*_B X^I = - X^I$, where $*_{B}$ denotes the Hodge star on $B$.\footnote{The $X^I$ also satisfy the algebraic relations
\hbox{$X^{I}{}_a{}^c X^{J}{}_c{}^b \ = \ -\delta^{IJ} \delta_a{}^b + \epsilon^{IJK} X^{K}{}_a{}^b$}. Here $a,b=1,\dots,4$ denote tangent space indices on the K\"ahler base.}

The gauge field of a supersymmetric solution in the timelike class is specified by the geometry and reads
\be\label{gaugefieldGG}
F \ =\ \frac{\sqrt{3}}{2}\,\diff \left[ f(\diff y + \omega) + \frac{\ell}{3} P\right].
\ee 
Moreover, denoting by $R_{B}$  the Ricci scalar of $\diff s^2_{B}$ and introducing  $G^{\pm} = \tfrac{f}{2}(\diff \omega \pm *_{B}\, \diff \omega)$,
supersymmetry also implies that
\be\label{relfR}
f^{-1} \ = \ - \frac{\ell^2}{24}R_{B}\,, 
\ee
and 
\be
G^+ \ = \ -\frac{\ell}{2}\left( {\cal R} - \frac{R_{B}}{4} X^1 \right) \label{G+isTracelessRicci},
\ee
while $G^-$ may be determined using the Maxwell equation.\footnote{Although the procedure of~\cite{GaGuAllMinimalGauged} is constructive once a 
four-dimensional K\"ahler basis is given, one should be aware that not all K\"ahler bases give rise to supersymmetric solutions. This was first noted in 
an example in~\cite{FiguerasHerdeiroPaccetti}. We can express the constraint to be satisfied by the K\"ahler geometry in a general form by observing that 
equation (3.24) in~\cite{GaGuAllMinimalGauged} has to satisfy the integrability condition $\overline\Omega \wedge (\diff - \ii\, P\wedge )\Theta = 0\,$, where $\Theta$
is also defined in~\cite{GaGuAllMinimalGauged}.
In the specific case of interest for us, this is automatically satisfied.} For more details we refer to \cite{GaGuAllMinimalGauged}. 

Similarly to~\cite{GutowskiReall}, we will consider the following  $SU(2)_{\rm left}\times U(1)_{\rm right}$ symmetric ansatz for the metric
\be
\diff s^2_{B} \ = \ \diff \rho^2 + a^2(\rho)(\hat{\sigma}_1^2+\hat{\sigma}_2^2) +b^2(\rho)\hat{\sigma}_3^2 \,,
\ee
where $\hat{\sigma}_1,\hat{\sigma}_2, \hat{\sigma}_3$ are $SU(2)$ left-invariant one-forms
parameterized by coordinates 
$\theta, \phi,\hat{\psi}$, and $\rho$ is a radial coordinate.\footnote{Although~\cite{GutowskiReall} used right-invariant forms, we will also have a different choice of orientation, so that the final supersymmetry equations will turn out the same.}
Specifically, we define
\bea
\hat{\sigma}_1 &=& -\sin\hat{\psi} \,\diff \theta + \cos\hat{\psi} \sin\theta \,\diff \phi\,,\nn\\ [1mm]
\hat{\sigma}_2 &=& \cos\hat{\psi}\, \diff \theta + \sin\hat{\psi} \sin\theta \,\diff \phi\,,\nn\\ [1mm]
\hat{\sigma}_3 &=& \diff \hat{\psi} + \cos\theta \,\diff  \phi\, .\label{DefLeftInv1forms}
\eea
Note that these satisfy $\diff \hat{\sigma}_1 = \hat{\sigma}_2 \wedge \hat{\sigma}_3\,$, $\diff \hat{\sigma}_2 = \hat{\sigma}_3 \wedge \hat{\sigma}_1\,$ and $\diff \hat{\sigma}_3 = \hat{\sigma}_1 \wedge \hat{\sigma}_2\,$. 
The hat on $\hat\psi$ (and the consequent one on the left-invariant one-forms) serves to distinguish this coordinate from a new one, $\psi$, that will be introduced later. While at this stage $\hat\psi$ should be regarded just as a local coordinate, $\psi$ will be an actual Euler angle on~$S^3$.
The main motivation for working with this ansatz is that the supersymmetry equations reduce to a system of ODE's, which ultimately can be expressed as a single sixth-order ODE for one function \cite{GutowskiReall}. As we are interested in AlAdS solutions, $\rho$ is related (but, as we will see, not
identical) to a Fefferman--Graham coordinate. The boundary geometry will necessarily have locally $SU(2)_{\rm left}\times U(1)_{\rm right}\times U(1)_y$ isometry, therefore supersymmetry implies that the asymptotic 
metric and gauge field (see section \ref{UVanalysis} for their expression) must be that of the example in section 4 of \cite{SuperWeylAnomaly_paper}.

We fix the  orientation on $B$ defining ${\rm vol}_{B} = a^2b \,\hat{\sigma}_1 \wedge \hat{\sigma}_2 \wedge \hat{\sigma}_3 \wedge \diff \rho$, while the five-dimensional 
orientation is given by ${\rm vol}_5 = f^{-1}\diff y \wedge {\rm vol}_B$. 
The $SU(2)_{\rm left}\times U(1)_{\rm right}$ invariant ansatz is completed  by taking
\bea
X^1 & = &  - a^2 (\rho) \hat{\sigma}_1 \wedge \hat{\sigma}_2 + b(\rho) \hat{\sigma}_3 \wedge \diff \rho \, , \nn\\ [1mm]
\Omega & = & a(\rho)(\hat{\sigma}_1 + \ii\, \hat{\sigma}_2)\wedge (\diff \rho + \ii\,  b(\rho)\hat{\sigma}_3 )\,,\nn\\ [1mm]
\omega & = & \Psi (\rho) \, \hat{\sigma}_3\,, \nn\\ [1mm]
P & = & p(\rho)\, \hat{\sigma}_3 \, ,
\eea
where the function $a(\rho)$ is chosen positive. 
The gauge field~\eqref{gaugefieldGG} becomes
\be\label{gaugefieldGR}
F \ =\ \frac{\sqrt{3}}{2} \,\diff \left[  f \diff y + \left(  f \Psi + \frac{\ell}{3}p \right)      \, \hat{\sigma}_3 \right].
\ee
The K\"ahler conditions then imply that $b  = 2aa'$ and  $p  =  4 a'^2 + 2 a a'' -1$, where a prime denotes derivative with respect to $\rho$, while eq.~\eqref{relfR} yields
\be
f^{-1} \ = \ \frac{\ell^2}{12 a^2 a'}[4 (a')^3 + 7 a\, a' a'' - a' + a^2 a'''] \,.\label{invfgeneral}
\ee
Imposing \eqref{G+isTracelessRicci} implies that $\Psi$  obeys the differential equation
\be
\frac{\Psi'}{2aa'} - \frac{\Psi}{a^2} \ =\ \frac{\ell g}{2f}  \,,
\label{psisusy}
\ee
where we defined
\be
g \ = \ -\frac{a'''}{a'} - 3 \frac{a''}{a} - \frac{1}{a^2} + 4 \frac{(a')^2}{a^2}\,.
\ee
Finally,  combining the supersymmetry condition~\eqref{psisusy} with the Maxwell equation~\eqref{MaxwellEq}, one can solve for $\Psi$ in terms of $a$,
\be\label{exprPsiGeneral}
\Psi\ = \ - \frac{\ell a^2}{4}\left( \nabla^2 f^{-1} + 8 \ell^{-2} f^{-2} - \frac{\ell^2g^2}{18} + f^{-1} g \right),
\ee
and eventually derive the equation governing $a$~\cite{GutowskiReall}: 
\be\label{SixthOrderEq}
\Big(\nabla^2 f^{-1} +8 \ell^{-2}f^{-2} - \frac{\ell^2 g^2}{18} + f^{-1} g \Big)' + \frac{4a'g}{af} \ = \ 0\,,
\ee
where $\nabla^2$ is the Laplacian on $B$. After using  the expressions for $f$ and $g$ given above, one obtains a non-linear, sixth-order equation for $a(\rho)$ with no explicit dependence on~$\rho$.

Any solution to eq.~\eqref{SixthOrderEq} gives a supersymmetric solution to minimal five-dimensional gauged supergravity, preserving at least one quarter of the supersymmetry, namely two real supercharges. In~\cite{GutowskiReall}, a simple family of solutions was found,
\be\label{GRsolution}
a(\rho) \ =\ \alpha \ell \sinh(\rho/\ell)\,,
\ee
where $\alpha$ is a parameter. For $\alpha = 1/2$ this yields just AdS$_5$ (of radius $\ell$)  with a vanishing Maxwell field, while for $\alpha > 1/2$ one obtains a charged, rotating black hole with a regular horizon. 
 In order to obtain a geometry with a horizon, the authors of~\cite{GutowskiReall} imposed specific boundary conditions on the function $a(\rho)$. Here we will impose different conditions, in particular we will require 
that in the interior the space closes off smoothly like $\mathbb{R}_t\times \mathbb{R}^4$, where the first factor is a time coordinate, precisely as in global AdS$_5$. Another key difference with respect
to the solution of  \cite{GutowskiReall} is that the latter is asymptotically AdS, with a conformally flat boundary $\mathbb{R}_t \times S^3_\mathrm{round}$, 
while our solution will be asymptotically \emph{locally} AdS, with a non conformally flat boundary comprising a squashed $S^3$.  Correspondingly, while the gauge field strength in  \cite{GutowskiReall} vanishes asymptotically, in our case it will remain non-trivial at the~boundary.

\section{The solution}
\label{newsol_section}

In this section we study solutions to the sixth-order equation~\eqref{SixthOrderEq}. We start by obtaining a general asymptotic solution satisfying the AlAdS condition. 
Then we analyse the equation at finite $\rho$ and impose that the solution closes off smoothly in the interior. This will give a new one-parameter family of solutions. 
We will show that a solution connecting the asymptotic and the interior regions exists both by presenting a linearised solution around the AdS$_5$ background and by 
providing numerical evidence.

We will set $\ell =1$ for simplicity; factors of $\ell$ can easily be restored by dimensional analysis (in particular by sending $\rho \to \rho/\ell$, $y\to y/\ell$ and $a\to a/\ell$).

\subsection{Solution in the UV}
\label{UVanalysis}

In the following, we study the sixth-order equation~\eqref{SixthOrderEq} perturbatively at large positive $\rho$ (i.e.\ in the ``UV region''). 
We assume an asymptotic expansion for the unknown function $a$ of the type
\bea
a(\trho) &=& a_0 e^{\trho} \Big[ 1 + \sum_{k \geq 1}\,\sum_{0\leq n \leq k}  a_{2k,n} \,\trho^{\,n} \big(a_0 e^{\trho}\big)^{-2k} \Big]\nn \\ [2mm]
&=& a_0 e^{\trho} \left[1 + \left(a_{2,0} + a_{2,1} \trho\right) \frac{e^{-2\trho}}{a_0^2} +  \left(a_{4,0} + a_{4,1} \trho + a_{4,2} \trho^2 \right) \frac{e^{-4\trho}}{a_0^4} + \ldots \right],\label{UVexpansion}
\eea
with $a_0 \neq 0$. Terms weighted by odd negative powers of $a_0 e^{\trho}$ could be included in the square bracket, but would be set to zero by the differential equation.
We solved the latter perturbatively up to order $\mathcal O(e^{-11\trho})$ (included), and found that there are five free coefficients that determine all the others. 
Renaming them for convenience, these free coefficients are $a_0$, $a_2 \equiv a_{2,0}$, $\cL  \equiv a_{2,1}$, $a_4  \equiv a_{4,0}$ and $a_6  \equiv a_{6,0}$. 
We display the first few terms of the solution:
\bea
a(\trho) &=&  a_0 e^{\trho} + (a_2 + \cL \trho)\frac{e^{-\trho}}{a_0} + \left(a_4 + \frac{2- 16a_2 -5\cL}{12}\cL\trho -\frac 23 \cL^2 \trho^2\right) \frac{e^{-3\trho}}{a_0^3}\nn \\[2mm]
&&+\, \Big( a_6 + \frac{12 - 282a_2 + 1488a_2^2 -1548 a_4 -54\cL +537 a_2\cL + 59\cL^2 }{972} \cL\trho  \nn \\ [2mm]
&& \;\,\quad -\,\frac{ 90 - 840a_2 - 197 \cL }{324}\,\cL^2\trho^2 + \frac{70}{81}\cL^3\trho^3\Big)\frac{e^{-5\trho}}{a_0^5} \,+\, \mathcal O(e^{-6\trho}) \,.\label{UVsola}
\eea
As discussed in the previous section, this determines completely the large-$\trho$ form of the five-dimensional metric and the gauge field $F$. 
In the following we provide their expression at leading order. 
Before presenting this, we introduce new coordinates $(t,\psi)$ that will prove particularly natural from a boundary perspective. These are given by 
\be\label{changepsi}
\hat{\psi} \ =\ \psi - \frac{2}{1-4\cL}\,t 
\,,\qquad y \ =\ t\,.
\ee
Then the five-dimensional metric takes the asymptotic form
\be\label{asymptoticmetric}
\diff s^2 \ = \ \diff \trho^2 + e^{2\trho} \, \diff s^2_{\rm bdry}\,+\, \ldots\,,
\ee
with boundary metric
\be\label{bdrymetric}
\diff s^2_{\rm bdry} \ = \  (2a_0)^2\left[ - \frac{1}{\sq^2} \diff t^2 + \frac{1}{4}\left(\sigma_1^{\,2} + \sigma_2^{\,2}  +   \sq^2\sigma_3^{\,2}\right) \right],
\ee
while the gauge field reads
\be\label{asymptoticA}
A \ = \ A_{\rm bdry} \,+\, \mathcal O(e^{-\trho})\,,
\ee
with boundary value
\be\label{bdryA}
A_{\rm bdry}\ =\  
\frac{1}{2\sqrt 3} \left[ \diff t + (\sq^2-1) \sigma_3\right]\,.
\ee
Here, we have introduced the parameter
\be\label{defsquashing}
\sq^2 = 1-4\cL\,.\ee
The one-forms $\sigma_1,\sigma_2,\sigma_3$ are defined in the same way as $\hat\sigma_1,\hat\sigma_2,\hat\sigma_3$ in~\eqref{DefLeftInv1forms}, with $\hat{\psi}$ replaced by the new coordinate $\psi $. We take $0\leq \theta \leq\pi $, $0\leq \phi < 2\pi$ and $0 \leq \psi  < 4\pi$, so that $\theta, \phi, \psi$ are Euler angles on $S^3$. Then from~\eqref{bdrymetric} we see that the boundary is a direct product geometry including a squashed~$S^3$, with radius $2a_0$ and squashing parameter $\sq$. Namely, if we regard $S^3$ 
as a Hopf fibration over $S^2$, we have that the radius of the $U(1)$ fiber generated by $\partial/\partial \psi$ is rescaled with respect to the $S^2$ radius $2a_0$ by a factor of $\sq$. We will denote this squashed three-sphere by $S^3_\sq$. So in these coordinates the boundary metric describes a direct product $\mathbb R \times S^3_\sq$; this will allow us to perform a simple Wick-rotation later on by just analytically continuing $t$. 

Note that we need $\cL < 1/4$ in order to have a Riemannian metric on $S^3$ and avoid closed timelike curves in the boundary. When $\cL = 0$, then $\sq^2=1$ and there is no squashing; namely, $\diff s^2_{\rm bdry}$ becomes conformally flat and describes an $\mathbb R \times S^3$ boundary with a round $S^3$ of radius $2a_0$. Correspondingly, all $\trho^n$ terms in~\eqref{UVsola} vanish in this case. However, these terms are crucial for our purposes, as we are interested in non conformally flat boundaries. The only other free parameter of the UV solution~\eqref{UVsola} 
appearing in the leading order expression of the supergravity fields is $a_0$, which controls the overall size of the boundary.
Since equation~\eqref{SixthOrderEq} has no explicit dependence on~$\rho$, we could set $a_0$ to any chosen non-zero value by performing a constant shift of $\rho$, followed by a redefinition of the other parameters in~\eqref{UVsola}.\footnote{This is a manifestation of the fact that changes of radial coordinate in the bulk induce conformal transformations on the boundary.} 
However, we will not fix this shift symmetry here as it will be convenient to use it in the next section,  when we will study the solution in the interior. 

An important comment about the gauge potential $A$ is in order. This is determined by supersymmetry only up to a gauge transformation: from eq.~\eqref{KillingSpEqDirac} we 
see that a gauge shift $ A \to A + \diff \Lambda$ just transforms the supersymmetry parameter as $\epsilon \to e^{-\ii \sqrt 3\Lambda/\ell}\,\epsilon\,$. In~\eqref{asymptoticA}, \eqref{bdryA}, we chose the gauge by imposing $A_\rho =0$ and that the $SU(2)\times U(1)\times U(1)$ symmetry is respected; then we fixed the residual freedom to shift $A$ by a term proportional to $\diff t$ by imposing that 
the spinor $\epsilon$ is \emph{independent of the time coordinate} $t$. This last property will be very important for us in the following.\footnote{One can see that in this gauge a spinor depending just on the radial coordinate $\rho$ solves equation~\eqref{KillingSpEqDirac}.
That the boundary spinor is constant with $A$ chosen as in~\eqref{bdryA} follows from the discussion in~\cite[sect.~4]{SuperWeylAnomaly_paper}. Our gauge
choice is different from 
the one of~\cite{GutowskiReall}, which is~$A_y = \frac{\sqrt 3}{2} f$.} It will prove crucial in section~\ref{HoloRenSection} when we will evaluate the supergravity 
action on-shell, as the Chern--Simons term~\eqref{sugraction} is not invariant under (large) gauge transformations on a space with boundary, hence the on-shell action changes when different gauge choices are made.

In appendix~\ref{DetailsSol} we provide more details about the five-dimensional UV solution, including  the first few sub-leading terms of the metric and the gauge field, in which the remaining free parameters $a_2$, $a_4$, $a_6$ appear. In the same appendix, we also show that the five-dimensional spacetime is AlAdS by recasting the five-dimensional fields in Fefferman--Graham form.

\subsection{Solution in the IR}

\label{sec:IRsolution}

Having determined the asymptotic behavior of the solution at large $\rho$, we should now study how this ends in the interior (i.e.\ the ``IR region''). 
We will require that the spacetime shrinks smoothly to zero size, with no horizon being formed. By exploiting the freedom to shift $\rho$, we can assume with no loss of generality that this happens at $\rho \to 0$. We also assume that $a(\rho)$ can be expanded in a Taylor series around $\rho =0$ as
\be\label{IRansatz}
a(\trho) \ = \ a_0^{\rm IR} + a_1^{\rm IR} \trho + a_2^{\rm IR} \trho^2 + a_3^{\rm IR} \trho^3 + \ldots \,.
\ee
Although more general ans\"atze may be considered, we will see that~\eqref{IRansatz} is enough to describe a new one-parameter family of regular solutions. Since we need $a \to 0$ as $\rho \to 0$, we will choose $a_0^{\rm IR} = 0$. Then, expanding equation~\eqref{SixthOrderEq} at small $\trho$, we find that this requires $a_1^{\rm IR} \neq 0$, $a_2^{\rm IR} = 0$ and
\be
\left[ 11 (a_1^{\rm IR})^2 -8\right] a_4^{\rm IR} \ = \ 0\,.
\ee
We will choose $a_4^{\rm IR} = 0$.\footnote{Choosing instead $(a_1^{\rm IR})^2 = 8/11$ and $a_4^{\rm IR}\neq 0$ leads to an expansion involving both odd and even powers of $\rho$. The corresponding first terms in the expansion of the function $f$ read
\be
f(\rho) \ = \ \frac{32}{7}\trho^2 - \frac{1424 \sqrt{22}}{49} a_3^{\rm IR} \trho^4 - \frac{176 \sqrt{22}}{3}a_4^{\rm IR} \trho^5 + O(\trho^6)\,,\nn
\ee
so $f \to 0$ when $\rho \to 0$. This is compatible with the presence of a horizon, because $f$ is the norm of the Killing vector $V$, and this should become null at the horizon. The fact that $g_{\theta\theta} = f^{-1}a^2$ remains finite is compatible with a horizon of finite size. We will not study this any further in the present paper.} At the following order in the $\rho$ expansion we get
\be
\left[ 10 \, a_1^{\rm IR} a_5^{\rm IR} -3 (a_3^{\rm IR})^2 \right] \left[ 4 (a_1^{\rm IR})^2 -1 \right]  \ = \ 0\,.
\ee
Vanishing of either the first or the second factor leads to two distinct families of solutions: if the first factor is zero then by going on in the perturbative expansion we reconstruct~\eqref{GRsolution}, namely the solution of~\cite{GutowskiReall}, while if it is the second factor that vanishes we obtain a new family of solutions. So in the following we fix $a_1^{\rm IR}= 1/2$ and focus on the latter.\footnote{One can see that the solution with $a_1^{\rm IR} = -1/2$ is straightforwardly related to the one with $a_1^{\rm IR} = +1/2$ by sending $\rho \to -\rho$, $a_3^{\rm IR} \to -a_3^{\rm IR}$, $a_5^{\rm IR} \to -a_5^{\rm IR}$. 
} At higher orders, we find that all even coefficients $a_{2k}^{\rm IR} $ are zero, while $a_3^{\rm IR}$ and $a_5^{\rm IR}$ are free parameters, and all the successive coefficients are determined (at least up to the order $\mathcal{O}(\rho^{25})$, where we stopped the analysis).  Actually, of the two free parameters $a_3^{\rm IR}$ and $a_5^{\rm IR}$, only one is physical. To see this, note~\cite{GutowskiReall} that under a rescaling of the coordinates $\rho = \lambda^{-1}\tilde \rho$, $\,y =  \lambda^2  \,\tilde y\,$, a solution $a(\rho)$ is transformed into another solution $\tilde a(\tilde \rho) =  \lambda\, a(\lambda^{-1}\tilde\rho)$. This also implies $\tilde f(\tilde \rho) = \lambda^2 \,f(\lambda^{-1}\tilde\rho)$.
The transformation has the effect of rescaling $a_3^{\rm IR}$ by $\lambda^{-2}$ and $a_5^{\rm IR}$ by $\lambda^{-4}$. So we could rescale e.g.\ $a_3^{\rm IR}$ at will without changing the five-dimensional solution. 
However, for the moment we keep $a_3^{\rm IR}$ arbitrary; later on we will tune it in order to match the UV solution described in section~\ref{UVanalysis}, where the freedom to rescale $\rho$ has already been fixed by assuming that asymptotically the solution goes like $e^{\trho}$ rather than $e^{\trho/\lambda}$ (this also gives $f\to 1$ in the UV).
 It will be convenient to trade $a_5^{\rm IR}$ for a new parameter, $\irp$, invariant under rescaling of the radial coordinate:
\be
a_5^{\rm IR} \ = \ \frac{3}{5}(1+ 3\irp)(a_3^{\rm IR})^2\,,
\ee
where the numerical factors are chosen for convenience. Thus $\irp$ parameterizes a one-parameter family of solutions.

We find that the IR solution is a double series in $\ttrho = (\sqrt{a_3^{\rm IR}}\,\trho)$ and $\irp$, of the form:
\be
a(\rho) \ = \ \frac{1}{\sqrt{a_3^{\rm IR}}}\, \sum_{n \geq 0}\,\sum_{0\leq k\leq \left[\frac{n}{2}\right]} c_{n,k} \;\irp^k\; \ttrho^{\,2n+1}\,.
\ee
We display the first few terms:
\be\label{IRsolution}
a(\rho) = \frac{1}{\sqrt{a_3^{\rm IR}}}\left[ \frac 12 \ttrho + \ttrho^{\,3} + \frac{3(1+3\irp)}{5}  \ttrho^{\,5} + \frac{6(1-3\irp)}{35}\ttrho^{\,7} +  \frac{3 +54 \irp + 119 \irp^2}{315}\ttrho^{\,9} + \mathcal{O}(\ttrho^{\,10}) \right].
\ee
If $\irp = 0$ (and choosing $a_3^{\rm IR}= 1/12$),~\eqref{IRsolution} matches the small $\rho$ expansion of the exact solution yielding AdS$_5$, which is given by~\eqref{GRsolution} with $\alpha = 1/2$. Hence switching $\irp$ on corresponds to a deformation of AdS$_5$, showing up in $a(\rho)$ at order $\mathcal{O}(\rho^5)$.

The IR solution for the function $f$ is
\be\label{solfIR}
f (\rho) \,=\, \frac{1}{12 a_3^{\rm IR}}\left[ 1- 6 \irp \ttrho^2 + 12 (2\irp+3\irp^2) \ttrho^4 -\frac{8}{5}(54\irp+167 \irp^2+  135 \irp^3) \ttrho^6 + \mathcal{O}(\ttrho^7)\right],
\ee
and takes the form of an alternate series with even powers only: for $\irp>0$, all the $\ttrho^{4n}$ terms are positive while all the $\ttrho^{4n+2}$ are negative. While we omit the expression of $\Psi$, it is instructive to look at the form of the five-dimensional metric in the IR:
\bea
\diff s^2 \!&=&\! 12\left(1+6\irp\, \ttrho^2\right)\diff\ttrho^2 + 12 \ttrho^2\frac{\sigma_1^2 +\sigma_2^2 + \sigma_3^2}{4} + \frac{(2+\irp)\sq^2 -24 a_3^{\rm IR}}{2 a_3^{\rm IR}\sq^2} \ttrho^2 \diff t\,\sigma_3\nn\\ [2mm]
\!\!&&\!\! - \left[\frac{1}{144(a_3^{\rm IR})^2} - \frac{144(a_3^{\rm IR})^2 + \irp \sq^4 - 12 (2+\irp) a_3^{\rm IR} \sq^2 }{12 (a_3^{\rm IR})^2\sq^4}\ttrho^2\right] \diff t^{\,2} + \ldots\,.
\label{favelesse}
\eea
Here, both the UV parameter $\sq$ and the IR parameter $a_3^{\rm IR}$ are functions of $\irp$. The former appears because we are performing the change of coordinates~\eqref{changepsi} setting the boundary metric in a direct product form. We know that when $\irp = 0$, then $\sq = 1$ and $a_3^{\rm IR} = 1/12$. In the next section we will determine both $\sq(\irp)$ and $a_3^{\rm IR}(\irp)$ at linear order in $\irp$. We will see that the relation is such that at first order in $\irp$ the mixed term $\diff t\, \sigma_3$ disappears.
We see that in the limit $\rho \to 0$, there is a \emph{round} $S^3$ shrinking to zero, while the $g_{tt}$  component of the metric remains finite. This implies that in a neighbourhood of $\rho=0$ the solution is non-singular and the topology of the space-time is 
$\mathbb R^{1,4}\simeq \mathbb{R}_t\times \mathbb{R}^4$.
The gauge field is also smooth as can be seen from its IR expansion 
\be
A \ = \ \frac{\sqrt 3}{2} \left[\left(\frac{1}{12 a_3^{\rm IR}} - \frac{2}{3\sq^2} + \frac{ 12a_3^{\rm IR} -
\sq^2}{2a_3^{\rm IR} \sq^2}\irp\ttrho^2  \right)\diff t - 3 \irp \ttrho^2 \sigma_3  + \mathcal O(\ttrho^3) \right]\,.
\label{favealburro}
\ee
In particular, notice that while the $A_t$ component is finite in the  $\rho \to 0$ limit, the $A_\psi$ component vanishes smoothly at the origin of $\mathbb{R}^4$. 

The linearised solution to be discussed next will demonstrate that the solution is smooth for any finite value of $\rho$, at least when the deformation parameter $\xi $ is small, while the numerical analysis of section~\ref{numerics_sec} will provide evidence that this persists for all the allowed values of the deformation parameter.

\subsection{Linearised solution}
\label{linearsec}

In this section, we study the solution analytically at first order in an expansion around global AdS$_5$. This establishes the existence of a regular solution connecting the UV and IR asymptotics, at least when this is a small perturbation around AdS$_5$.

For small values of the parameter $\irp$, our solution is a perturbation of the AdS$_5$ solution $a(\rho) = \frac{1}{2}\sinh{\trho}$. We can thus linearise the sixth-order equation around this background and solve for the perturbation. Of the six integration constants that arise, we find that three are fixed by regularity at $\rho = 0$ and by the requirement $a(0) = 0$. A fourth one is determined by imposing that at large $\rho$ the solution diverges as $a \sim e^{\trho} $ (specifically, this removes a term going as $\trho\, e^{\trho}$). So at the linearised level we obtain a two-parameter family of solutions deforming AdS$_5$. However, the IR analysis in the previous section demonstrates that this cannot hold at the non-linear level, where only two distinct one-parameter families of solutions survive, the first being the one of~\cite{GutowskiReall}, and the other being the new family parameterized by $\irp$. In order to determine the latter analytically 
at the linearised level, we match the two-parameter linearised solution with the IR solution parameterized by $\irp$. As we have imposed that the linearised solution goes like $a \sim e^{\trho}$ at large $\rho$, the comparison also fixes the IR parameter $a_3^{\rm IR}$ (see the discussion in the previous section). We find:
\be\label{rela3IRwithxi}
a_3^{\rm IR} = \frac{1}{12}(1 - \irp) + \mathcal{O}(\irp^2)\,.
\ee
In this way, we obtain the solution
\be\label{solaFirstOrderxi}
a(\trho) \ = \ \frac{1}{2}\left(1+ \frac{\irp}{4}\right)\sinh\trho + \frac{1}{2}\irp\cosh\trho\left[  \coth^3\trho\:\log\cosh\trho - \trho - \frac{1}{\sinh(2\trho)}\right] + \mathcal{O}(\irp^2)\,.
\ee
It follows a particularly simple expression for $f$:
\be
f \ = \ 1 + \frac{2\irp \log\cosh\trho}{\sinh^2\trho} + \mathcal{O}(\irp^2)\,,
\ee
while the one for $\Psi$ is:
\be
\Psi =  -\frac{1}{2}\left(1+\frac{\irp}{2}\right) \sinh^2\trho - \irp\left[ 1 - \coth^2\trho\left(2 -\sinh^2\trho\right)\log\cosh\trho - \frac 12\trho \sinh(2\trho) \right] + \mathcal O(\irp^2).
\ee
We also provide the corresponding expression of the five-dimensional supergravity fields. In the coordinates $(t,\theta,\phi,\psi,\rho)$ bringing the boundary metric to the direct product form $\mathbb R \times S^3_{\sq}$, the bulk metric is
\bea\label{metric5dlinearxi}
\diff s^2 \!\!\!&=&\!\!\! \left[ 1 - 2\irp \frac{\log \cosh\trho}{\sinh^2\trho} \right] \diff \trho^2 - \cosh^2 \trho \left[ 1 + 2\irp  \left( 1+ \log\cosh\trho - \trho \tanh\trho \right)\right]\diff t^2\nn \\ [2mm]
\!\!\!&+&\!\!\!\!   \frac{1}{4}\!\left[ \left(1+\frac{\irp}{2}\right)\sinh^2\!\trho - \irp\left( 1 - 2 \left( \cosh^2\!\trho + \frac{1}{\sinh^{2}\!\trho} \right) \log\cosh\trho   + \trho \sinh(2\trho)\right) \right]\!\!\left(\sigma_1^2+\sigma_2^2\right)\nn \\ [2mm]
\!\!\!&+&\!\!\!   \frac{1}{4}\!\left[ \left(1- \irp \right)\sinh^2\trho - \irp \left( -2 - 2 \left( \cosh^2\!\trho -\frac{2}{\sinh^{2}\!\trho} \right) \log\cosh\trho   + \trho \sinh(2\trho)\right) \right]\sigma_3^2 \nn \\ [2mm]
\!\!\!&+&\!\!\!  \mathcal O(\irp^2)\,,
\eea
while the gauge field reads
\be\label{A5dlinearxi}
A \ = \ \frac{1}{2\sqrt 3}\diff t - \frac{\sqrt 3}{4}\,\irp \left[ 1 -\frac{  2 \log\cosh \trho}{\sinh^2 \trho}\right] \sigma_3 \,+\, \mathcal O(\irp^2)\,.
\ee
These solve the equations of motion \eqref{EinsteinEq} and \eqref{MaxwellEq} at order $\mathcal O(\irp)$. Moreover, one can see that near $\rho \to 0 $ they
reduce to the metric (\ref{favelesse}) and gauge field (\ref{favealburro}), as required.

For $\irp =0$, we correctly retrieve the AdS$_5$ metric in global coordinates,
\be
\diff s^2_{\rm AdS}\ = \ \diff\rho^2 - \cosh^2\!\rho \,\diff t^2 + \frac{1}{4}\sinh^2\! \rho \left(\sigma_1^2 +\sigma_2^2 + \sigma_3^2\right)\,,
\ee
together with a flat gauge field $A= \tfrac{1}{2\sqrt 3}\diff t $.\footnote{Note that it is important to keep track of this gauge field, because it is for this choice of gauge \emph{only} that the Killing 
spinors are independent of $t$.} Since $F = \mathcal O(\irp)$, from the Einstein equation~\eqref{EinsteinEq} we see that the Ricci tensor of the metric~\eqref{metric5dlinearxi} satisfies
$R_{\mu\nu} = -4 g_{\mu\nu} + \mathcal O(\irp^2)\,,$
so at linear order in $\irp$ this is still an Einstein space, of the same Einstein constant as AdS$_5$. However, the Riemann tensor is different from the one of AdS$_5$, and matches it only asymptotically, as required for an AlAdS space. We also checked that the five-dimensional space is not conformally flat, as the Weyl tensor of~\eqref{metric5dlinearxi} is proportional to $\irp$.
 
The boundary metric reads
\bea
\diff s^2_{\rm bdry} &=& - \frac{1}{4} \left[ 1+ 2 \irp ( 1-\log 2) \right]\diff t^2 + \frac{1}{16} \left[1+\frac{\irp}{2}( 1-4\log 2)\right]\left(\sigma_1^2+\sigma_2^2\right) \nn \\[2mm]
&&\!\!\!\!+\, \frac{1}{16}\left[1- \irp (1+2\log 2)\right]\sigma_3^2 \,+\, \mathcal O(\irp^2)\,.
\eea

Finally, expanding~\eqref{solaFirstOrderxi} at large $\trho$ and comparing with the UV perturbative solution~\eqref{UVsola}, we can determine the relation between the UV parameters $a_0, a_2, a_4, a_6, c$ and the IR parameter $\irp$, at linear order in $\irp$. We find:
\bea
a_0 \!&=&\! \frac 14 +\frac{\irp}{16}\left(1 - 4 \log 2 \right) + \mathcal{O}(\irp^2)\,, \qquad a_2 \ =\ - \frac{1}{16} - \frac{3\,\irp}{32}\left( 1 + 4 \log 2 \right) + \mathcal{O}(\irp^2)\,,\nn\\ [2mm]
\qquad a_4 \!&=&\! \frac{3\,\irp}{32}\left( \frac{3}{16} - \log 2 \right)+ \mathcal{O}(\irp^2)\,,\qquad\quad\, a_6 \ = \ \frac{\irp}{512}\left( \frac{113}{48} - 7\log 2 \right)+ \mathcal{O}(\irp^2)\,,\nn \\
c \!&=&\! \frac{3}{8}\,\irp + \mathcal{O}(\irp^2)\,. \label{UVparFromIRpar}
\eea

It might be possible to solve the sixth-order equation beyond linear order, for example expressing the higher order terms in the perturbative expansion in a recursive fashion, similarly to the expansion of the solution of~\cite{Butti:2004pk} obtained in~\cite{Gaillard:2010qg}. In this respect, the linearised solution of this section is analogous the solution found in~\cite{Gubser:2004qj}.


\subsection{Numerics}

\label{numerics_sec}

In this section we present a numerical study of the one-parameter family of solutions. Our primary scope is to show the existence of a regular solution connecting the IR and UV asymptotics beyond linear level in the deformation. Secondly, we wish to determine how the IR parameter $\irp$ and the UV parameters $a_0$, $a_2$, $a_4$, $a_6$, $\sq^2 = 1-4\cL$ are related beyond the small $\irp$ approximation leading to~\eqref{UVparFromIRpar}.

Let us briefly describe our procedure. We fix the IR initial conditions around $\rho = 0$ as in section~\ref{sec:IRsolution} and integrate the sixth-order equation numerically towards the UV. To do this, we have to assign a value to the IR parameters $\irp$ and $a^{\rm IR}_3$. As discussed in section~\ref{sec:IRsolution}, only $\irp$ is physical, while $a^{\rm IR}_3$ can be changed by rescaling the radial coordinate, and will be fixed by requiring that asymptotically the solution goes like $a\sim e^{\rho}$. At linear order in $\irp$ this is achieved by imposing~\eqref{rela3IRwithxi}, but beyond that we do not have an analytic expression and need to resort to numerics. Recalling the discussion in section~\ref{sec:IRsolution}, $a^{\rm IR}_3$ can be fixed as follows: for any choice of $\irp$, we integrate a first time choosing the reference value $a^{\rm IR}_3=1/12$;\footnote{This is the value appropriate to the AdS solution $\xi = 0$, yielding $a = \frac{1}{2} \sinh(\rho)$ and therefore $a \sim e^{\rho}$ 
asymptotically.} then 
from the UV behavior of the numerical solution we read off the rescaling $\lambda$ to be performed so that $a \sim e^\rho$; this is most easily achieved by evaluating the function $f$ and finding the rescaling $\lambda^2$ ensuring that it goes to 1 asymptotically. Then $a^{\rm IR}_3=1/{(12\lambda^2)}$ will produce the wanted UV behavior; thus we fix this value and repeat the numerical integration. 
The results we will present have this procedure implemented. Recalling~\eqref{solfIR}, the final value of $a_3^{\rm IR}$ can be read in the plots below from the behavior of $f$ at $\rho = 0$.

\begin{figure}[tp]
\centering
\includegraphics[width=0.6\textwidth]{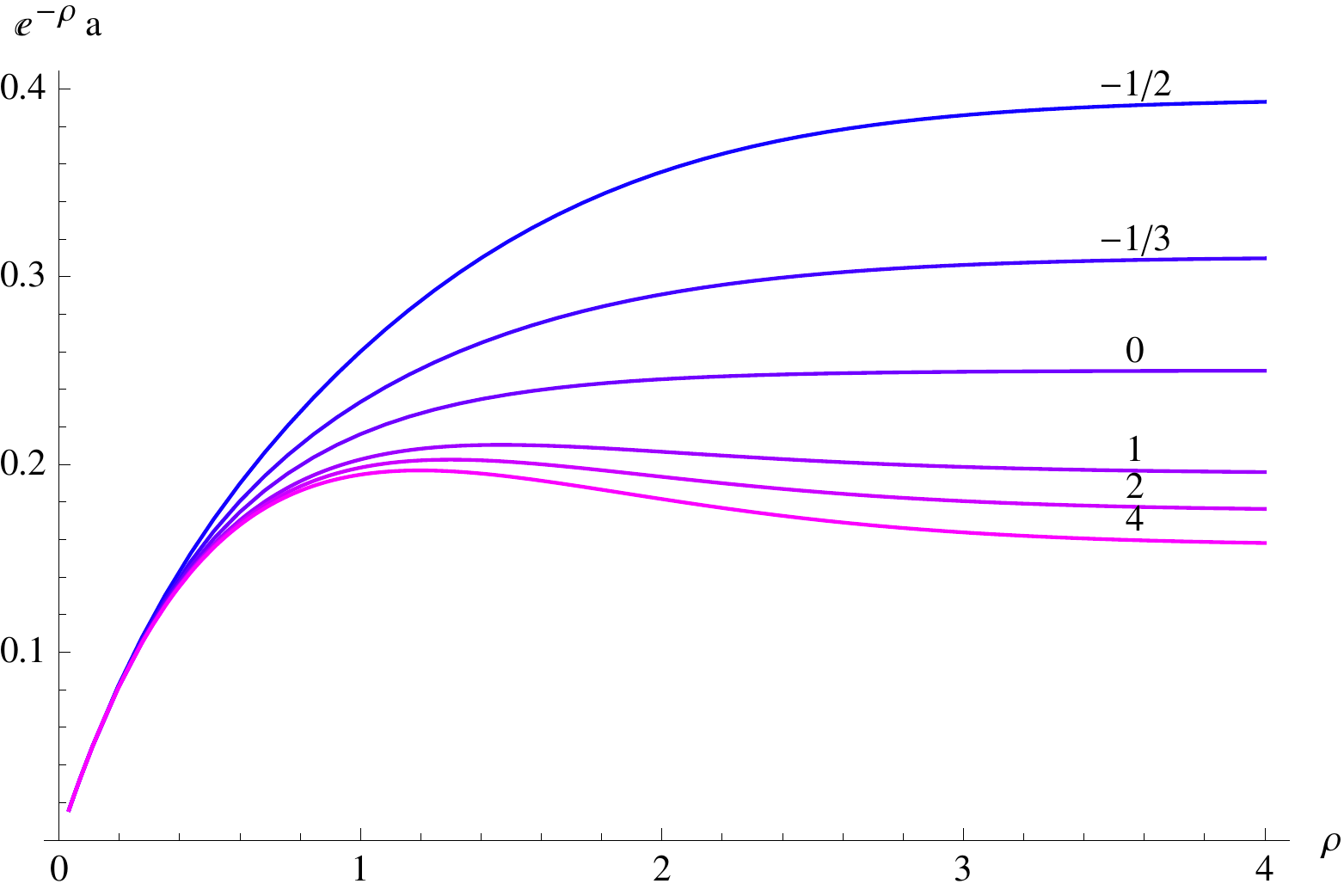}
\caption{The solution $a$ to the sixth-order equation, rescaled by $e^{-\rho}$. Its square equals $g_{\theta\theta}/g_{\rho\rho}\,$. Asymptotically, it gives the parameter $a_0$, controlling the overall size of the boundary. The different values of the IR parameter $\irp$ are indicated on the curves.}\label{fig:aPlot}
\end{figure}
\begin{figure}[htp]
\centering
\includegraphics[width=0.6\textwidth]{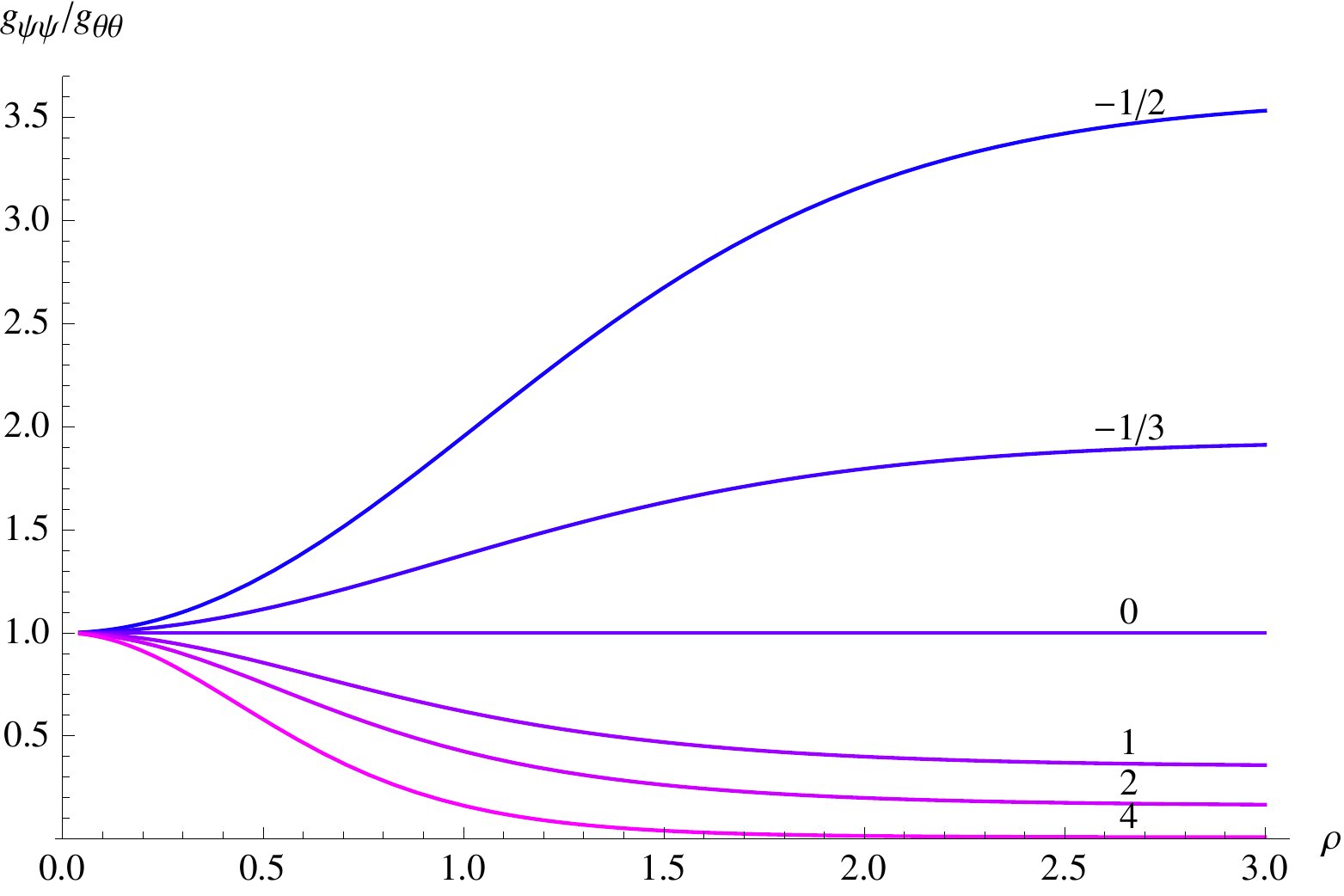}
\caption{Ratio between the $g_{\psi\psi}$ and $g_{\theta\theta}$ components of the metric. Asymptotically, this gives the value of the parameter $\sq^2$, controlling the squashing of the boundary $S^3$.}\label{fig:squashingPlot}
\end{figure}

\begin{figure}
     \centering
     \subfigure[The function $f$. This is equal to $g_{\rho\rho}^{-1}$.]{
          \label{fig:fPlot}
          \includegraphics[width=.48\textwidth]{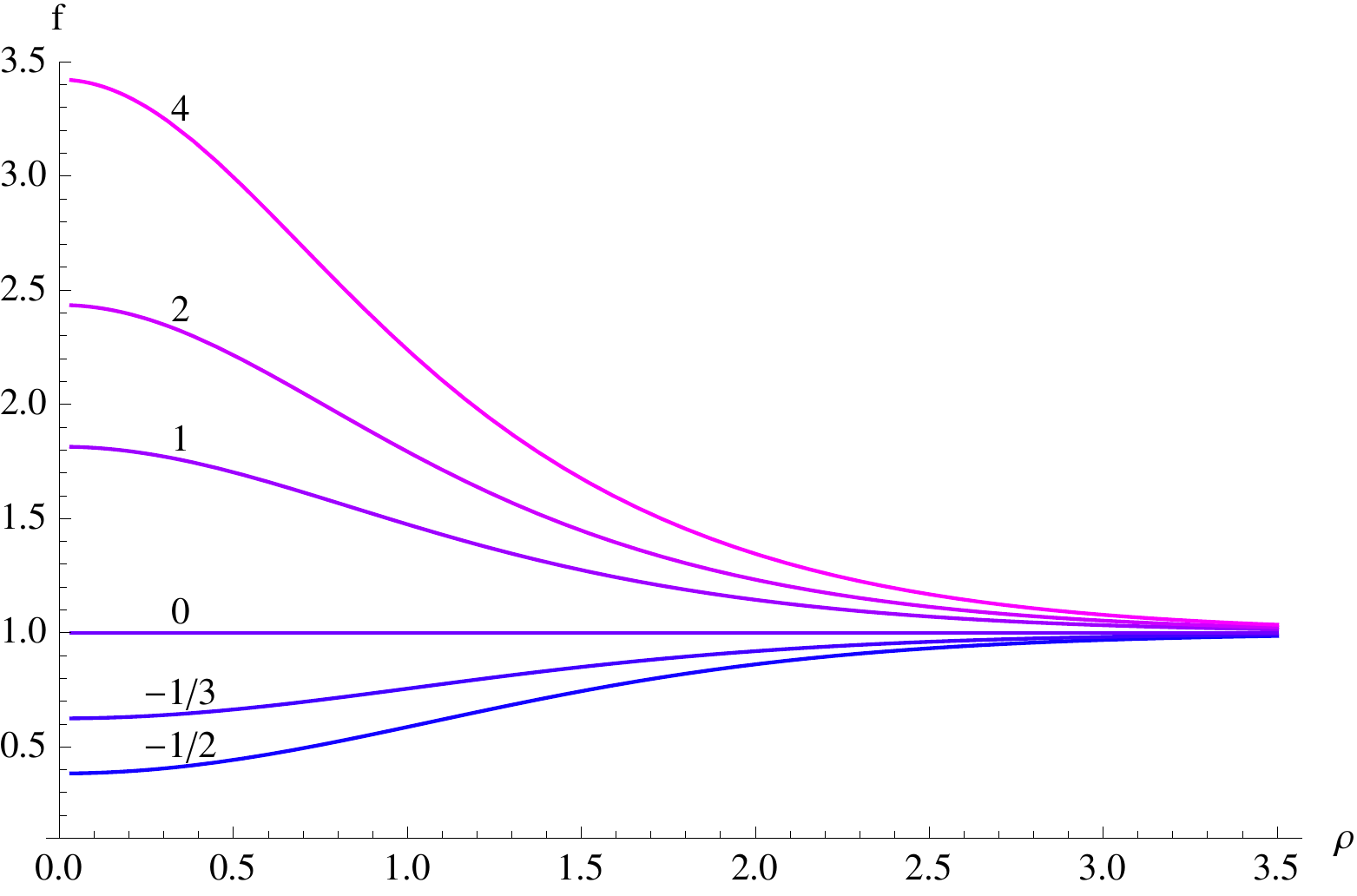}}
     \hspace{1mm}
     \subfigure[$g_{tt}$ component, rescaled by $e^{-2\rho}$.]{
          \label{fig:gttPlot}
          \includegraphics[width=.48\textwidth]{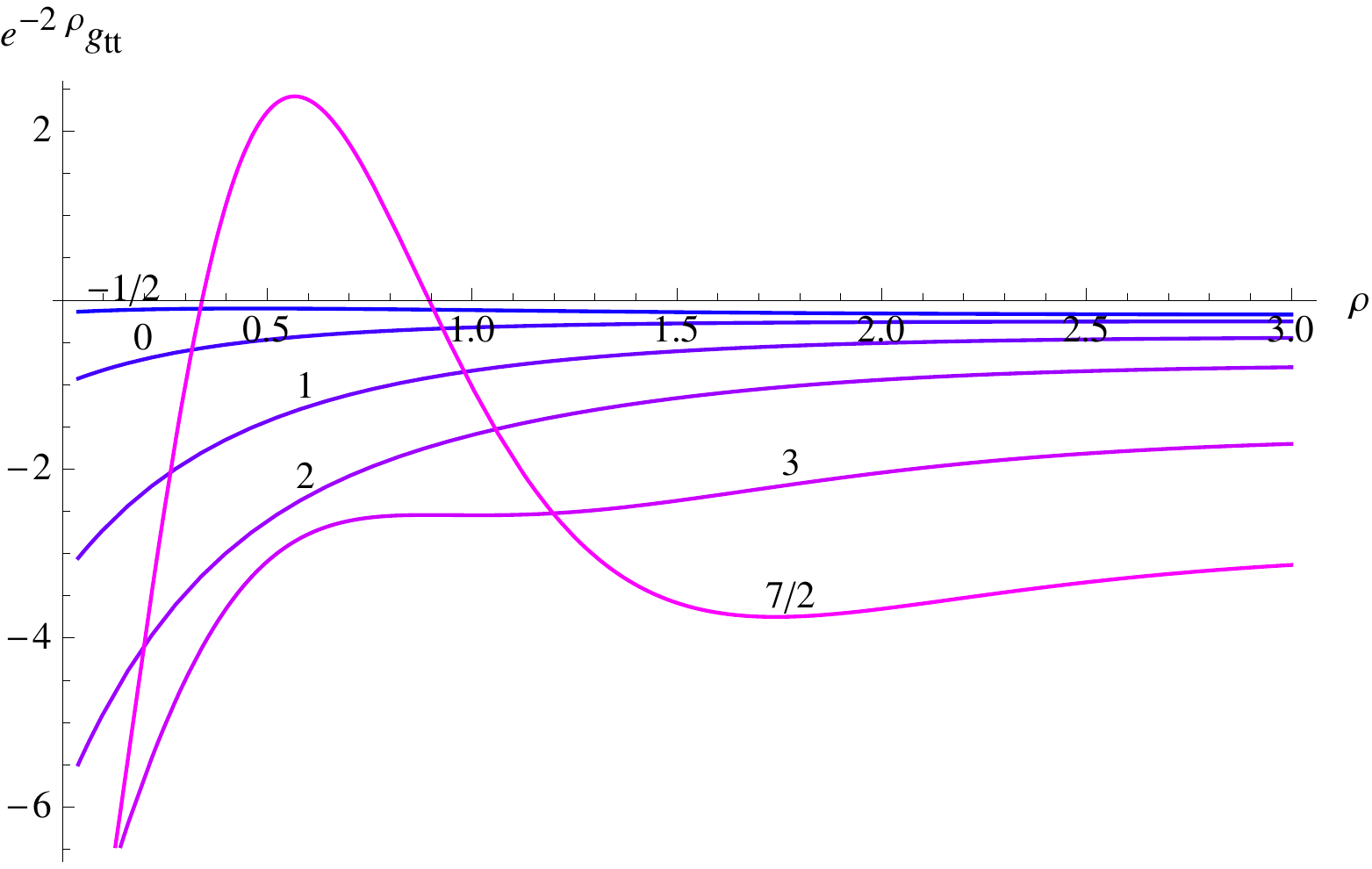}}
 \\    \vspace{5mm}
     \subfigure[$A_\psi$ component.]{
          \label{fig:ApsiPlot}
          \includegraphics[width=.48\textwidth]{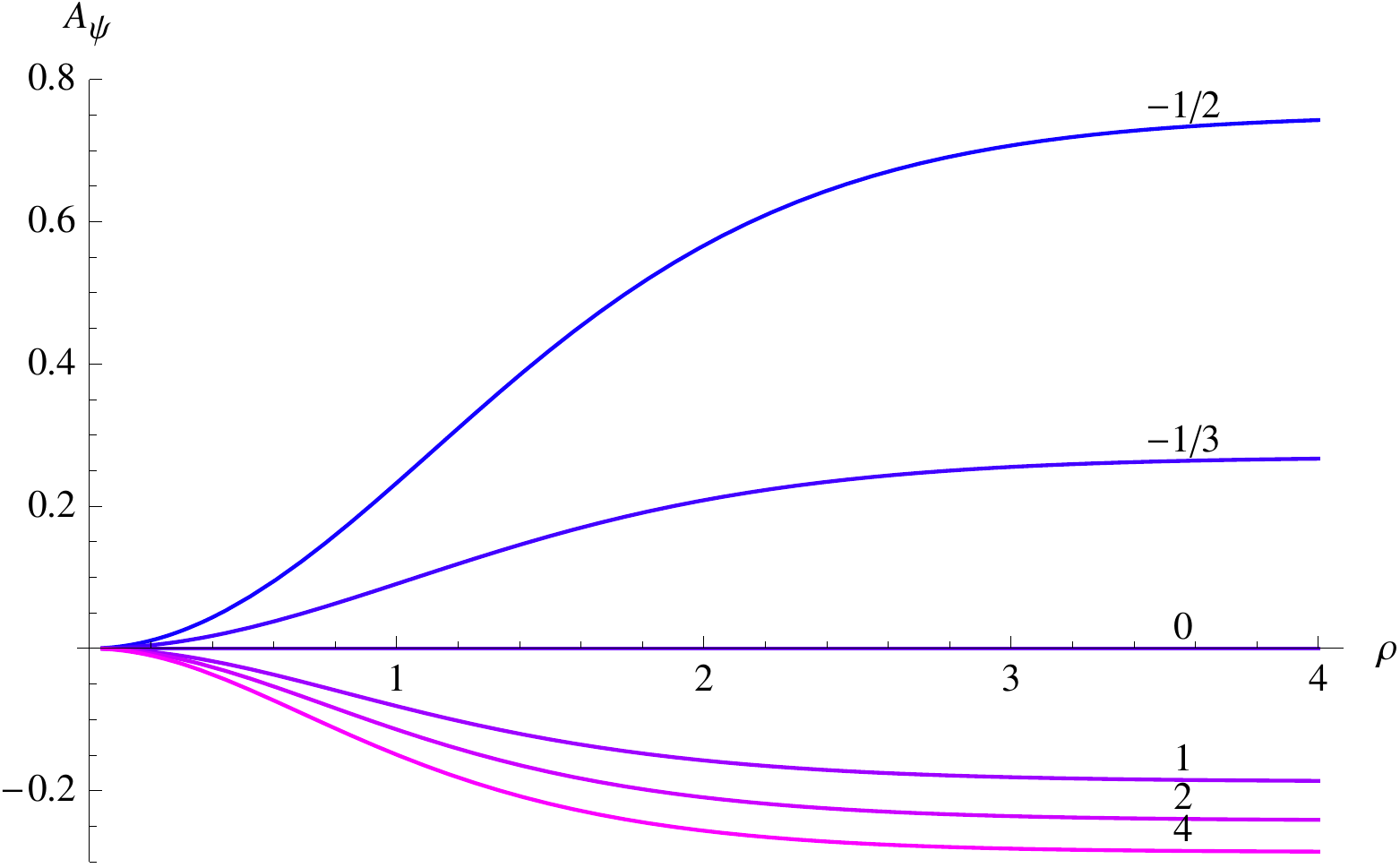}}
     \hspace{1mm}
     \subfigure[$A_t$ component.]{
           \label{fig:AtPlot}
           \includegraphics[width=.48\textwidth]{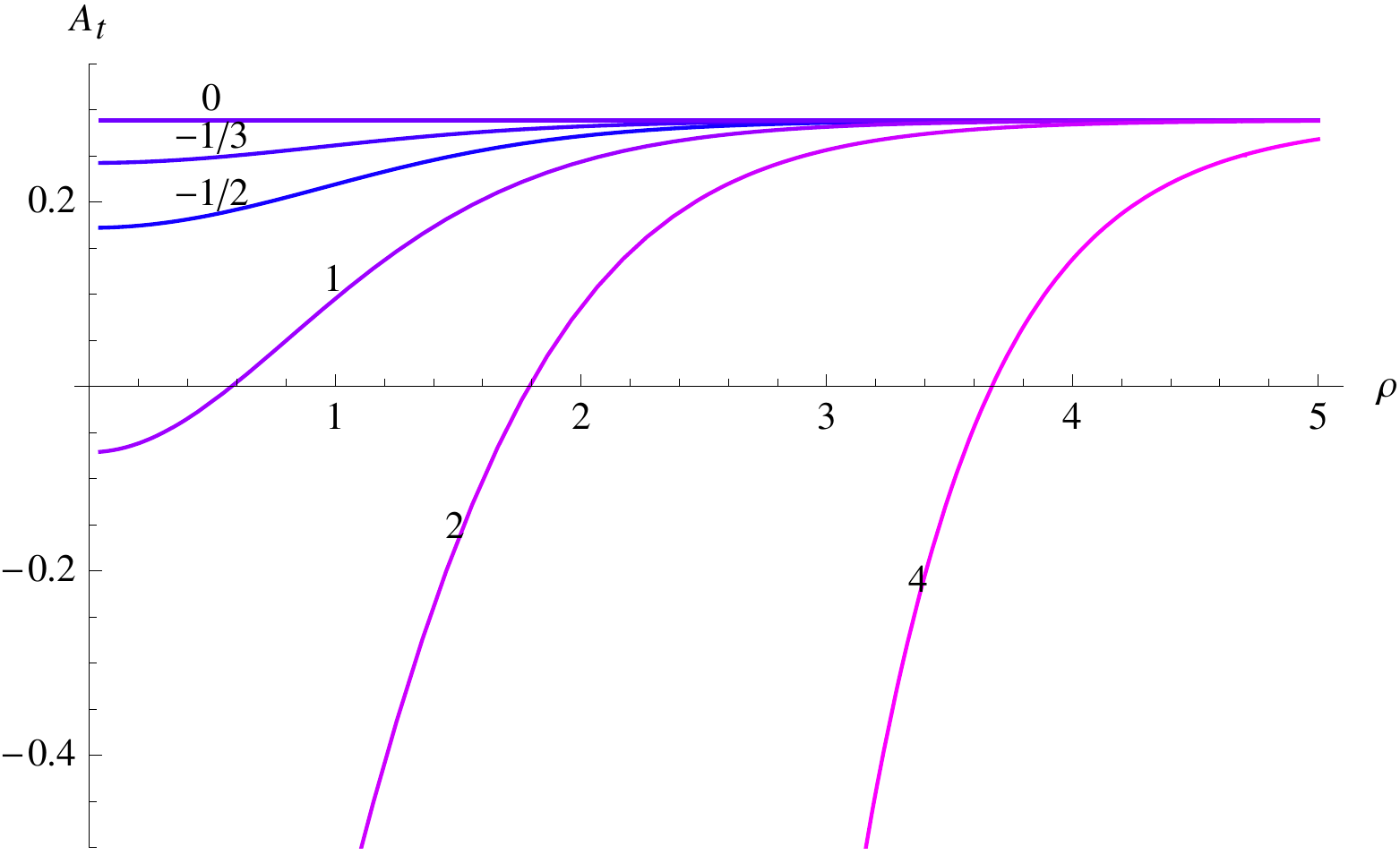}}
     \caption{Other metric components and the gauge field $A$.}
     \label{fig:differentplotsMetricAndA}
\end{figure}

\begin{figure}[htp]
\centering
\includegraphics[width=0.58\textwidth]{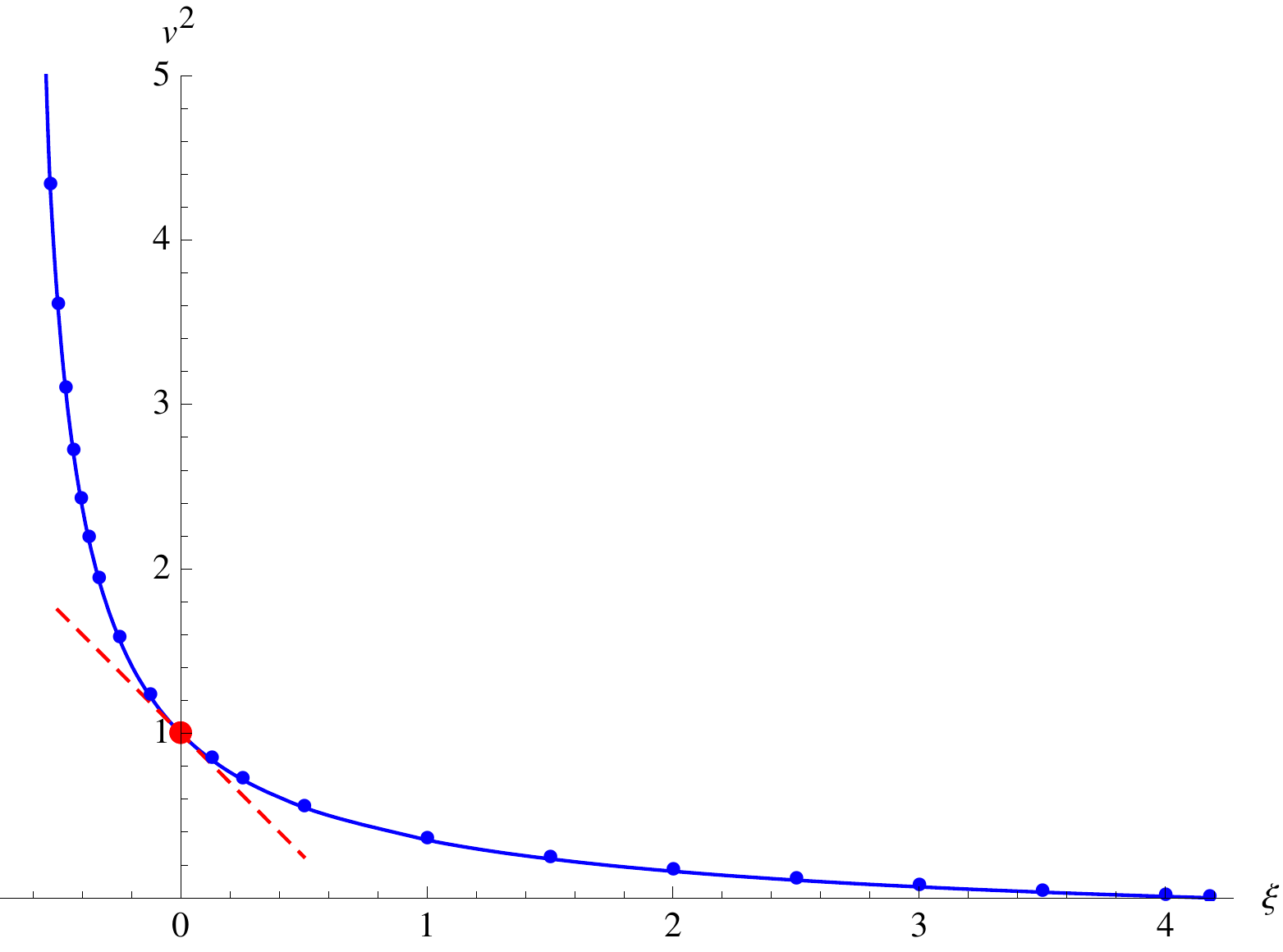}
\caption{Relation between the IR parameter $\irp$ and the UV parameter $\sq^2$ controlling the squashing of the boundary $S^3$. The squashing ranges between 0 and $\infty$ for \hbox{$4.2 \gtrsim \xi \gtrsim -0.7$.} The dots are effectively calculated values while the blue, continuous line is an interpolation. The red, dashed line represents the relation \eqref{UVparFromIRpar} obtained from the linearised analysis around the AdS solution at $\irp =0$ (which is denoted by the slightly larger, red dot).}\label{fig:PlotVsquaredVsXi}
\end{figure}

\begin{figure}
     \centering
     \subfigure[The boundary size parameter $a_0$.]{
          \label{fig:Plota0VsVsquared}
          \includegraphics[width=.48\textwidth]{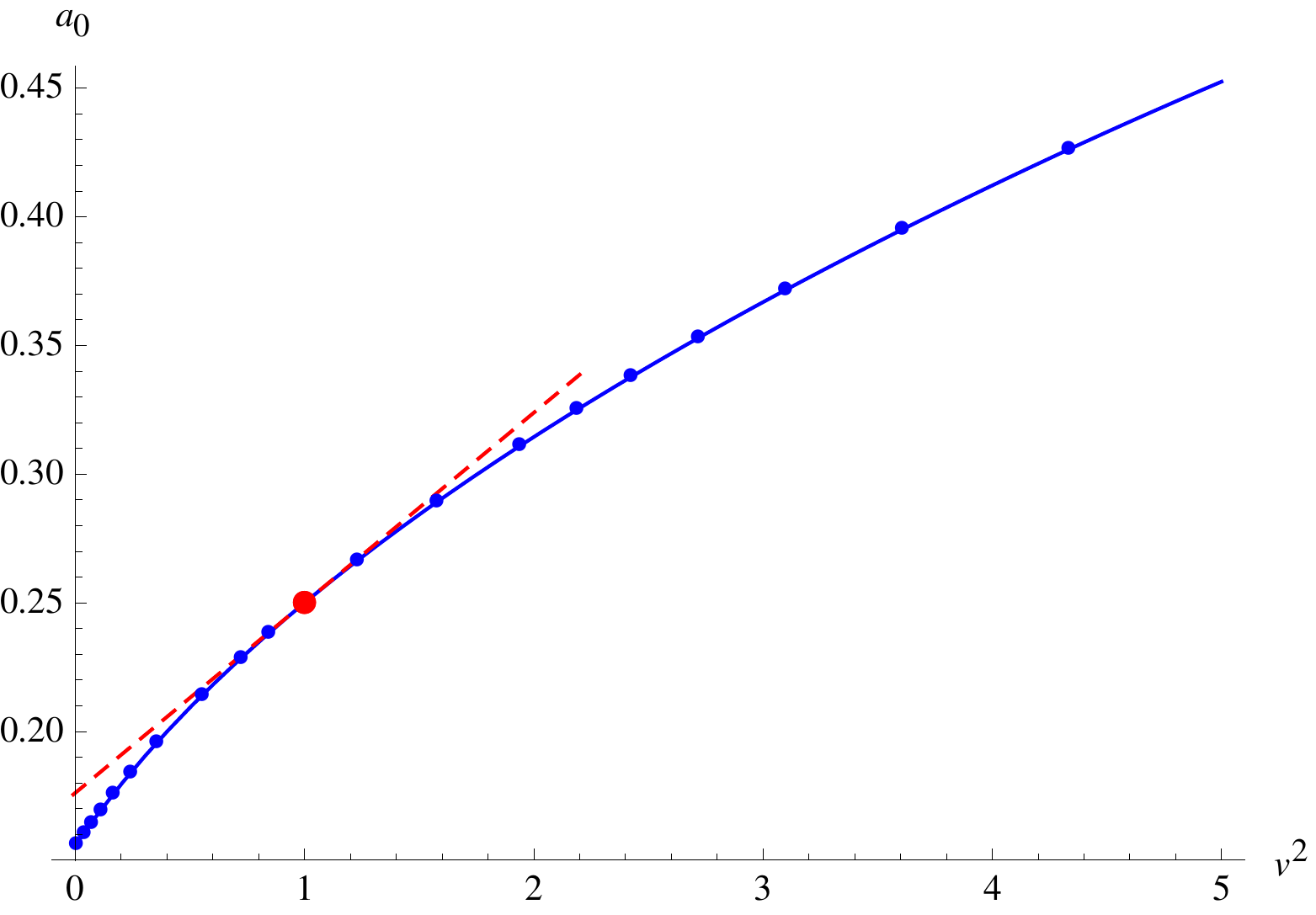}}
     \hspace{1mm}
     \subfigure[The parameter $a_2$.]{
          \label{fig:Plota2VsVsquared}
          \includegraphics[width=.48\textwidth]{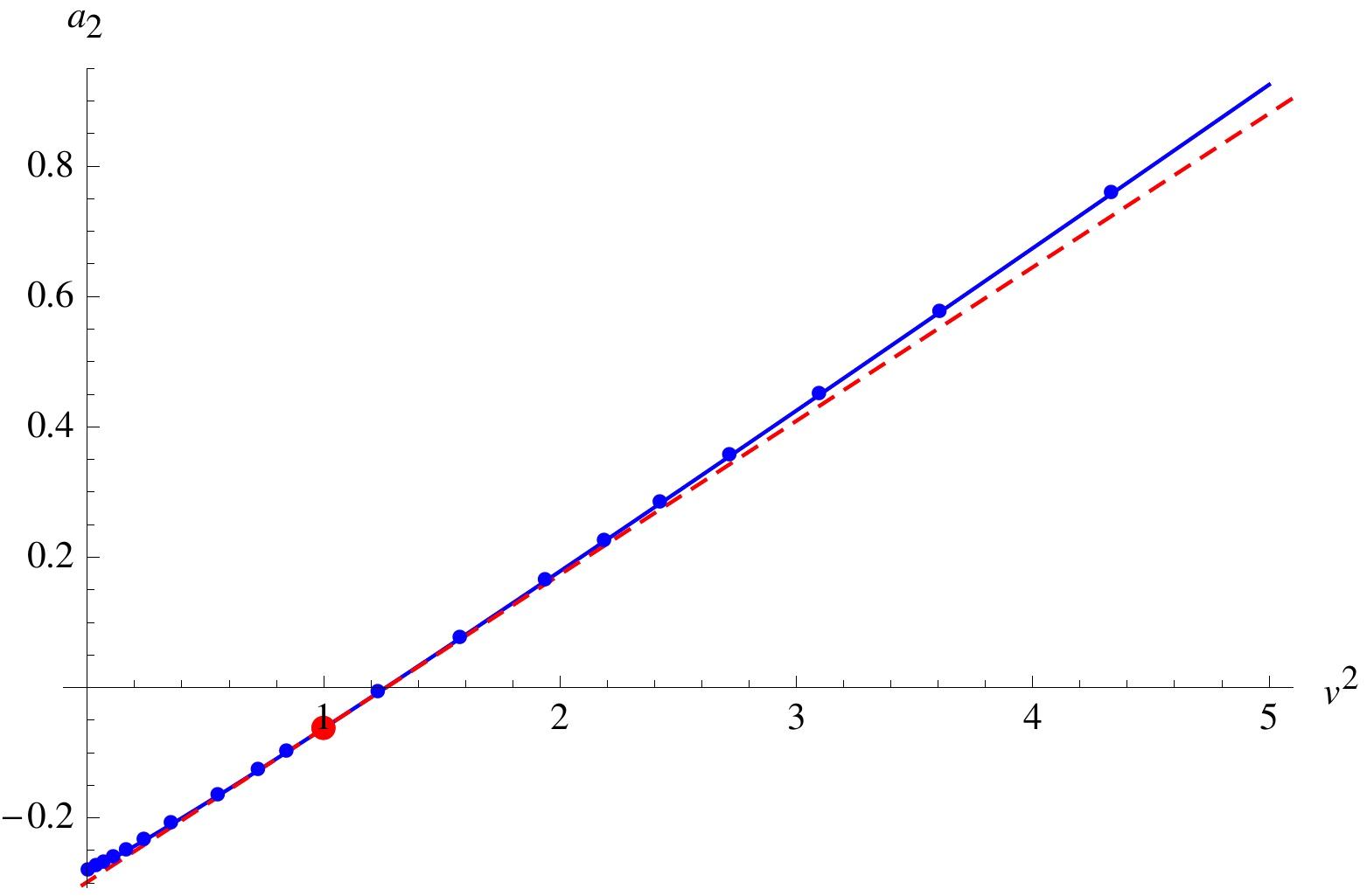}}\\
     \vspace{5mm}
     \subfigure[The parameter $a_4$.]{
           \label{fig:Plota4VsVsquared}
           \includegraphics[width=.48\textwidth]
                {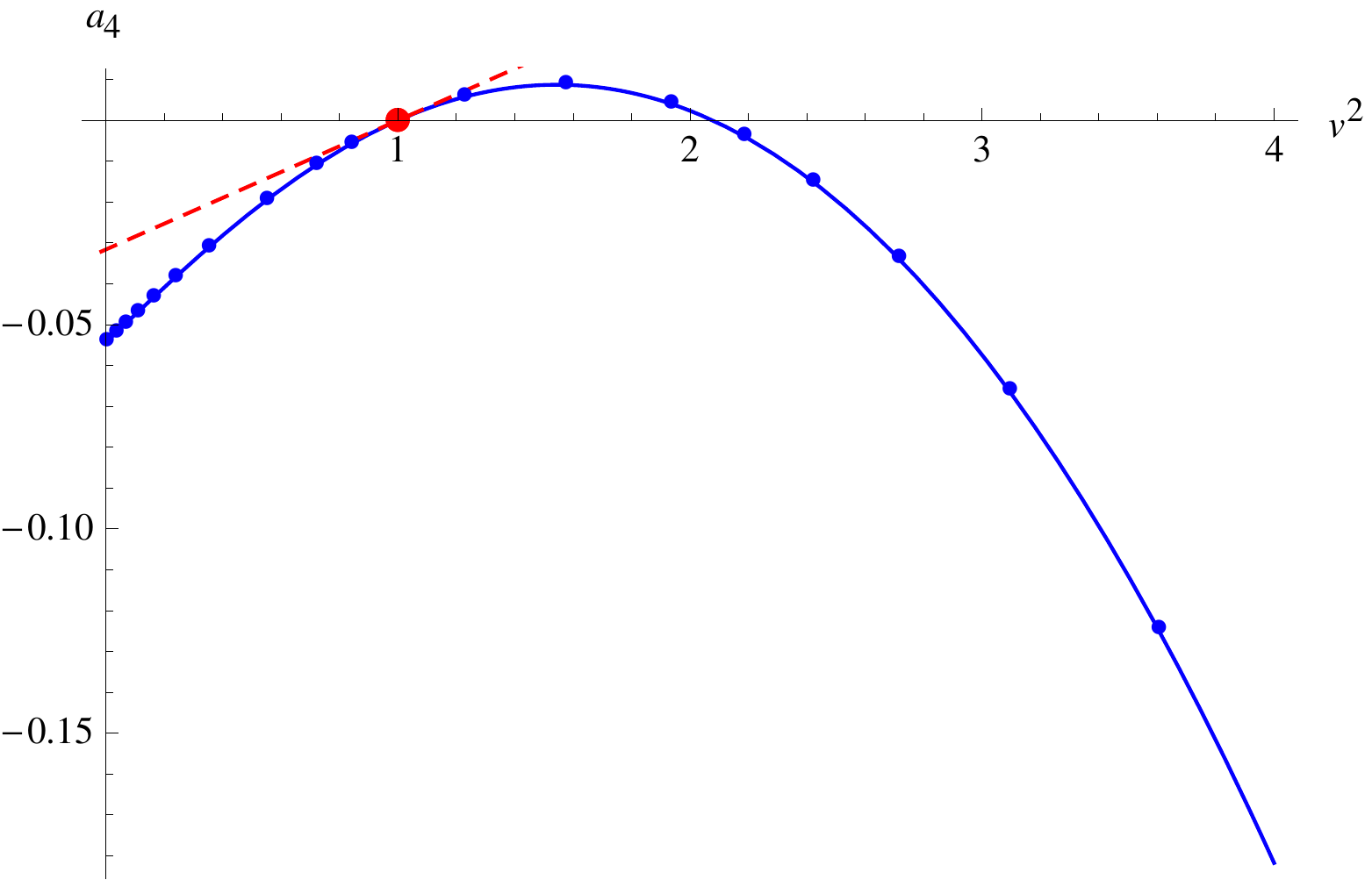}}
      \hspace{1mm}
     \subfigure[The parameter $a_6$.]{
           \label{fig:Plota6VsVsquared}
          \includegraphics[width=.48\textwidth]{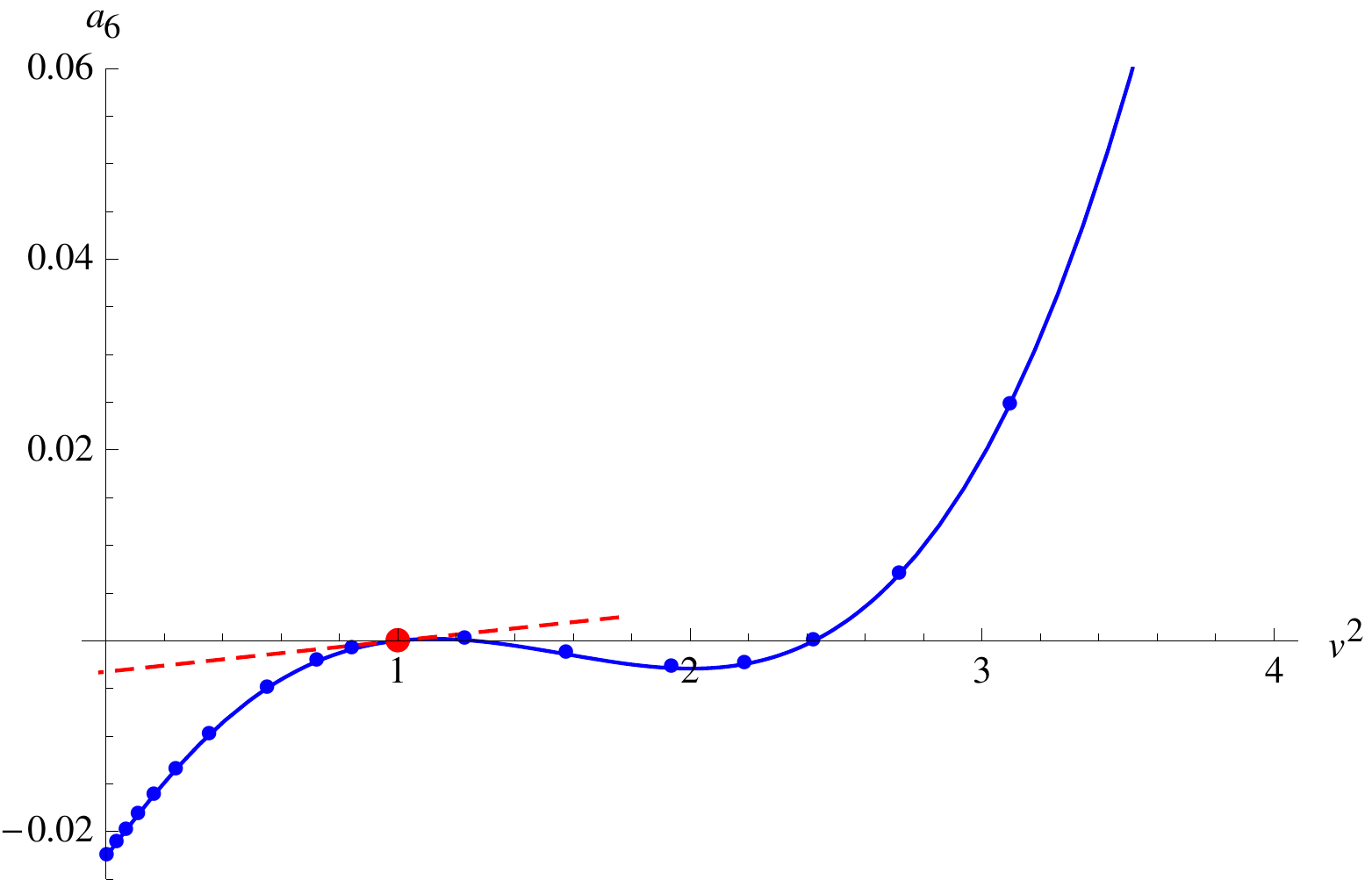}}
     \caption{The other UV parameters in terms of the squashing $\sq^2$.}
     \label{fig:otherUVparametersPlots}
\end{figure}

In figure~\ref{fig:aPlot} we display the solution $a(\rho)$ to the sixth-order equation, for different choices of the IR parameter~$\irp$. Figures~\ref{fig:squashingPlot} and \ref{fig:differentplotsMetricAndA} provide further plots representing components of the metric, written in the form 
\be
\diff s^2 \ = \ g_{\rho\rho} \diff \rho^2 + g_{\theta\theta}\left( \sigma_1^{\,2} + \sigma_2^{\,2} \right) + g_{\psi\psi}\sigma_3^{\,2} + g_{tt}\diff t^2 + 2 g_{t\psi} \,\sigma_3\, \diff t\,,
\ee
and of the gauge field
\be
A \ = \ A_t \,\diff t + A_\psi\, \sigma_3\,.
\ee
These are evaluated using~\eqref{MetricCompsImplicit} and~\eqref{AcompImplicit}.\footnote{In order to evaluate $g_{tt}$ and $A_t$ we also need to know the value of the UV parameter $\sq^2$ as a function of $\irp$; this is determined as we will explain.} As one can see, the solution appears perfectly smooth within the numerical approximation.

Given a value of the IR parameter $\irp$, we then proceed to compute the parameters $a_0$, $a_2$, $a_4$, $a_6$ and $\sq^2 =1-4\cL$ controlling the UV solution. This is done by matching the numerical solution $a$ obtained integrating from the IR with the UV solution given in~\eqref{UVsola}, for some reasonably large value of the radial coordinate $\rho$. 
In practice, we compare the numerical and the UV solutions at several points in the interval $4 < \rho  <8 $, and  determine the UV parameters via a best fit.
The results are presented in figures~\ref{fig:PlotVsquaredVsXi} and~\ref{fig:otherUVparametersPlots}, where a comparison with the expressions \eqref{UVparFromIRpar} from the linearised analysis around the AdS solution at $\irp=0$ is also made.
Figure~\ref{fig:PlotVsquaredVsXi} displays the relation between the squashing parameter $\sq^2$ and $\irp$. One can see that for $\irp$ running between $\xi \sim 4.2$ and $\xi \sim -0.7$, the squashing spans all values between $0$ and $\infty$. From an AdS/CFT perspective, it is actually more natural to regard the family of solutions as parameterized by the boundary parameter $\sq^2$ rather than by the IR parameter~$\irp$.
For this reason, figure~\ref{fig:otherUVparametersPlots} gives the other UV parameters as functions of $\sq^2$. We recall that although here $a_0$ is determined by $\sq^2$, this is only because for ease of analysis we imposed that the solution ends at $\rho = 0$; in fact, $a_0$ can be set to any desired non-zero value simply by shifting~$\rho$.

Relevant physical properties of the solution can be inferred from the figures.
Fig.~\ref{fig:squashingPlot} allows to exclude the existence of closed timelike curves. These appear whenever the $g_{\psi\psi}$ component of the metric, evaluated from~\eqref{MetricCompsImplicit}, becomes negative. However, from fig.~\ref{fig:squashingPlot} we see that as long as $g_{\psi\psi}$ is positive asymptotically,  it remains positive in the interior as well.\footnote{The figure actually represents the ratio $g_{\psi\psi}/g_{\theta\theta}$, but $g_{\theta\theta}$ cannot be negative, see~\eqref{MetricCompsImplicit}.} Imposing this UV boundary condition is the same as requiring that the boundary metric has one negative eigenvalue only, and is very natural from an AdS/CFT perspective. We conclude that within this assumption the solution is free from closed timelike curves.

Moreover, note from fig.~\ref{fig:gttPlot} that for sufficiently large $\irp$ there is a region in the bulk where the vector $\partial/\partial t$, generating time translations with respect to the asymptotic rest frame, becomes spacelike. Hence for sufficiently large $\irp$ the solution presents an ergoregion although there is no horizon.\footnote{See e.g.~\cite{Giusto:2004id,Jejjala:2005yu} for examples of supersymmetric and non-supersymmetric asymptotically flat geometries with an ergoregion and no horizon.} On the other hand, we should note  that the supersymmetric vector
\be\label{susyvector}
V\ =\ \frac{\partial}{\partial y} \ =\ \frac{\partial}{\partial t} + \frac{2}{\sq^2}\frac{\partial}{\partial \psi }\,,
\ee associated to time translations with respect to a boosted asymptotic frame, is everywhere timelike in the bulk (this follows from eq.~\eqref{generalggmetric} and fig.~\ref{fig:fPlot}); so  if $V$ generates time translations then there is no ergoregion. However, $V$ is null with respect to the boundary metric, so from an AdS/CFT perspective one would rather take $\partial/\partial t$ as the generator of time translations both in the boundary and in the bulk. Similar features appear in the asymptotically AdS black hole of~\cite{GutowskiReall}.


\subsection{A global constraint}\label{sec:globalconstraint}

The IR analysis has established a one-parameter family of solutions, labeled by the IR parameter $\irp$. The UV parameters controlling the asymptotic behavior of the solution cannot remain arbitrary, and should be related between them and to~$\irp$. In section~\ref{linearsec} we worked out this relation analytically at linear order in $\irp$, while in section~\ref{numerics_sec} we studied it numerically. In the following, we make analytic progress beyond linear level by showing that a global property satisfied by our solution allows to determine one of the UV parameters in terms of the others. 

Let us consider the following integral on the three-sphere at infinity, $S^3_{\rm bdry}$,
\be\label{PageCharge}
\int_{S^3_{\rm bdry}} \Big(*_5 \! F + \frac{2}{\sqrt 3} A \wedge F\Big)\,.
\ee
Note that  $S^3_{\rm bdry}$ is the boundary of a Cauchy surface (namely, a hypersurface at constant $t$) $\mathcal C\simeq \mathbb{R}^4$ that closes off smoothly in the interior; using the IR expansions of section~\ref{sec:IRsolution} one can also check that the integrand goes to zero smoothly 
 on $\mathcal C$ when $\rho\to 0$. Also recalling that $A$ is globally defined, we can apply Stokes' theorem and write
\be
\int_{S^3_{\rm bdry}} \Big(*_5 \! F + \frac{2}{\sqrt 3} A \wedge F\Big)\ = \ \int_{\mathcal C} \diff \Big(*_5 \! F + \frac{2}{\sqrt 3} A \wedge F\Big) \ = \ 0\,,
\ee
where we used the Maxwell equation~\eqref{MaxwellEq}. We conclude that in our solution we have
\be\label{starFtoAwedgeF}
\int_{S^3_{\rm bdry}} *_5 F  \ =\ - \frac{2}{\sqrt 3} \int_{S^3_{\rm bdry}}  A \wedge F \ =\ - \frac{128}{3\sqrt 3} \pi^2 \cL^2 \ = \ - \frac{8\pi^2}{3\sqrt 3}  (1-\sq^2)^2\,,
\ee
where in the second equality we used the boundary gauge field given in~\eqref{bdryA}, and in the third just the definition~\eqref{defsquashing}.\footnote{In our conventions, the positive orientation on $S^3$ is specified by taking the 
volume form of the unit, round three-sphere to be
${\rm vol}(S^3_{\rm round})= \frac{1}{8}{\sigma}_1 \wedge {\sigma}_2\wedge {\sigma}_3 = -\frac 18 \sin \theta \diff \theta \wedge \diff \phi \wedge \diff \psi$. Although this is a slightly unusual orientation, of course  we have $\int_{S^3} \!{\rm vol}(S^3_{\rm round}) = + 2\pi^2$.
}

On the other hand, starting from the UV expansion of $*_5 F$ given in~\eqref{starF_UV}, we obtain an expression for $\int_{S^3_{\rm bdry}} \!\!*_5 F$ that -- once compared with \eqref{starFtoAwedgeF} -- allows to solve for the parameter $a_4$ in terms of $a_2$ and $\cL\,$. We obtain
\be\label{sola4}
a_4 \ = \ \frac{1}{384}\left( 5 + 64 a_2 - 256 a_2^2 + 32 \cL - 160 a_2 \cL - 104 \cL^2 \right)\,.
\ee
Later on we will be able to increase our analytic control on the UV behavior of the solution by also determining the parameter $a_6$.

We note that the integral~\eqref{PageCharge} is  a \emph{Page charge} and therefore should be quantized.
It is thus a natural notion of electric charge to be used when discussing black holes and other solitonic objects. 
In the next section we will see that this is \emph{different} from the notion of holographic charge obtained from integrating the holographic $R$-current, \emph{cf.}
(\ref{defElCharge}) below.

\section{Holographic renormalisation}
\label{HoloRenSection}

In this section we perform holographic renormalisation and compute the on-shell action as well as the holographic charges in terms of the squashing parameter~$\sq$.

\subsection{The on-shell action}

The on-shell supergravity action is divergent due both to the infinite volume of the spacetime and to terms going to infinity while approaching the boundary. 
The divergences can be removed implementing holographic renormalisation \cite{WittenAdS,HenningsonSkenderis,BalasubramanianKraus,deHaroSoloSkenderis} (see~\cite{Taylor:2000xw,BianchiFreedmanSkenderis,PapadimitriouSkenderisAlAdS} for the inclusion of a Maxwell field). Our AlAdS spacetime $\Mfive$ can be seen as a foliation of timelike hypersurfaces homeomorphic to the boundary $\partial \Mfive$, labeled by the radial coordinate $\rho$. We denote by $\partial \Mfive_\rho$ the hypersurface at fixed $\rho$ and by $\Mfive_\rho$ its interior region. If we write the five-dimensional metric as
\be\label{bulkmetric}
\diff s^2 \ = \ f^{-1}(\rho) \diff\rho^2 +  h_{ij}(\rho,x) \diff x^i\diff x^j\,,
\ee
with $i,j = 0,\ldots 3$, then $h_{ij}(x,\rho)$, is the induced metric on $\partial \Mfive_{\rho}$.\footnote{The positive orientation on $\partial \Mfive_\rho$ is obtained by contracting the bulk volume form with the unit, outward pointing vector $f^{1/2}\partial / \partial \rho\,$, which yields 
${\rm vol}_4 \equiv 
 \sqrt{h} \,\diff^4 x = - \sqrt{h} \,\diff t \wedge \diff \theta \wedge\diff \phi\wedge \diff \psi\,$, 
with $h = |\det h_{ij}|$.
\label{OrientationHypersurface}}
 Holographic renormalisation regulates the action by evaluating it on $\Mfive_\rho$, and prescribes the addition of suitable counterterms which cancel the divergences arising when $\partial \Mfive_\rho$ is sent to the conformal boundary $\partial \Mfive$, so that the action remains finite in the limit.
 The renormalised on-shell action is given by
\be\label{RenormalizedAction}
S_{\textrm{ren}} \ =\  \lim_{\rho\to\infty} \left( S_{\rm bulk} + S_{\rm GH} + S_{\rm ct}\right),
\ee
where $S_{\rm bulk}$ is the bulk supergravity action~\eqref{sugraction} evaluated on-shell on $\Mfive_\rho$. $S_{\rm GH}$ is the Gibbons--Hawking boundary term, needed to make the variational problem with Dirichlet boundary conditions well-definite; it reads
 \be
S_{\rm GH}\ = \  \frac{1}{8\pi G}\!\int_{\partial \Mfive_\rho}\!\!\!\! \diff^4x\sqrt{h}\, K\,,
\ee
where  $h = |\det h_{ij}|$ and $K$ is the trace of the extrinsic curvature $K_{ij} \,=\, \frac{1}{2} f^{1/2} \, \frac{\partial h_{ij}}{\partial\rho}$ of the hypersurface $\partial \Mfive_\rho$.  
Finally, the counterterm action $S_{\rm ct}$ is given by
\be\label{Counterterm}
S_{\rm ct}\ = \ -\frac{1}{8\pi G}\int_{\partial \Mfive_\rho} \!\!\diff^4x\sqrt{h} \left[\frac 3\ell + \frac{\ell}{4} R + \frac{\rho}{\ell}\frac{\ell^3}{8} \left( R_{ij}R^{ij} - \frac{1}{3}R^2 - \frac{4}{\ell^2} F_{ij}F^{ij} \right)\right],
\ee
where in this formula the Ricci tensor $R_{ij}$ and the Ricci scalar $R$ are those of $h_{ij}$, and the indices are raised with $h^{ij}$. Note that in this section we are reinstating the AdS radius~$\ell$.
The first two terms~\cite{BalasubramanianKraus} are local covariant expressions on $\partial \Mfive_\rho$, designed to cancel terms that diverge with a power-law (when the solution is written in Fefferman--Graham coordinates). The term quadratic in the curvatures~\cite{HenningsonSkenderis,Taylor:2000xw} only depends on the metric and gauge field at the conformal boundary and cancels possible ``logarithmic'' divergences,\footnote{The logarithm appears when the cutoff is written as $\rho/\ell \sim \log r + $ sub-leading terms.} which are proportional to the Weyl anomaly of the dual conformal field theory~\cite{HenningsonSkenderis}.
However, the results of~\cite{SuperWeylAnomaly_paper} imply that this anomaly --- and correspondingly the logarithmic divergence in the on-shell supergravity action --- vanishes in the case of interest for us (see section 4 therein).
This can also be checked directly: from the expressions~\eqref{asymptoticmetric}--\eqref{bdryA} we find that 
asymptotically\footnote{Note that the gauge field used here is related to the one in \cite{Cassani:2012ri,SuperWeylAnomaly_paper} as $A_{\rm here} = - \frac{\ell}{\sqrt{3}} A_{\rm there}$, which accounts for the different numerical factor in the equation below with respect to \emph{e.g.} equation (1.4) in \cite{SuperWeylAnomaly_paper}.}
\be\label{ContribConformalAnomaly}
\sqrt h \left( R_{ij}R^{ij} - \frac{1}{3}R^2 \right) \ = \ \frac{4}{\ell^2}\sqrt h\, F_{ij}F^{ij} \ = \  \frac{4}{3\ell} (1-\sq^2)^2 \sin\theta\,,
\ee
hence the gravitational and the gauge field contributions to the logarithmic term, though non-vanishing, cancel against each other in~\eqref{Counterterm}. 
 The full expression~\eqref{Counterterm} is nevertheless needed in order to derive the counterterms renormalising the energy-momentum tensor and the R-current to be introduced below. The counterterms~\eqref{Counterterm} provide a ``minimal subtraction'' scheme ensuring that all divergences are cancelled. However, one could include additional \emph{finite} counterterms in $S_{\rm ct}$, and these would affect the result for the on-shell action. We postpone a discussion of these extra counterterms and the ensuing ambiguities to section \ref{fielddisc}, when we will compare our gravity results with the field theory side.

Let us now manipulate the five-dimensional integral computing the bulk action. Although we do not have full analytic control on the solution, we will be able to evaluate $S_{\rm bulk}$ exactly, by showing that it reduces to a boundary term. Using the Maxwell equation~\eqref{MaxwellEq} to eliminate the Chern--Simons term and then the trace of the Einstein equation~\eqref{EinsteinEq}, the bulk action~\eqref{sugraction} can be written as
\be
S_{\rm bulk} \ =\  -\frac{1}{2\pi G \ell^2}\int_{\Mfive_\rho}\!\! \diff^5 x \sqrt g \, -\, \frac{1}{12\pi G}\int_{ \Mfive_\rho} \diff (A\wedge *_5F)\,.
\ee
Therefore, when the solution is non-singular and $A\wedge *_5F$ goes to zero sufficiently fast 
 in the IR, the last term reduces to an integral on the boundary $\partial \Mfive_\rho$. 
Remarkably, we find that this is true for the first integral as well. 
To see this, we need to recall some notions introduced in section~\ref{susysec} and exploit the following chain of equalities showing that the five-dimensional volume form is exact:
\be
\diff^5 x \sqrt g  \, =\, \frac{\ell^2}{48} R_{B} \,\diff t \wedge X^1\wedge X^1 \,= \, \frac{\ell^2}{12}\diff t \wedge  \mathcal R \wedge X^1 \,= \, -\frac{\ell^2}{12} \diff (\diff t \wedge P \wedge X^1)\,.
\ee
In the first equality we used the general form of the five-dimensional metric~\eqref{generalggmetric} together with~\eqref{relfR} and the fact that the volume form of the four-dimensional K\"ahler metric is related to the K\"ahler form $X^1$ by ${\rm vol}_{B} = -\frac 12 X^1 \wedge X^1$. The second equality is obtained noting that $\frac{1}{4}R_{B} X^1$ is the primitive part of the Ricci form $\mathcal R$, namely the piece of $\mathcal R$ that has non-zero wedging with $X^1$. The last expression follows from $\mathcal R = \diff P$ and $\diff X^1=0$. If the solution is regular and $P \wedge X^1$ goes to zero in the IR then we can apply Stokes' theorem and write the bulk on-shell action as
\be\label{SbulkAsBoundaryIntegral}
S_{\rm bulk}\  =\   \frac{1}{24\pi G} \int_{\partial \Mfive_{\rho}} \left(\diff y \wedge P \wedge X^1 - 2 A\wedge *_5F \right)\,.
\ee
This formula holds for any regular supersymmetric solution to minimal five-dimensional gauged supergravity of the time-like class satisfying the conditions to apply Stokes' theorem.
This is the case for our solution, hence we can pass to evaluate $S_{\rm bulk}$ explicitly.
The intermediate steps of the computation are more readable when presented in the Fefferman--Graham radial coordinate $r$ introduced in appendix~\ref{DetailsSol}. We will also implement the change of coordinates~\eqref{changepsi}. Plugging the expansions in, we find for the contribution from the five-dimensional volume (up to terms vanishing in the limit $\partial \Mfive_\rho \to \partial \Mfive$):
\bea
\frac{1}{24\pi G} \hspace{-10mm}&&\int_{\partial \Mfive_{\rho}}\diff t \wedge P \wedge X^1 \ =\ -\frac{2\pi}{3G} a^2(\rho) \left[4 a'^2(\rho) + 2 a(\rho) a''(\rho) -1\right] \int \diff t
\nn \\ [2mm] 
&=& -\frac{\pi\ell^2}{G}\left[ 4 a_0^4 r^4 - \left(1+\frac{4}{3}\cL\right)a_0^2r^2 - \frac{32}{9}\cL^2 \log r - \frac{32}{9}a_2 \cL + \frac{3}{32} + \frac{\cL}{36} + \frac{19}{18}\cL^2 \right ]\int \diff t,\nn \\  \label{Sbulk_vol5_FG}
\eea
where in the first equality we used the expressions of the Ricci potential $P$ and K\"ahler form $X^1$ given in section~\ref{susysec}, while in the second equality we plugged the UV solution~\eqref{UVsola} in and passed to the Fefferman--Graham coordinate $r$ using~\eqref{changeradialcoord}. Since time translations are a symmetry of our solution, integration over time just gives an overall factor. Using the expansions~\eqref{formofA}--\eqref{starF_UV}, the gauge field contribution yields 
\be\label{AFcontribSbulk}
-\frac{1}{12\pi G} \int_{\partial \Mfive_{\rho}}  A\wedge *_5F \ = \ -\frac{\pi\ell^2}{G} \left[ \frac{32}{9}\cL^2 \log r + \frac{32}{9}a_2\cL + \frac{8\cL}{27(1-4\cL)} - \frac{2}{27}\cL - \frac{8}{3}\cL^2 \right]\int \diff t\,.
\ee
In both equations~\eqref{Sbulk_vol5_FG} and \eqref{AFcontribSbulk} we invoked the global constraint discussed in section~\ref{sec:globalconstraint} and used~\eqref{sola4} to eliminate $a_4$. Note that both the logarithmic divergences and the terms proportional to $a_2$ cancel out when the two contributions to $S_{\rm bulk}$ are added up, leaving just power-law divergences and finite terms depending on the boundary parameter~$c$.  

The evaluation of the remaining contributions to the on-shell action just involves the UV solution and is thus straightforward. The Gibbons--Hawking term is found to contain only divergences,
\be
S_{\rm GH} \ = \ -\frac{\pi\ell^2}{G} \left[ \left(1+\frac{4}{3}\cL \right) a_0^2 \, r^2 - 16\, a_0^4\, r^4 \right] \int \diff t,
\ee
while the counterterms~\eqref{Counterterm} contain both a divergence and a finite piece,
\be\label{Sct_FG}
S_{\rm ct} \ = \ -\frac{\pi\ell^2}{G} \left(\frac{8}{3}\cL^2 + 12 a_0^4 r^4 \right) \int \diff t\,.
\ee
Adding up the contributions~\eqref{Sbulk_vol5_FG}--\eqref{Sct_FG}, and writing the result in terms of the squashing parameter $\sq^2 = 1-4\cL$, we obtain for the renormalised on-shell action
\be\label{resultSonshell}
S_{\rm ren} \ = \ -\frac{\pi\ell^2}{G}\left[ \frac{2}{27 \sq^2} + \frac{2}{27} - \frac{13}{108}\sq^2 +\frac{19}{288}\sq^4 \right]\int \diff t\,.
\ee

It is crucial to stress that the action is gauge-dependent, so this result is sensitive to the gauge chosen. From~\eqref{SbulkAsBoundaryIntegral} and \eqref{starFtoAwedgeF} we see that under a gauge transformation $\delta A = \delta A_t\,\diff t$, where $\delta A_t$ is a constant (possibly depending on $\sq$), the on-shell action changes by
\be\label{gaugeshiftSonshell}
\delta S_{\rm ren} \ =\ -\frac{\delta A_t}{12\pi G} \int \diff t  \int_{S^3_{\rm bdry}} *_5F \ = \  \frac{2\pi \ell^2}{9\sqrt 3G} \,\delta A_t \, (\sq^2-1)^2\int \diff t\,.
\ee
Our motivation for fixing the gauge as in~\eqref{bdryA} is that with this choice the supersymmetry parameter does not depend on the time coordinate $t$ (neither on the boundary, nor in the bulk) when the obvious left-invariant frame following from~\eqref{bdrymetric} is used. In particular, note that the term in $1/\sq^2$ in~\eqref{resultSonshell}, which will play an important role in section~\ref{fielddisc}, directly comes from the gauge field contribution~\eqref{AFcontribSbulk}.

Of course, the Lorentzian time should be non-compact and therefore $\int \diff t$ in the formulae above is just a formal writing. However after performing an analytic continuation $t \to \ii\, \tE$ we can compactify the time coordinate. Then the boundary topology becomes $S^1 \times S^3$, the boundary metric~\eqref{bdrymetric} becomes Euclidean and the boundary gauge field~\eqref{bdryA} acquires an imaginary, flat component. We remark that both the bulk metric and the bulk gauge field become complex, 
however this does not affect the on-shell action, which remains real.
Denoting by $\onshE$ the analytically continued on-shell action, and by $\Delta_\tE$ the finite period of $\tE$, we obtain\footnote{We define $\onshE$ by requiring that this is related to the  analytic continuation of the Lorentzian action as
 $ \exp({-\onshE}) = \exp({\ii \,S_{{\rm Lorentz},\, t \to \ii \tE}})$. Note that we regard the Lorentzian time as $x^0$ and the analytically continued time as $x^4$, so the Lorentzian volume form on $\partial \Mfive_\rho$ is related to the analytically continued one as $ 
 \sqrt{h}\, \diff x^{0123} = - \ii \sqrt{h}\, \diff x^{1234}$.}
\be\label{resultSonshellEucl}
\onshE \ = \ \frac{\pi\ell^2\Delta_\tE}{G}\left[ \frac{2}{27 \sq^2} + \frac{2}{27} - \frac{13 }{108}\sq^2 +\frac{19}{288}\sq^4 \right].
\ee
This exact expression for the renormalised on-shell action in terms of the squashing of the boundary $S^3_\sq$ is a main result of this paper.

\subsection{Holographic charges}
\label{sec:HoloCharges}

The regularised action $S_{\rm reg} = S_{\rm bulk} + S_{\rm GH} + S_{\rm ct}$ gives rise to the quasi-local energy-momentum tensor
\be\label{defEnMom}
T_{ij} \ = \ -\frac{2}{\sqrt{h}}\frac{\delta S_{\rm reg}}{\delta h^{ij}}\,.
\ee
In order to obtain the dual field theory energy-momentum tensor $\langle T_{ij}\rangle$, one needs to rescale this before taking the limit $\rho \to \infty$, because in AlAdS space the metric of the background on which the field theory lives is multiplied by the divergent conformal factor $e^{2\rho/\ell}$. 
This gives the formula 
\bea\label{HoloEnMomTensor}
\langle T_{ij}\rangle & = & -\frac{1}{8\pi G}\lim_{\rho \to \infty} {e^{2\rho/\ell} } \bigg[ K_{ij} - K h_{ij} + \frac{3}{\ell} h_{ij}  - \frac{\ell}{2}\left( R_{ij}-\frac 12 R\, h_{ij}\right) \nn \\ [2mm]
&&\qquad\qquad\qquad\, -\,\frac{\rho}{\ell}\frac{\ell^3}{2}\! \left( -\frac{1}{2}B_{ij} - \frac{4}{\ell^2} F_{ik}F_j{}^k + \frac{1}{\ell^2} h_{ij} F_{kl}F^{kl} \right) \bigg] \ , 
\eea
where the curvatures are computed and the indices are raised with the induced metric $h_{ij}$, and $B_{ij}$ denotes the Bach tensor (see eq.~\eqref{BachTensor}).\footnote{The Bach tensor arises from the variation of the square of the Weyl tensor $C_{ijkl}$ with respect to the metric. This appears noting that in~\eqref{Counterterm} $R_{kl}R^{kl}-\frac{1}{3}R^2 = \frac{1}{2}\left(C_{ijkl}C^{ijkl} - \mathcal E\right)$, and recalling that the variation of the Euler density $\mathcal E$ with respect to the metric vanishes. One may expect the term $R_{ik}R^k{}_j$ to appear from the variation of $R_{kl}R^{kl}$; this indeed shows up by using the identity $R_{ikjl}R^{kl} = R_{ik}R^k{}_j - \nabla_k \nabla_{(i} R^k{}_{j)} + \frac{1}{2} \nabla_i \nabla_j R$.} The first line yields a finite contribution, while the second line just cancels ``logarithmic'' divergences, here going as $\rho/\ell$.

In addition to the energy-momentum tensor, we can also introduce a current by varying the regularised action with respect to 
the gauge field $A_i\,$ at the boundary $\partial \Mfive_{\rho}$:
\be
j^i \ = \  \frac{1}{\sqrt{h}} \frac{\delta S_{\rm reg}}{\delta A_i}\,. 
\ee
The possible contributions arise from the bulk action~\eqref{sugraction} and from the counterterm action~\eqref{Counterterm}. Varying the bulk action with respect to $A$ and using the Maxwell equation~\eqref{MaxwellEq} to eliminate the bulk integral we are left with the boundary integral
\be
\delta S_{\rm bulk} \ = \ -\frac{1}{4\pi G} \int_{\partial \Mfive_\rho} \delta A \wedge \left(*_5 F + \frac{4}{3\sqrt 3} A \wedge F\right).
\ee
Note that here we are assuming that the spacetime is smooth and that $\partial \Mfive_\rho$ is the only boundary. The counterterm contribution cancels a ``logarithmic'' divergence arising from~$*_5 F$.
Since the graviphoton $A$ couples to the dual field theory R-symmetry current, sending $\rho \to \infty$ after appropriate rescaling yields a finite expectation value $\langle j^i \rangle$ for the latter.  
Thus we arrive at the expression
\be\label{ExpressionRcurrent}
\langle j^i \rangle \ = \ \lim_{\rho \to \infty} \frac{e^{4\rho/\ell}}{\sqrt{h}} \frac{\delta S_{\rm reg}}{\delta A_i} \ = \ \frac{1}{4\pi G} \lim_{\rho \to \infty} e^{4\rho/\ell}\left\{ *_4\! \left[ \diff x^i \wedge \left(*_5 F + \frac{4}{3\sqrt 3} A \wedge F\right) \right]  - \rho\, \nabla_j F^{ji}\right\}
\,,
\ee
where $*_4$ is computed using the induced metric $h_{ij}$ and it is understood that $*_5 F + \frac{4}{3\sqrt 3} A \wedge F$ is restricted to the boundary $\partial \Mfive_\rho\,$. The last term comes from the variation of $S_{\rm ct}$ in~\eqref{Counterterm}.

From the holographic energy-momentum tensor and R-current one can construct a set of conserved charges. In order to do so, one chooses a spacelike, compact hypersurface $\Sigma$ in the boundary $\partial \Mfive$, with metric $\gamma_{\alpha\beta}$, $\alpha,\beta =1,2,3$, and writes the boundary metric in ADM form:
\be
\diff s^2_{\rm bdry} \ = \ - N^2\diff t^2 + \gamma_{\alpha\beta}(\diff x^\alpha + N^\alpha \diff t) (\diff x^\beta + N^\beta  \diff t)\,.\label{bdrymetricADM}
\ee
Let $u$ be the unit, timelike vector orthogonal to $\Sigma$.  
 As a one-form, it reads $u = - N \diff t$.
  Then from any Killing vector $Z$ of the boundary metric one can construct a conserved charge as
\be\label{conservedQgeneral}
Q_{Z} \ = \ \int_\Sigma \diff^3x \sqrt{\gamma} \,u^i \, \langle T_{ij} \rangle Z^j \,.
\ee
Moreover, from the current $\langle j\rangle$ one obtains an electric charge $Q$, corresponding to the vev of the dual field theory R-charge operator, by evaluating the integral
\be
Q \ = \ \int_\Sigma \diff^3 x \sqrt \gamma\,u_i \langle j^i\rangle \,.
\ee
 Assuming that $\nabla_j F^{ji}u_i = 0$ on $\partial \Mfive_\rho$, so that the divergent pieces 
 drop from~\eqref{ExpressionRcurrent} (this is satisfied in the case of interest for us), it is straightforward 
 to see that\footnote{In the absence of a Chern--Simons term, this was shown in~\cite{PapadimitriouSkenderisAlAdS}.}
\bea
Q &=&  \int_{\Sigma} \diff^3 x \sqrt \gamma\, u_i \langle j^i \rangle \ = \ \frac{1}{4\pi G} \int_\Sigma   \left(*_5 F + \frac{4}{3\sqrt 3} A \wedge F\right).\label{defElCharge}
\eea
The charge $Q$ is invariant under small gauge transformations of $A$, and it is conserved provided the R-current is non-anomalous, which is true in the case we are studying here.\footnote{In Lorentzian signature, or in Euclidean signature when two supercharges of opposite R-charge are preserved, the chiral anomaly of the R-current vanishes whenever the background is supersymmetric and $F \wedge F = 0$ on the boundary~\cite{SuperWeylAnomaly_paper}.}
 However, we remark that this is different from the quantized Page charge~\eqref{PageCharge}, as we already remarked in section \ref{sec:globalconstraint}. The two definitions coincide only 
 when the Chern--Simons contribution  $\int_\Sigma  A \wedge F $ vanishes.

We now compute the conserved charges for our solution. The spacelike boundary hypersurface $\Sigma$ is the squashed three-sphere $S^3_{\rm bdry}$, and from the boundary metric~\eqref{bdrymetric} we see that $u \; = \;  \frac{\sq}{2a_0} \frac{\partial}{\partial t}\,.$
To compute the holographic R-charge we can take advantage of formula~\eqref{starFtoAwedgeF}; then~\eqref{defElCharge} gives
\be\label{ResultQ}
Q \ = \ - \frac{1}{4\pi G}\frac{2}{3\sqrt 3} \int_{S^3_{\rm bdry}} A \wedge F \ = \ -\frac{2\pi\ell^2}{9\sqrt 3G}\left( \sq^2 - 1\right)^2 .
\ee

The other relevant charges are the \emph{total energy}
$\Egrav$, associated with the generator of time translations $\partial/\partial t$,
\be\label{FormulaForMass}
\Egrav \ = \ \int_{S^3_{\rm bdry}} u^i\langle T_{it}\rangle \, {\rm vol}(S^3_{\rm bdry})\,,
\ee
and the angular momentum $J$ for the $U(1)$ right-isometry generated by $\partial/\partial\psi$,
\be\label{DefAngMom}
J \ =\ \int_{S^3_{\rm bdry}} u^i\langle T_{i\psi}\rangle \, {\rm vol}(S^3_{\rm bdry}) \,,
\ee
while one can check that the charges associated with the $SU(2)$ left-isometries vanish. Note that $E$ and $J$ are defined also on non conformally flat boundaries, 
differently from the Ashtekar--Das conserved quantities~\cite{AshtekarDas}.
$\Egrav$ and $J$ can be evaluated by inserting the UV solution in eq.~\eqref{HoloEnMomTensor} for $\langle T_{ij}\rangle $, and performing the integrals above.
This yields expressions depending on the UV parameters on  $a_2$, $a_6$ and $\sq^2$ (recall that $a_4$ was determined in~\eqref{sola4} as a consequence of the smoothness of the solution). Ultimately all charges have to be functions of a single independent parameter of the solution, that we identify with the squashing $\sq$. Although we do not have full analytical control on the relations between these UV parameters, in a moment we will introduce an extra information allowing to deduce the exact expressions of the charges in terms of $\sq$. 
Already at this stage, by using the values of the on-shell 
action in (\ref{resultSonshellEucl}), the charge $Q$ in (\ref{ResultQ}), and the energy-momentum tensor obtained 
from (\ref{HoloEnMomTensor}), one can check that the following relation is satisfied:\footnote{From (\ref{HoloEnMomTensor}) one obtains $\langle T_{ij} \rangle$ as a function of $v,a_2,a_4,a_6$ (and $a_0$) that we have 
not presented. The point is that after using~\eqref{sola4} the dependence on $a_0,a_2,a_4,a_6$ drops out of the combination $E+ \frac{2}{\ell \sq^2}J$. In Lorentzian signature
$\Delta_\tE$ denotes simply a time interval.}
\be\label{BPSrelation}
\Egrav +\frac{2}{\ell\sq^2} J + \frac{3\sq^2 -2}{2\sqrt 3\sq^2} \,Q \ = \ \frac{\onshE}{\Delta_\tE}\,.
\ee
We note that the combination $\Egrav +\frac{2}{\ell\sq^2} J $ is the holographic charge $Q_V$ associated with the supersymmetric Killing vector $V$ given in~\eqref{susyvector}, and that the coefficient in front of $Q$ is the boundary value of $V^\mu A_\mu$. So \eqref{BPSrelation} can be written as 
\be\label{SmarrGeneral}
Q_{V} + (V^\mu A_\mu^{\rm bdry}) \, Q \ =\ \frac{\onshE}{\Delta_\tE}\ .
\ee
Recalling~\eqref{gaugeshiftSonshell} and~\eqref{ResultQ}, we observe that under a gauge transformation 
the on-shell action (that we can define formally in any gauge, provided we do not compactify the time direction) 
shifts by a term proportional to the electric charge:
$\delta\onshE / \Delta_\tE  \,= \,   \delta A_t\,  Q\,$.
This implies that the relation above remains valid independently of the gauge chosen.
Note that in the gauge $V^\mu A_\mu = \frac{\sqrt 3}{2} f$, the relation~\eqref{SmarrGeneral} is satisfied by the black hole of~\cite{GutowskiReall}, 
with $M = E- \onshE / \Delta_\tE$ identified with the Ashtekar--Das mass \cite{AshtekarDas}.

In order to obtain $\Egrav$ and $J$ as functions of $\sq$ we will exploit the fact that for any continuous  
parameter $\mu$ of the \emph{boundary} geometry,\footnote{For example, the horizon radius of the supersymmetric black hole of \cite{GutowskiReall} 
does not yield such a relation.}  one can derive a corresponding ``Ward identity'' by simply applying the chain rule \cite{Martelli:2012sz} 
\bea
\frac{\diff }{\diff \mu} S_{\rm ren} &=& \int_{\partial \Mfive} \diff^4 x \sqrt{\hat h} \left( -\frac{1}{2} \langle T_{ij}\rangle \frac{\diff \hat h^{ij}}{\diff \mu}  +  \langle j^i\rangle 
\frac{\diff A_i}{\diff \mu}   \right),\label{WardIdentityGeneral}
\eea
where $\hat h_{ij}$ is the finite metric on the conformal boundary: $\hat h_{ij} = \lim_{\rho\to\infty}e^{-2\rho/\ell} h_{ij}$. 
The parameter we will to consider is the one that can be introduced in the 
boundary metric~\eqref{bdrymetric} and gauge field~\eqref{bdryA} by rescaling the coordinate $t$ as $t = b\, t'$. Varying with respect to $b$ we get 
\be
\langle T_{ij}\rangle \frac{\diff \hat h^{ij}}{\diff b} 
\;=\; \frac{\sq}{a_0 b}u^i \langle T_{it}\rangle  \,, \quad\qquad j^i\frac{\diff A_i}{\diff b} 
\;=\; -\frac{\sq}{2a_0 b} u_i j^{i} A_{t}\,.
\ee
So~\eqref{WardIdentityGeneral} gives
\be
\frac{\diff }{\diff b}S_{\rm ren} 
= \int \diff t' \int_{S^3_{\rm bdry}} {\rm vol}(S^3_{\rm bdry}) \left( -u^i \langle T_{it} \rangle  - u_i j^i A_t   \right)
= -\left(\Egrav + A_t\, Q \right)\int \diff t' \,.
\ee
Recalling that $A_t = \frac{1}{2\sqrt 3}\,$, and turning to the analytically continued on-shell
action (so that $\frac{\diff }{\diff b}S_{\rm ren}$ becomes $- \frac{\onshE}{\Delta_\tE} \int \diff t'$), we arrive at the relation 
\be\label{RelFromChainRule}
\Egrav + \frac{1}{2\sqrt 3} Q \ = \ \frac{\onshE}{\Delta_\tE}\,.
\ee
Denoting $U={\de}/{\de t}$,  this can be re-written in the covariant form 
\be\label{RelFromChainRuleAbstract}
Q_{U} + (U^\mu A_\mu^{\rm bdry}) \, Q \ =\ \frac{\onshE}{\Delta_\tE}\,,
\ee
which is manifestly gauge-invariant.   One can also consider a variation of the on-shell action with respect to the parameter that can be introduced in the background fields by rescaling the original time coordinate $y$. 
This yields precisely the relation~\eqref{BPSrelation}, thus providing a general way to derive it. In section \ref{algebsection}
we will discuss the field theory interpretation of the conserved charges and of this relation. 
Finally, one can check that varying with respect to the squashing $\sq^2$ does not yield new information.

Inserting the expressions~\eqref{ResultQ} for $Q$ and~\eqref{resultSonshell} for the action in~\eqref{RelFromChainRule}, we obtain the total energy as a function of the squashing parameter only:
\be\label{resultMass}
\Egrav \ =\  \left( \frac{2}{27\sq^2} + \frac{1}{9} - \frac{7}{36}\sq^2 + \frac{89}{864}\sq^4 \right)\frac{\pi \ell^2}{G} \,.
\ee
We can compare this expression with the one obtained from the definition~\eqref{FormulaForMass} using the holographic energy-momentum tensor~\eqref{HoloEnMomTensor}, which depends on the UV parameters $a_2$, $a_6$ and $\sq^2$. In this way, we obtain an equation that determines $a_6$, see eq.~\eqref{sola6} in the appendix for the explicit expression.
The angular momentum can now be deduced from~\eqref{BPSrelation}, which yields the following simple expression:
\be\label{ResultJ}
J \ =\ \left(\sq^2 -1\right)^3 \frac{\pi \ell^3}{27 G}\,. 
\ee
We have thus obtained all relevant charges as exact expressions of the squashing parameter~$\sq$.

A check of our results so far is provided by the limit $\sq \to 1$, in which the boundary three-sphere $S^3_\sq$ becomes round, and the bulk solution is just AdS$_5$. In this case, $Q$ and $J$ vanish, while $\Egrav$ and  $\onshE$ remain finite and reproduce the known expressions for pure AdS$_5$, namely $\Egrav = \frac{3\pi\ell^2}{32G}$ and  $\onshE = \Egrav\Delta_\tE$ \cite{BalasubramanianKraus}.
 For completeness, in appendix~\ref{EnMomTandRcurrent} we give the explicit expressions for the energy-momentum tensor and for the R-symmetry current.

\subsection{Summary of the solution}\label{SummarySol}

Before moving on to discussing the dual field theories, it may be useful to provide a summary of the one-parameter family of solutions that we constructed. 

In section~\ref{UVanalysis} (see also appendix~\ref{DetailsSol}), we determined the general asymptotic expansion of an AlAdS solution to the ODE (\ref{SixthOrderEq}) governing  $SU(2)\times U(1)\times U(1)$ invariant supersymmetric solutions to minimal five-dimensional gauged supergravity.
 This expansion is characterized by five parameters $a_0$, $a_2$, $a_4$, $a_6$ and $\sq$. Two of them, $a_0$ and $\sq$, appear explicitly 
 in  the boundary metric~\eqref{bdrymetric}; this describes the direct product $\mathbb R_t\times S^3_v$ with overall size proportional to $a_0$, where $S^3_v$ is a three-sphere with $SU(2)\times U(1)$ isometry and squashing parameter $\sq$. For $\sq \neq 1$ this is a non conformally flat boundary and, correspondingly, the boundary gauge field~\eqref{bdryA} is non-trivial. The three remaining UV parameters $a_2$, $a_4$, $a_6$ appear in sub-leading components of the metric and of the gauge field.

The IR analysis of section~\ref{sec:IRsolution} established the existence of a one-parameter family of solutions closing off smoothly in the interior. 
By means of a linearised and a numerical analysis (sections~\ref{linearsec} and \ref{numerics_sec} respectively) we showed that this parameter can be directly related to the squashing~$\sq$ at the boundary (see fig.$\:$\ref{fig:PlotVsquaredVsXi}), and that for any value of \hbox{$\sq^2 > 0$} the solution is everywhere regular and free from closed time-like curves (figs.$\:$\ref{fig:aPlot}--\ref{fig:differentplotsMetricAndA}). In addition, we determined the relations between the different UV parameters corresponding to our one-parameter family (fig.$\:$\ref{fig:otherUVparametersPlots}). For two of the UV parameters, namely $a_4$ and $a_6$, we obtained an analytic expression in terms of $\sq$ and $a_2$.\footnote{We also compared the analytic expressions for $a_4(v^2,a_2)$ and $a_6(v^2,a_2)$, given in (\ref{sola6}),
with the numerical results, finding perfect agreement.} Recalling that $a_0$ can be chosen freely by shifting the radial coordinate, the only UV parameter that we could not determine analytically as a function of $v^2$ is $a_2$; however, this drops out of the on-shell action as well as of all the holographic charges.

Despite the lack of full analytic control on the solution, in the present section we evaluated the on-shell action as well as the holographic conserved charges exactly, in terms of~$\sq$. The on-shell action could be evaluated by showing that it reduces to a boundary integral, and is given in eq.~\eqref{resultSonshell}. We remarked that this is a gauge-dependent result, and gave it in a gauge such that the supersymmetry parameter is independent of the time coordinate $t$. On the other hand, the holographic charges are gauge-independent, and the relevant ones are the energy $\Egrav$, the angular momentum $J$ and the electric charge $Q$. Their evaluation involved a Ward identity (following simply from applying the chain rule to the variation of the renormalised on-shell action), and their 
expressions are given in~\eqref{ResultQ},~\eqref{resultMass} and~\eqref{ResultJ}.

\section{Field theory duals}
\label{fielddisc}

In this section we discuss the field theory duals to our supergravity solution. We first recall some aspects of rigid superymmetry at the conformal boundary and its relation
with supersymmetry in the bulk, and then discuss the relation of the localised partition function computed in our  background with the supersymmetric index and the Casimir energy. We will end comparing the latter with the holographically renormalised on-shell gravity action.

Here we will assume that the five-dimensional solution is uplifted \cite{Buchel:2006gb} to a ten-dimensional solution $M_5\times Y_5$ of type IIB supergravity, where $Y_5$ is a Sasaki--Einstein manifold. In this case, when the squashing parameter is trivial so that $M_5=$ AdS$_5$ (in global coordinates), the field theory dual is generically an ${\cal N}=1$ superconformal quiver gauge theory, with matter in bi-fundamental or adjoint representations of the gauge group. 
For example, one can consider ${\cal N}=4$ super Yang--Mills \cite{Maldacena:1997re}, the Klebanov--Witten quiver gauge theory \cite{Klebanov:1998hh}, or the $Y^{p,q}$ family of quivers~\cite{Benvenuti:2004dy}. 
 Our solution should then correspond to a relevant deformation of the flat-space Lagrangian of the given field theory, obtained through couplings to a non-trivial background metric and a gauge field sourcing the R-symmetry current~\cite{FestucciaSeiberg}.

\subsection{Rigid supersymmetry at the boundary}
\label{algebsection}

In the following we discuss some general properties of the superconformal field theories living at the non conformally flat boundary of our solution. We already mentioned that the Weyl invariance and R-symmetry are preserved also at the quantum level, since the associated anomalies vanish on our background~\cite{SuperWeylAnomaly_paper}. Here we focus on the algebra associated with the supersymmetry being preserved. For definiteness we work in Lorentzian signature, however most of what we will discuss can be repeated in Euclidean signature. 

A superconformal field theory can be coupled to curved space by considering conformal supergravity and freezing the background.\footnote{For an introduction to conformal supergravity we refer to e.g.~\cite{FradkinTseytlin,FreedmanVanProeyenBook}.} Though this is a general procedure, here we are interested in the $\mathcal N=1$ case with a minimal set of background fields.\footnote{See~\cite{Klare:2013dka} for the $\mathcal N=2$ case.} The field theory energy-momentum tensor couples to the background metric while the $U(1)$ R-current couples to the gauge field sitting in the gravity multiplet of conformal supergravity, that we will denote by $\Acks$.
For the background to be bosonic the gravitino must be set to zero, and imposing that its supersymmetry variation be zero too yields the differential equation~\cite[eq.$\:$(16.10)]{FreedmanVanProeyenBook}
\be
 \nabla_i^A \zeta_+ - \gamma_i\, \eta_-  \ = \ 0\,,
\ee
where $\nabla_i^A \zeta_+ = \left(\nabla_i - \ii \Acks_i\right) \zeta_+ $, and $\nabla_i$ is the Levi-Civita connection constructed with the background metric.  
 Here, $\zeta_+$ is the positive-chirality part of the Majorana spinor parameter associated with the $Q$-supersymmetry transformations of conformal supergravity, while $\eta_-$ is the negative-chirality part of the parameter of special $S$-transformations. These can be disentangled by taking the gamma-trace of the equation above:
\be\label{susycond_CSG}
\nabla_i^A \zeta_+ \,=\, \frac{1}{4} \gamma_i \gamma^j\nabla_j^A \zeta_+    \,,\qquad \eta_- \,=\, \frac{1}{4} \gamma^i\nabla_i^A \zeta_+\,.
\ee
Therefore we see that the supersymmetry preserved on a bosonic background of conformal supergravity generically is a combination of a $Q$- and an $S$-transformation. The first equation in~\eqref{susycond_CSG} is a charged version of the {\it conformal Killing spinor (CKS) equation}.
In Lorentzian signature, it admits a solution with no zeros if and only if the four-dimensional background has a null conformal Killing vector $z$~\cite{Cassani:2012ri}. This can be written as a spinor bilinear as
\be\label{defz}
z^i \ =\ \frac 14\, \overline{\zeta_+} \gamma^i \zeta_+\,.
\ee
In Euclidean signature, the charged CKS equation has been studied in~\cite{KTZ} and leads to the requirement that the manifold be Hermitian.

Precisely the same conditions are retrieved in a holographic setup~\cite{KTZ,Cassani:2012ri}. In fact, one can show that on AlAdS spaces the supersymmetry transformations of five-dimensional minimal gauged supergravity reduce at the boundary to the transformations of conformal supergravity \cite{Balasubramanian:2000pq}. So the fields and supersymmetry parameter of any supersymmetric AlAdS solution satisfy \eqref{susycond_CSG} asymptotically, with $\zeta$ and $\eta$ corresponding to the leading and first sub-leading components of the asymptotic five-dimensional supersymmetry parameter. 
The conformal supergravity gauge field $\Acks$ is identified with the boundary value of the bulk gauge field $A$ appearing in the previous sections; in our normalizations, we have $\Acks = - \frac{\sqrt{3}}{\ell} A^{\rm bdry}$. The bulk supersymmetry time-like Killing vector $V$ introduced in section~\ref{susysec} reduces to the null (conformal) Killing vector $z$ 
 on the boundary; see~\cite{Cassani:2012ri} for more details on the dictionary between bulk and boundary quantities.

Given a solution $\zeta_+$ to the CKS equation and its charge conjugate $\zeta_- = (\zeta_+)^c$, the associated superalgebra can be obtained from the conformal supergravity algebra, which gives\footnote{This can be seen from the transformations given e.g.\ in~\cite{FreedmanVanProeyenBook} taking into account that the supersymmetry being preserved by a charged conformal Killing spinor is a combination 
of an ordinary supersymmetry $Q$ and a special supersymmetry~$S$. Our conformal supergravity gauge field is related to the one of~\cite{FreedmanVanProeyenBook} as $\Acks = \frac 32 A^{\rm there}$.}
\be\label{superalgebraCKS}
 [\delta_{\zeta_+} ,\delta_{\zeta_-}]\Phi \ = \ 2\ii \Big[\, \mathcal L_z  -  \ii\, r z^i \Big( \Acks_i + \frac 32 {\rm Re}\vcks_i\Big)  \Big] \Phi + \cdots\,,
\ee
where $\Phi$ is a generic field in the theory, $r$ is its R-charge, $\mathcal L_z$ is the Lie derivative along $z$ and the ellipsis denotes a scale transformation and a special conformal transformation proportional to the imaginary part of the complex one-form $\vcks$. The latter is defined through
\be\label{defVnm}
\gamma^i\, \nabla_i^A \zeta_+ \ = \ 2 \ii\, \vcks_i \gamma^i \zeta_+\,,
\ee
where we are assuming that the spinor has no zeros so that the definition holds everywhere.  
In fact, the imaginary part of $\vcks$ is related to the failure of the conformal Killing vector 
$z$ to be Killing~\cite{Cassani:2012ri}. So when $z$ satisfies the Killing condition $\nabla_{(i} z_{j)} =0$, $\mathbb V$ is real and the terms in the ellipsis in~\eqref{superalgebraCKS} vanish. 
 Locally, a conformal Killing vector can always be made Killing by a conformal transformation of the 
metric (which in a holographic setup is implemented by a change of radial coordinate). 
 
When $z$ is a Killing vector, one obtains the same algebra by coupling the field theory to the \emph{new minimal} formulation of off-shell supergravity.
Indeed, modulo the conformal transformation just discussed, the CKS equation in~\eqref{susycond_CSG} is equivalent to the supersymmetry condition that is
obtained from the rigid limit \cite{FestucciaSeiberg} of new minimal supergravity \cite{Sohnius:1981tp}. This reads 
\be
\left(\nabla_i - \ii\, \Anm_i + \ii\, \Vnm_i + \frac{\ii}{2} \Vnm_j \gamma^j{}_i\right) \zeta_+ \ = \ 0\,,
\label{nmkse}
\ee
where the gauge field $\Anm_i$ and the global one-form $\Vnm_i$, satisfying $\diff * \Vnm = 0$, 
constitute the new minimal set of auxiliary fields. 
 The equivalence is easily seen by identifying 
\be
\vcks\ = \ \Vnm\,,\qquad \Acks \ = \ \Anm - \frac{3}{2}\Vnm \,,
\ee
see~\cite{KTZ,DFS,Cassani:2012ri} for more details. The 
definition~\eqref{defVnm} leaves the component of $\vcks$ along $z$ undetermined, and this can be used to arrange for $\diff *\Vnm =0$. In the new minimal variables, (\ref{superalgebraCKS}) reads simply
\be\label{superalgebranm}
 [\delta_{\zeta_+} ,\delta_{\zeta_-}]\Phi \ = \ 2\ii \left( \mathcal L_z  -  \ii\, r z^i \Anm_i \right) \Phi \,,
\ee
in agreement with \cite{Cassani:2012ri}. This equivalence with new minimal supergravity may be useful for concrete field theory computations in  
AdS/CFT context, as typically the superconformal field theories dual to gravity solutions are defined as the strongly coupled IR fixed point
of a Lagrangian theory in the UV, which is non-conformal. 
Therefore, as long as one is interested in supersymmetric quantities that are independent of the coupling, one may take advantage of the new minimal formulation of rigid supersymmetry in order to couple the UV 
Lagrangian to the background of interest.

Let us now focus on the $\mathbb R \times S^3_v$ background defined at the boundary of our bulk supergravity solution.
Working in the frame
\be
e^0 = \frac{2a_0}{\sq} \diff t\,,\qquad e^1 = a_0\ell \, \sigma_1 \,,\qquad e^2 = a_0\ell \,\sigma_2 \,,\qquad e^3 = a_0\ell\, \sq \,\sigma_3\,,
\ee
for the boundary metric~\eqref{bdrymetric}, 
one can see that the CKS equation is solved by a constant spinor $\zeta_+$ satisfying the projection $\gamma^{12} \zeta_+ = \ii\, \zeta_+\,$, with (recall~\eqref{bdryA})
\be\label{Acs}
\Acks \ =\  -\frac{\sqrt 3}{\ell} A_{\rm bdry} \ = \ -\frac{1}{2\ell} \diff t -\frac{1}{2}(\sq^2-1)\, \sigma_3 \,.
\ee
The new minimal equation is solved by the same $\zeta_+$, fixing the background fields as
\bea
\Vnm &=&  \frac{\sq^2}{2}\sigma_3 + \kappa \,z,\nn\\ [2mm]
\Anm &=& \Acks + \frac{3}{2}\Vnm \ = \ - \frac{1}{2\ell}\diff t + \frac{1}{4}\left(2+\sq^2 \right)\sigma_3 + \frac{3}{2}\kappa\, z\,,\label{VnmAnm}
\eea
where $\kappa$ parameterizes the part of $\Vnm$ and $\Anm$ left undetermined by the equation, and can be any function consistent with $\diff * \Vnm = 0$.
From~\eqref{defz} it follows that the vector $z$ is Killing and can be written as
\be
z \ = \ \frac{\partial}{\partial t} + \frac{2}{\ell\sq^2}\frac{\partial}{\partial \psi}\,,
\ee 
where we have conveniently fixed the complex constant parameterizing $\zeta_+$.
As a one-form, $z$ reads
\be
z \ =\ (2a_0)^2\left( -\frac{1}{\sq^2}\diff t + \frac{\ell}{2}\sigma_3 \right)\,.
\ee 

Noting that $z \cdot \Vnm = {1}/{\ell}$, we see that the supersymmetry algebra
depends manifestly on the gauge choice of $\Acks$ (but does not depend on $\kappa$), 
through the term $ r (z \cdot \Acks ) $. We now illustrate some distinguished choices of gauge.
The following parallels a discussion in~\cite{FestucciaSeiberg}, extending it from the round case to our $S^1\times S^3_v$ geometry. 
We start from the gauge adopted so far, in which $\Acks$ is given by~\eqref{Acs}.
Denoting by $\mathcal Q_+$, $\mathcal Q_-$ the supercharges associated with $\zeta_+$, $\zeta_-$, and by $H$, $J$ and $R$ the abstract operators 
associated to $ {\cal L}_\frac{\de}{\de t}$,  $ {\cal L}_\frac{\de}{\de \psi}$ and R-symmetry transformations,  respectively, 
 in this case  the algebra takes the form

\be\label{abstractSuperalgebra}
\{ \mathcal Q_+ , \mathcal Q_- \} \ = \ H + \frac{2}{\ell \sq^2} J -  \frac{1}{\ell \sq^2} R\,.
\ee
The commutators of $\mathcal Q_\pm$ with $H$, $J$ and $R$ can be inferred from the fact that the conformal Killing spinor $\zeta_+$ has R-charge $+1$ and satisfies\footnote{The  Lie derivative of a spinor field $\zeta$ along a vector $X$ is defined by $\mathcal L_X \zeta \,=\, X^i \nabla_i \zeta - \frac{1}{4} \nabla_i X_j \gamma^{ij} \zeta\,.$}
\be
\mathcal L_{\frac{\partial}{\partial t}} \zeta_+ \ = \ 0\,,\qquad \mathcal L_{\frac{\partial}{\partial \psi}} \zeta_+ \ = \ \frac{\ii}{2} \zeta_+\,.
\ee
Therefore we have $[R,\mathcal Q_\pm] = \pm\mathcal Q_\pm\,$,  $[J,\mathcal Q_\pm] = \mp\frac{1}{2} \mathcal Q_\pm\,$ and $[H,\mathcal Q_\pm] = 0\,$. 
These clearly identify $J$ as the generator of right $U(1)$ angular momentum, and $R$ as the generator of $U(1)_R\,$. The fact that $H$ commutes with the supercharges  identifies this uniquely as the operator relevant for the Euclidean path integral. In the round limit $\sq =1$, this is a two supercharges sub-algebra of the one given in eq.~(5.9) of~\cite{FestucciaSeiberg}. 

A second special gauge is $z\cdot \Acks =0$, which  can be reached by implementing the gauge transformation  $\Acks \to \Acks +  \frac{3\sq^2 -2}{2 \ell\sq^2}\diff t$. 
The conformal Killing spinor acquires a phase $\zeta_+ \to e^{\ii t \frac{3\sq^2-2}{2\ell\sq^2}}\zeta_+$, inducing the same dependence
 of the supercharges on $t$. This modifies the commutator with the generator of time translations 
 and changes the factor in front of $R$ in~\eqref{abstractSuperalgebra}. 
 Namely, denoting now by $\Delta$ the operator associated with $ {\cal L}_{\frac{\de}{\de t}}$, the  
 algebra becomes $[\Delta , \mathcal Q_\pm] = \mp \frac{3\sq^2 - 2}{2\sq^2} \mathcal Q_\pm$ and 
\be
\{ \mathcal Q_+ , \mathcal Q_- \} \ = \ \Delta + \frac{2}{\ell \sq^2} J -  \frac{3}{2\ell} R\, ,
\label{abstractSuperalgebra2}
\ee
with the other commutation relations unchanged. For $\sq = 1$, this reduces to a sub-algebra of the one in eq.~(6.11) of~\cite{FestucciaSeiberg}, and can be embedded in
the superconformal algebra on  $\mathbb R \times S^3_{\rm round}$ given \emph{e.g.}\ in~\cite{Romelsberger:2005eg,Kinney:2005ej}, where $\Delta$ is identified with the conformal 
Hamiltonian generating dilatations.

A third special gauge choice is  $z\cdot \Anm = 0$  (that is $z\cdot \Acks = - \frac 32 z\cdot \Vnm$), 
which can be reached by making the gauge transformation  $\Acks \to \Acks -  \frac{1}{\ell \sq^2}\diff t$, starting from (\ref{Acs}). 
Then the  spinor acquires a phase $\zeta_+ \to e^{-\frac{\ii t}{\ell\sq^2}}\zeta_+$; denoting by $P_0$ the generator associated
with ${\cal L}_{\frac{\de}{\de t}}$, the modified
commutation relations are $[P_0 , \mathcal Q_\pm] =  \pm\frac{1}{\sq^2}\mathcal Q_\pm$ and
\be
\{ \mathcal Q_+ , \mathcal Q_- \} \ = \ P_0  + \frac{2}{\ell \sq^2} J\, , 
\label{abstractSuperalgebra3}
\ee
which for $\sq =1$ reduces to a sub-algebra of the one in  eq.~(5.6) of~\cite{FestucciaSeiberg}. 

As all the anticommutators in (\ref{abstractSuperalgebra}), (\ref{abstractSuperalgebra2}), (\ref{abstractSuperalgebra3}) vanish on a supersymmetric state, each of them yields
a BPS relation. However, noting that the algebras can be mapped into each other by identifying 
\be\label{RelationsHamiltonians}
H -\frac{1}{\ell v^2}R \ =\ P_0 \ =\ \Delta - \frac{3}{2\ell}R\,,
\ee 
we see that these are all equivalent, for example to
\bea
 \ H + \frac{2}{\ell \sq^2} J -  \frac{1}{\ell \sq^2} R & = & 0\, .
 \label{bpscond}
\eea 
When $\sq=1$, this is precisely the shortening condition obeyed by the multiplets contributing to the supersymmetric index \cite{Romelsberger:2005eg,Kinney:2005ej,Gadde:2010en}.
 Inserting either of the anticommutators in the path integral one notices that (at least formally), the BPS condition (\ref{bpscond}) is obeyed by the vacuum expectation values of the operators, 
\emph{i.e.}  $\langle  H \rangle  + \frac{2}{\ell \sq^2} \langle J \rangle -  \frac{1}{\ell \sq^2} \langle R \rangle  = 0$, etcetera. We can therefore attempt to compare this relation with the relation among the holographic charges that we derived in section \ref{sec:HoloCharges}.

Since here we are interpreting the holographic charges as expectation values of operators in the field theory, we will denote them as
\bea
\langle J \rangle \ =  \ \frac{\pi \ell^3}{27 G}  \left(\sq^2 -1\right)^3 \,, \qquad  
\langle R \rangle \ =  \ - \frac{\ell }{\sqrt{3}} Q  \ = \ \frac{2\pi\ell^3}{27G}\left( \sq^2 - 1\right)^2 \, .
 \eea
  The interpretation of the holographic total energy $E$ is somewhat more subtle, as this contains the expectation value of 
a Hamiltonian operator plus a contribution from the vacuum energy. As noted around equation (\ref{SmarrGeneral}),
 this is most familiar in the context of  asymptotically AdS black holes \cite{GutowskiReall}, where the non-vacuum energy can be interpreted as the Ashtekar--Das mass. Although for general AlAdS solutions with non conformally flat boundary we do not have such a definition of mass, we will use the 
black hole example as a guide for interpreting the total energy  $E$ in our case. 

We begin writing the relation (\ref{BPSrelation}) in the form
\be
 E + \frac{2}{\sq^2\ell} \langle J \rangle   \ = \ \frac{\onshE}{\Delta_\tE}   - (z \cdot  \Acks)  \langle R \rangle  \,, 
 \label{comparetm}
\ee
 where the left hand side is manifestly gauge-invariant, and so is the right hand side (albeit less explicitly). Since 
 $E -\ \frac{\onshE}{\Delta_\tE}   + (z \cdot  \Acks)  \langle R \rangle $ is a gauge-invariant energy, and does not contain the contribution of the on-shell action 
 (which will be discussed in the next subsection), it is natural to identify it with the vacuum expectation value of a Hamiltonian operator. 
Comparing  (\ref{comparetm}) with the field theory BPS relation, we see that the natural identification is
\be
\langle P_0 \rangle  \ = \ E - \frac{I}{\Delta_t}  + (z \cdot \Acks) \, \langle R \rangle \, ,
\ee
or equivalently 
\be
\langle \Delta \rangle  \ = \ E - \frac{I}{\Delta_t} + (z \cdot \Anm) \, \langle R \rangle \, .
\ee
With this identification the field theory BPS relation and relation (\ref{BPSrelation}) match. 
The reason why we have not referred to as ``BPS relation'' the latter is that we have not derived it from the supersymmetry algebra of gauged supergravity. The black hole solution \cite{GutowskiReall} obeys the BPS relation 
\bea
M_\mathrm{AD} + \frac{2}{\ell} \langle J\rangle  - \frac{3}{2\ell} \langle R \rangle & = & 0\, , 
\eea
extracted  from the ordinary AdS$_5$ superconformal algebra, which is presumably valid for any asymptotically AdS solution.
Therefore the Ashtekar--Das mass $M_\mathrm{AD}$ may be identified as the vev of the conformal Hamiltonian $\langle \Delta \rangle$. 
Although we noted that relation~\eqref{BPSrelation}
reduces to this in the limit $v=1$ and in the gauge $z \cdot \Anm=0$, this is only formal, as in the case $v=1$ our solution reduces to pure AdS$_5$, so that 
$   \langle \Delta \rangle  =  \langle J \rangle =  \langle R\rangle =0$. Nevertheless, we expect that the 
generalized (\emph{i.e.} $v$-deformed)  BPS relation (\ref{bpscond})
can be derived from the supersymmetry algebra of five-dimensional gauged supergravity, analysed in the context of AlAdS solutions. 

The explicit expression  of $\langle \Delta \rangle$ in terms of $\sq$ reads 
\be
\langle \Delta \rangle\ =\ \frac{\pi \ell^2}{27 G\,\sq^2}(\sq^2-1)^2(2+\sq^2)~,
\ee
with the expressions for $\langle H \rangle$ and $\langle P_0 \rangle$ following from \eqref{RelationsHamiltonians}. These, however, are predictions for (universal)  one-point functions in
strongly coupled superconformal field theories at large $N$, which are not easily 
computable. 
In the next section we will turn our attention to the on-shell action in the specific
 gauge leading to the algebra \eqref{abstractSuperalgebra}; we will interpret it as a Casimir energy, and then compare it with a corresponding quantity on the field theory side.

\subsection{Supersymmetric index and Casimir energy}\label{IndexAndCasimir}

In this section we wish to discuss the field theories from the point of view of the path integral, therefore 
we will concentrate on the Euclidean signature version of the background, with periodically identified time.  
In order to have well-defined Killing spinors 
(at the boundary as well as in the bulk) we must fix the gauge in which these are time-independent, namely  $\Acks$ is the one in~\eqref{Acs}, thus justifying \emph{a posteriori} the choice of gauge for the graviphoton field $A_\mu$ made in section~\ref{UVanalysis}.
In this case the generator of time translations in the field theory $H$ commutes with the supercharges, $[\eqft, \mathcal Q_\pm]=0\,$.

The AdS/CFT master equation \cite{Gubser:1998bc,WittenAdS}  relates the holographically renormalised 
supergravity action evaluated on a classical five-dimensional geometry $M_5$ 
to the  Euclidean path integral in the field theory defined  on the four-dimensional boundary $M_4=\de M_5$, namely
\bea
e^{-S_\mathrm{gravity} [M_5]} &  = & Z_\mathrm{QFT} [M_4 ] \qquad \mathrm{for}\qquad  N \to \infty~.
\label{master}
\eea
In particular, when $M_5\simeq S^1\times \mathbb{R}^4$ (topologically) 
the path integral on $M_4 \simeq S^1 \times S^3$  is performed with \emph{periodic} boundary conditions for the fermions on $S^1$, 
and therefore it is related to the trace over all the states of the field theory \cite{WittenAdS}  as
\bea
Z_\mathrm{QFT} [M_4 ] & \simeq & \mathrm{Tr}[(-1)^F e^{-\beta H}]\ \equiv\  {\cal I}(\beta) \, , 
\label{notquite}
\eea
where $F$ is the fermion number, $\beta$ is the circumference of $S^1$, and $H$ is in general an operator 
commuting with the supercharges preserved by the background.

Let us first consider the case  $M_4=S^1_\beta \times S^3_\mathrm{round}$.
Then $H$ is the Hamiltonian commuting with the supercharges defined on  $S^1\times S^3_\mathrm{round}$, 
and the trace is the supersymmetric index, which may also be defined in radial quantisation 
\cite{Romelsberger:2005eg,Kinney:2005ej}. The \emph{Casimir energy} of a quantum field theory on $S^3$ may be defined from 
the  limit of large  $S^1_\beta$ radius of  the partition function   as
\bea
\eind & \equiv & - \lim_{\beta \to \infty}  \frac{\diff }{\diff \beta}  \log Z_\mathrm{QFT} [S^1_\beta \times S^3_\mathrm{round}]~.
\label{usualcasimir}
\eea
In the semiclassical gravity approximation, when we can use (\ref{master}), the relation  $S_\mathrm{gravity} [M_5]= \beta \eind$
 is  valid for any $\beta$; when one introduces additional background fields, 
 as long as  translation invariance along the $S^1$ is preserved, the renormalised  
on-shell action is guaranteed to remain linear in $\beta$ \cite{WittenAdS}.  However, while in the large $N$ limit $S_\mathrm{gravity} = {\cal O}(N^2)$, 
the supersymmetric index ${\cal I}(\beta)$ is of order ${\cal O}(N^0)$ \cite{Kinney:2005ej}, 
which appears to contradict the relation (\ref{master}) together with (\ref{notquite}). 
The caveat is that the precise  version of (\ref{notquite}) includes a factor  $\prefac$, often omitted in the literature 
\bea
Z_\mathrm{QFT} [S^1_\beta\times S^3_\mathrm{round}] & = &  e^{-\prefac} \cdot  {\cal I} (\beta )~.
\eea 
This implies that, at least for field theories with a Freund--Rubin gravity dual in type IIB supergravity with $N$ units of five-form flux, we must have $\prefac = {\cal O}(N^2)$, and  
 using  (\ref{master}) we see that\footnote{When a gravity dual exists,  the Casimir energy  (on the round $S^3$) has been argued 
in \cite{Herzog:2013ed} to be proportional to the anomaly coefficient $\aano=\cano = {\cal O}(N^2)$.} 
\bea
\prefac \, = \, S_\mathrm{gravity} \, =\,  \beta\, \eind    \qquad \quad \mathrm{for}\qquad N \to \infty~.
\eea
In a related context, in a calculation of the partition function of a weakly coupled gauge theory on $S^1_\beta \times S^3_\mathrm{round}$with \emph{anti-periodic}  boundary conditions on $S^1$,   in \cite{Aharony:2003sx}  it was shown that $\prefac$ is the usual Casimir 
energy of the field theory on $S^3$. We shall return to the interpretation of $\prefac $ for the path integral with supersymmetric boundary conditions momentarily.

More generally, one can introduce a refined supersymmetric  index with  \emph{fugacities} turned on
\cite{Romelsberger:2005eg,Kinney:2005ej,Gadde:2010en,Eager:2012hx,CDFK}, which takes the form 
 \bea
 {\cal I} (p,q) & = & \mathrm{Tr}[(-1)^F p^{J+J'-\frac{R}{2}} q^{J-J'-\frac{R}{2}}]   \, ,
 \eea 
where $p, q$ are complex parameters and $J,J'$ denote the Cartan generators of $SU(2)$ and $SU(2)'$, respectively, and $R$ is the R-charge. This receives contributions only from states obeying 
the relation  $H  =  2J - R$, and hence  it reduces to the unrefined index  ${\cal I}(\beta)$ upon setting $p=q=e^{-\beta}$. 
In \cite{CDFK} it has been argued that  this index is computed by 
the path integral on a compact Hermitian manifold $\Hopf_{p,q}$ homeomorphic to $S^1 \times S^3$, namely a (primary)  
\emph{Hopf surface} \cite{kodaira}. Thus 
\bea
Z_\mathrm{QFT} [ \Hopf_{p,q} ] & = &  e^{-\prefac}   \cdot  {\cal I} (p,q)\, ,
\eea 
where $\prefac$ may depend on the complex structure of the manifold, as well as on counterterms local in the  background fields.   

Let us briefly recall the definition of a Hopf surface $\Hopf_{p,q}$ and make contact with our boundary four-dimensional geometry. 
A complex manifold homeomorphic to $S^1\times S^3$  is necessarily a primary Hopf surface \cite{kodaira}, namely a quotient of 
$\mathbb{C}^2 - (0,0)$ by an infinite cyclic group. In particular, denoting as $(z_1,z_2)$ the complex 
coordinates on $\mathbb{C}^2 - (0 ,0)$, the quotient acts as
\be
(z_1,z_2) \sim (p z_1,\,q z_2)~,
\ee
where $(p,q)$ are complex numbers with $0< |p|\leq  |q|<1$. The complex structure of the resulting manifold 
$\Hopf_{p,q}$ depends on the two parameters $(p,q)$. For more details we refer to  \cite{CDFK} and references therein. 
It is straightforward to determine
the complex structure $(p,q)$ compatible with a general metric on $S^1\times S^3$ with   $SU(2)\times U(1)\times U(1)$ isometry, namely
\be\label{metric4d}
\diff s^2 \ = \ r^2 \left[\tfrac{1}{4} (\sigma_1^2 +\sigma_2^2 + \sq^2 \sigma_3^2 ) + ( b  \,\diff t + \tfrac{k}{2}\, \sigma_3)^2\right]\, ,
\ee
where we have included a mixed term in the $S^1$ fibration over $S^3$, parameterised by the constant $k$. We will then set $k=0$ to make contact with our 
solution. 
We introduce complex coordinates
\be\label{complcoords}
z_1 \ =\ e^{\frac{\sq + \ii k}{\sq^2 +  k^2} b\,  t}\cos\frac{\theta}{2}\,
e^{\frac{\ii}{2}(\psi + \phi)} \,,\qquad z_2 \ =\ e^{\frac{\sq + \ii k}{\sq^2 +  k^2} b \, t}\sin\frac{\theta}{2}\, e^{\frac{\ii}{2}(\psi-\phi)} \,,
\ee
with  respect to the complex structure $\complstr^i{}_j= h^{ik} \complstr_{kj}$ associated to the fundamental two-form 
\bea
\complstr & = & \frac{r^2}{2} \left(b\, v \,\diff t \wedge \sigma_3  + \frac{1}{2}\sigma_1 \wedge \sigma_2 \right)~. 
\eea
Assuming the identification $t \sim t+\Delta_t $, from~\eqref{complcoords} we read off 
\be
p \ =\ q \ = \ e^{\frac{\sq + \ii k}{\sq^2 +  k^2} b  \Delta_t} ~.
\ee
In particular,  taking  $k=0$ and $b=-\frac{1}{\sq\ell}$ as in our solution, we have that  $p=q = e^{-\tfrac{\Delta_t}{\ell v^2}}\, \equiv \, e^{-\beta}$. 
This is precisely the standard  complex structure of the \emph{round} $S^1_\beta \times S^3_\mathrm{round}$, 
with ratio of radii given by $\beta =  \frac{\Delta_t}{\ell v^2}$, albeit the metric is  \emph{not} the round 
one. In the complex coordinates~\eqref{complcoords}, the metric~\eqref{metric4d} reads\footnote{We observe that the second term vanishes if and only if the metric is conformally flat and thus locally related by a coordinate transformation to the metric on the round $S^1 \times S^3$. Note also that this metric is different from the metric (4.7) in~\cite{CDFK} and coincides with it 
only in the conformally flat limit.}
\be
\diff s^2  \ = \ r^2\left[ (\sq^2 + k^2)\frac{ \diff z_1 \diff \overline z_1 + \diff z_2\diff \overline z_2}{{|z_1|^2 + |z_2|^2} } + (1 - v^2 - k^2) \frac{(z_1 \diff  z_2 - z_2 \diff z_1)(\overline z_1 \diff  \overline z_2 - \overline z_2 \diff \overline z_1)}{{\left(|z_1|^2 + |z_2|^2\right)^2}} \right] ,
\ee
with our metric corresponding to $k=0$, $b = -\frac{1}{\sq\ell}$ and $r=2a_0\ell$. According to \cite{CDFK}, the localised partition function of a supersymmetric 
gauge theory on our background must be proportional to the \emph{unrefined} index,  depending on the parameter $\beta=\tfrac{\Delta_t}{\ell\sq^2}\,$.

Let us now return to the interpretation of $\prefac$ in this case.\footnote{As discussed above, we expect that in general $\prefac$ should  contain information on a supersymmetric Casimir energy of the theory placed on an arbitrary Hopf surface $\Hopf_{p,q}$. This point will be addressed elsewhere \cite{ACM}.} 
  The authors of \cite{Kim:2012ava}  have  conjectured\footnote{In three dimensions,  a detailed derivation of the analogous quantity from the path integral on $S^1\times S^2$ 
 was presented in   \cite{Kim:2009wb}. We thank S.~Kim for clarifying comments.}  that
\bea
Z_\mathrm{QFT}[S^1_\beta\times S^3_{\rm round}] & = & x^{\epsilon_0} \cdot  {\cal I} (\beta)
\label{ztoind}
\eea
where $x= e^{-\tfrac{2}{3}\beta}$ and
\bea
\epsilon_0 & = & \mathrm{tr}[(-1)^F H ] \, = \, \mathrm{tr}_\mathrm{bosons}[H] - \mathrm{tr}_\mathrm{fermions}[H]\,.
\label{eindex}
\eea
When a  free field theory limit exists  this is just the sum of zero-point ``energies'', namely eigenvalues of $H$, 
weighted by a sign, and it was therefore  referred as to \emph{index Casimir energy} in  \cite{Kim:2009wb,Kim:2012ava}. This expression is divergent and must be regulated, with the result depending on the choice of regularisation. 
 As $H$ commutes with the supercharge(s), a natural regularisation \cite{Kim:2012ava} is to weight the terms in the infinite sum 
  with a factor of   $x^H$,  sending $x\to 1^-$ at the end.   One obtains
   \bea
\epsilon_0 & = & \lim_{x\to 1^-} \mathrm{tr}[(-1)^F H x^{H}] \ =\ \frac{1}{2} \lim_{x\to 1^-} x \frac{\diff}{\diff x}\sum_\mathrm{all~fields} 
\left(f_\mathrm{chiral} (x) + f_\mathrm{vector}  (x)  \right)\, , 
\label{baggio}
\eea
 where the letter indices  for a chiral multiplet with R-charge $r$ and a vector multiplet  are given by 
 \bea
f_\mathrm{chiral} (x) \, = \, \frac{ x^{3r/2} - x^{3(2-r)/2}}     {(1-x^{3/2})^2}\, , \qquad f_\mathrm{vector}  (x)\, = \,  \frac{2 x^3 - 2x^{3/2}}  {(1-x^{3/2})^2} \, ,
\eea
 respectively  \cite{Romelsberger:2005eg,Gadde:2010en}. 
In the case of interest to  us the symmetry of the theory is broken to $SU(2)\times U(1)\times U(1)_R$ explicitly by the squashing, therefore there is no compelling reason for using a regulator that would preserve $SO(4)$ rotation symmetry. 
See \cite{Kim:2013nva} for a related discussion in five dimensions.   

Using  (\ref{baggio}) and recalling that for a superconformal field theory the trace anomaly coefficients are 
\bea
\aano \, = \, \frac{3}{32} \left( 3\,\mathrm{tr}\,\rano^3 - \mathrm{tr}\,\rano \right) \, , \qquad \cano \, = \, \frac{1}{32} \left( 9\,\mathrm{tr}\,\rano^3 -  5 \,\mathrm{tr}\,\rano \right)\, , 
\eea 
where $\mathrm{tr}\,\rano^\alpha$ denotes the sum over the (R-charge)$^\alpha$ of all fermionic fields,
for a quiver gauge theory  one obtains  \cite{Kim:2012ava}
\bea
\frac{x}{2}\frac{\diff f(x)}{\diff x} & = & - \frac{24}{\beta^2} (\aano- \cano) + \frac{2}{9}(\aano+3\,\cano) + {\cal O}(\beta)\, .
\eea
After subtracting the first term, which is divergent in the $\beta \to 0$ limit 
(in any case for superconformal quivers this term vanishes \cite{Benvenuti:2004dw}), one obtains
\bea
\prefac & = &  \frac{2}{3}\beta \epsilon_0 \, = \,   \frac{4}{27}\beta\,(\aano+3\,\cano) \, .
\eea
Note that although we have taken the limit $\beta\to 0$  in (\ref{baggio}) to derive $\epsilon_0$, there is no contradiction with the definition of the Casimir energy given  in (\ref{usualcasimir}), and we see that from the latter we have
$\eind = \frac{2\epsilon_0}{3} - \lim_{\beta \to \infty} \frac{1}{\beta}\log {\cal I}(\beta) \,=\, \frac{2\epsilon_0}{3} + {\cal O}(1)$, in the limit $N\to \infty$.
For a superconformal quiver we can therefore compare  $\eind$   with the gravity side writing 
 \bea 
\eind & = & \frac{16}{27}\,\aano  \ =\  \frac{2}{27}  \frac{\pi\ell^3}{G}\, ,
\label{aspetta}
\eea
where we  used the holographic  relation $\aano  =  \frac{\pi\ell^3}{8G}$.

\subsection{Comparison with the on-shell action}

In section \ref{HoloRenSection} we computed the on-shell gravity action with supersymmetric $S^1\times S^3_v$ boundary conditions. After removing all divergences using the holographic counterterms, this reads 
\be
\label{copyon}
\onshE \ = \ \frac{\Delta_\tE}{\ell \sq^2} \, \frac{\pi\ell^3}{G} \,  \left[ \frac{2}{27} + \frac{2}{27}\sq^2 - \frac{13 }{108}\sq^4 +\frac{19}{288}\sq^6 \right].
\ee
Notice that the first term in the square bracket depends linearly  on the complex structure parameter~$\beta=\frac{\Delta_\tE}{\ell \sq^2}$, while the remainder is  
a polynomial in the squashing parameter $\sq^2$. The former agrees precisely 
with the field theory expectation (\ref{aspetta}) and we would like to interpret this as the 
(index) Casimir energy, while the remainder polynomial in $\sq^2$ should be identified with a local counterterm.

In the limit $v^2=1$  the solution becomes AdS$_5$ and (\ref{copyon}) reduces to $\frac{I}{\Delta_\tE} = E =  \frac{3}{32} \frac{\pi\ell^2}{G}$, which in 
\cite{BalasubramanianKraus} was interpreted as the Casimir energy on $S^3$. On the other hand, the $v^2=1$ limit of our proposal for the  Casimir energy is 
$ \frac{2}{27}\frac{\pi\ell^2}{G}$. The ratio between the former and the latter is $\frac{27}{32}\frac{3}{2}$, 
in agreement\footnote{Noticing the $\beta$ in  \cite{Kim:2012ava} and our $\beta$ differ by a factor of  $3/2$.} with 
\cite{Kim:2012ava}. In that reference it was argued that the difference with respect to the usual Casimir energy arises from the use of a regularisation not respecting the full symmetry of $S^1 \times S^3$.

Let us now discuss  the polynomial in $\sq^2$ in (\ref{copyon}).
In five dimensions, holographic renormalisation suffers from ambiguities associated to local, scale invariant counterterms 
that remain finite when the UV cut-off is sent to infinity. This means that after having removed  all the divergences, one may still add these finite terms to the action, 
with arbitrary coefficients. 
Although it is believed that different choices of counterterms should correspond
to different choices of renormalisation scheme in the dual field theory~\cite{BirrellDavies}, 
a complete understanding of the details is currently lacking. 
For a bulk action comprising only the Einstein--Hilbert term and the cosmological constant, 
generically one can construct three inequivalent  
finite counterterms using the metric and covariant derivatives \cite{BalasubramanianKraus}.
These are given by integrals of $R^2$, $R_{ij}R^{ij}$ and   $R_{ijkl}R^{ijkl}$,  where $R$, 
$R_{ij}$ and $R_{ijkl}$, are the Ricci scalar, Ricci tensor and Riemann tensor of the boundary metric, respectively. In the presence of a gauge field,
a fourth finite local invariant is $F_{ij}F^{ij}$. Equivalently, these may be parametrized by a basis given by 
$\mathcal E$,  $C_{ijkl}C^{ijkl}$, $R^2$ and $F_{ij}F^{ij}$, where $\mathcal E$ is the Euler density
$\mathcal E  = R_{ijkl}R^{ijkl} - 4 R_{ij}R^{ij} + R^2$,
and $C_{ijkl}C^{ijkl}$ is the square of the Weyl tensor,
$C_{ijkl}C^{ijkl}  = R_{ijkl}R^{ijkl} - 2 R_{ij}R^{ij} + \frac 13 R^2 $.
In particular, we have  the following standard finite counterterm for the on-shell action:
\be\label{ambij}
\Delta S \ =\ \frac{\ell^3}{8\pi G} \int_{\partial\Mfive}\diff^4x \sqrt h \left( \alpha\, \mathcal E + \beta\, C_{ijkl}C^{ijkl} + \gamma\, R^2 - \frac{\delta}{\ell^2} F_{ij}F^{ij} \right),
\ee
where $\alpha$, $\beta$, $\gamma$, $\delta$ are arbitrary numerical constants. 
For our background, $\mathcal E = 0$ and \eqref{ambij} gives a quadratic polynomial in $\sq^2$:
\be
\Delta S \,\ \propto \,\ \frac{\gamma}{4}\left(4-\sq^2 \right)^2 + \frac{1}{6}\left(8\beta - \delta\right) \left(1-\sq^2\right)^2 ~.
\ee
Tuning the coefficients $\beta, \gamma, \delta$ we can cancel some of the polynomial terms in (\ref{copyon}), suggesting that indeed these are ambiguous, but there is no choice cancelling all terms simultaneously. 
The main issue here is whether there exist \emph{new} types of counterterms that may be used  to modify the on-shell action (\ref{copyon}). Below we will discuss how new counterterms  may emerge naturally taking into account the additional structure arising from supersymmetry. 

The terms in (\ref{ambij}) are, apart for $R^2$, those appearing in the bosonic action of conformal supergravity. Therefore
  the only combination that is supersymmetric, with respect to \emph{rigid supersymmetry}, 
is that with $\gamma=0$ and $\delta = 8 \beta$, which vanishes~\cite{SuperWeylAnomaly_paper}.
However, the correct notion of supersymmetry in the context of holography is that of the bulk, and should involve also the divergent counterterms;
taking this into account it should be possible to formulate a supersymmetric version of holographic renormalisation.
From this point of view, it is very natural to expect that the supersymmetric counterterms should be derived 
by expanding asymptotically the $G$-structure encoding the bulk supersymmetry. This was done at leading order in Ref.~\cite{Cassani:2012ri}, where the $\mathbb{R}^2$-structure on the Lorentzian boundary was obtained from the
$SU(2)$-structure in the bulk.  
In Euclidean signature,\footnote{In the following we consider the  Euclidean version of our background, but similar considerations can be done in Lorentzian signature.} the relevant $G$-structure is a metric-compatible complex structure $\complstr_i{}^j$. 
Since on our background $\mathcal E = 0$ and $C_{ijkl}C^{ijkl} \propto F_{ij}F^{ij}$, the ambiguity so far may be parameterized by $R^2$ and $F_{ij}F^{ij}$. 
Using the boundary 
complex structure tensor $\complstr_i{}^j$ one can construct additional scale-invariant  counterterms. 
Here we will not attempt to classify systematically all these terms and to study how supersymmetry constrains them. We simply notice that using the complex structure we can 
introduce the \emph{Ricci form} of the boundary geometry, $\mathcal R_{ij} = \frac{1}{2} R_{ijkl} \complstr^{kl}$, which is invariant under global rescaling of the metric. Then a new finite counterterm is given by
\be\label{RicciFormTerm}
\int \diff^4 x \sqrt h\, \mathcal R_{ij}\mathcal R^{ij} \ =\   4 \pi^2  \left(4 - 3 \sq^2 \right)^2 \frac{\Delta_t}{\ell} \,.
\ee
This polynomial in $\sq^2$ is independent of those obtained by evaluating the $R^2$ and $F_{ij}F^{ij}$ terms. Other obvious terms such as $(\mathcal R_{ij}\complstr^{ij})^2$ and  $\mathcal R_{ij}F^{ij}$ are not independent, hence we do not need to consider them. 
 Combining  these three counterterms with suitably chosen coefficients, we can remove the remainder polynomial in $\sq^2$ in~\eqref{copyon}. Namely, we find that
\be\label{OnShellWithoutPolynomial}
\onshE  -\frac{1}{108}\frac{\ell^3}{8\pi G} \int_{\partial M}\diff^4x \sqrt h \left( \frac{7}{24} R^2 + \frac{17}{\ell^2} F_{ij}F^{ij} - \mathcal R_{ij}\mathcal R^{ij}\right) \ = \ \frac{2}{27} \frac{\Delta_\tE}{\ell\sq^2}\frac{\pi\ell^3}{G} \,.
\ee
In appendix \ref{AmbiguousApp}  we discuss how the addition of these counterterms affects the holographic energy-momentum tensor and the R-symmetry current. One can see that the value of the R-charge $Q$ and the angular momentum $J$ are not modified, while the total energy $E$ is shifted by the same amount that changes the on-shell action. Therefore, the relation (\ref{SmarrGeneral}) remains valid as well as~\eqref{RelFromChainRule}, and 
the vevs of the gauge-invariant hamiltonians introduced in section~\ref{algebsection}
are  not affected. 

In conclusion, we have shown that the supergravity on-shell action contains a term with the correct linear dependence on the complex structure modulus $\beta$,
that reproduces  the  (index) Casimir energy expected from a supersymmetric field theory on our $S^1\times S_v^3$ background. 
However, it also contains other terms that we suggested to be removable by a choice of local counterterms constructed with the background metric, the background gauge field and the complex structure.

From the point of view of the boundary field theory, in addition to the counterterms constructed
with the metric and the conformal supergravity gauge field $A^{\rm cs} = A^{\rm nm} - \frac{3}{2} V^{\rm nm}$, it is natural to also include counterterms obtained from other combinations of the new minimal auxiliary fields $A^{\rm nm}$ and $V^{\rm nm}$.\footnote{We thank Z. Komargodski for emphasizing this to us. See \cite{Closset:2012vg,Closset:2012vp} for a detailed discussion of counterterms in three dimensions, and \cite{Cecotti:1987qe} for four-dimensional counterterms from new minimal supergravity.} It would be interesting to study how these can arise from supersymmetry in the bulk and thus be employed in holographic renormalization.
  
Recently the authors of \cite{Bobev:2013cja} constructed  a five-dimensional supersymmetric solution and presented evidence that this  should be dual to a massive deformation of ${\cal N}=2$ super Yang--Mills theory on $S^4$. In particular, they showed that the renormalised on-shell action agrees with the logarithm of the localised partition function on $S^4$ \cite{Pestun:2007rz}, but only up to terms argued to be ambiguous. In this respect, our construction is somewhat analogous to theirs.

\section{Conclusions}
\label{disc}

In this paper we have constructed a new supersymmetric AlAdS one-parameter family of solutions to minimal five-dimensional gauged supergravity. This can be uplifted to type~IIB or to eleven-dimensional supergravity by using the consistent truncations of~\cite{Buchel:2006gb,Gauntlett:2007ma,GauntlettColgainVarela}.
The parameter $\sq^2$ deforming AdS$_5$ into our solution may be thought of as a squashing of the boundary metric, together with the addition of 
a non-trivial graviphoton field, therefore it corresponds to a relevant deformation of the Lagrangian in the dual superconformal field theory. 
The solution is smooth and has the topology of global AdS$_5$, namely $\mathbb{R}_t \times \mathbb{R}^4$;  for 
generic values of the parameter it  preserves an $SU(2)\times U(1)\times U(1)$ subgroup of the isometry group $SO(2,4)$ of AdS$_5$.  
Despite we constructed the solution using a combination of perturbative expansions and numerical integration of the relevant ODE, many 
quantities of physical interest were computed analytically as a function of the parameter~$\sq^2$. These include the on-shell action, the total energy, 
the angular momentum and the electric charge. We also provided numerical evidence that the solution is free from closed timelike curves.
After a Wick rotation of the time coordinate, at the  boundary the metric remains real while the background gauge field becomes complex.  
In the bulk both the gauge field and the metric become \emph{complex}, but the appearance of complex metrics in quantum gravity should not be 
surprising (see \emph{e.g.} \cite{complexmetric1,complexmetric2}).

Our main motivation for studying this supergravity solution is that via the gauge/gravity correspondence it provides the dual to a class of superconformal field theories defined on a four-dimensional curved manifold with certain background fields. In particular, after the analytic continuation, 
the Riemannian manifold is homeomorphic to $S^1\times S^3$, equipped with a non conformally flat Hermitian metric, namely it is a particular Hopf surface \cite{kodaira}. 
It has been argued in \cite{CDFK} that the localised partition function of supersymmetric field theories on a Hopf surface with complex structure specified by two 
parameters $(p,q)$ is proportional to the refined supersymmetric index with $(p,q)$ fugacities.   
The proportionality factor is expected to capture a supersymmetric \emph{Casimir energy} of the field theory, which 
may be expressed as a linear combination of the $\aano$ and $\cano$ anomaly coefficients~\cite{Kim:2012ava}. 
Since our background has complex structure $p = q \in \mathbb R$, we have compared the renormalised on-shell supergravity 
action\footnote{Here we have assumed that our solution is the unique supersymmetric filling of the boundary data, 
but we cannot exclude that there exist other fillings.
Note that a solution with the topology of  $\mathbb{R}^2\times S^3$, leads to the antiperiodic spin structure 
 on the boundary, and therefore it cannot contribute $\mathrm{Tr} (-1)^F e^{-\beta H}$ \cite{WittenAdS}.
 See \cite{Martelli:2012sz} for a related discussion in the context of asymptotically locally AdS$_4$ solutions.} to the partition function associated with the unrefined index,  obtaining  agreement with the field theory expectation, up to terms that may be removed by a finite local counterterm. We plan to come back to a more systematic analysis of the allowed supersymmetric counterterms in the future.

In \cite{Kinney:2005ej} it was noticed that the large $N$ behavior of the index could not be matched with the entropy of the supersymmetric black hole of  \cite{GutowskiReall}, which must scale like $S_\mathrm{BH}\sim {\cal O}(N^2)$. It may be possible to revisit this puzzling feature in the light of the discussion in the present paper. 

There are a number of possible extensions of this work that it is natural to study.
For simplicity, in this paper we have restricted our attention to a four-dimensional boundary with direct product  metric, while as discussed in  \cite{SuperWeylAnomaly_paper}
by a simple change of local coordinates it is straightforward to include a mixed term (a non-zero parameter $k$ in the metric (\ref{metric4d})).  As seen in section~\ref{IndexAndCasimir} this deformation corresponds to complexifying the complex structure parameter $p=q$ of the Hopf surface. 
It would be nice to study this modification 
of our solution in more detail (notice that this applies also to the unsquashed case $\sq=1$). 
 A more complicated generalisation consists in reducing the symmetry of the ansatz, for example down to a $U(1)^3$ sub-group. 
 In this case the supersymmetry conditions will lead to partial differential equations, making the problem of finding 
solutions much more difficult. 

A further interesting direction is that of constructing gravity solutions dual to supersymmetric field theories on four-dimensional Hermitian manifolds with topology different from $S^1\times S^3$,  preserving  generically only one supercharge. This is a very hard problem for a number of reasons. Firstly, while in the presence of 
two supercharges with opposite R-charges (that we considered presently) there exist al least two 
commuting
Killing vectors, a generic Hermitian background is not guaranteed to possess any isometry.
 We expect that in order to construct solutions with only one supercharge it will be necessary to start from Euclidean supergravity with general complex gauge field,
and possibly complex metric. Moreover, there exist topological obstructions for constructing smooth five-dimensional manifolds filling  a four-dimensional boundary; 
for example, it is not possible to construct a smooth five-dimensional solution  whose boundary is (topologically) $\mathbb{C}P^2$, while the first del Pezzo 
surface $dP_1$ is not obstructed \cite{SuperWeylAnomaly_paper}.

\subsection*{Acknowledgments}

We would like to thank B.~Assel, Z.~Komargodski, S.~Murthy and P.~West for insightful 
discussions, as well as  S.~Ferrara, S.~Giusto, S.~Kim, J.~Sparks, A.~Tomasiello and A.~Tseytlin
for useful correspondence.  D.C. is supported by the STFC grant ST/J002798/1. 
D.M. is supported by the ERC Starting Grant N. 304806,  
``The Gauge/Gravity Duality and Geometry in String Theory'', 
and also acknowledges partial support from the STFC grant ST/J002798/1.

\appendix

\section{Details on the asymptotic solution}
\label{DetailsSol}

In this appendix we provide more details on the UV solution. We set $\ell=1$.

\subsection{More terms in the UV expansion}

Here we give some more details about the solution at large $\trho$. From the discussion in section~\ref{susysec}, the five-dimensional metric has the form
\be
\diff s^2  = - f^2 (\diff y + \Psi\hat\sigma_3)^2 + f^{-1}\left[\diff \rho^2 + a^2(\hat{\sigma}_1^2+\hat{\sigma}_2^2) + (2aa')^2\hat{\sigma}_3^2 \right],
\ee
while the gauge field is given by
\be\label{formulaFappendix}
F \ =\ \frac{\sqrt{3}}{2} \,\diff \left[  f \diff y + \left(f\Psi +\frac{4 a'^2 + 2 a a'' -1}{3}\right) \hat{\sigma}_3 \right].
\ee

The UV expansion of the functions $f$ and $\Psi$ is obtained by plugging the UV solution~\eqref{UVsola} for $a$ into~\eqref{invfgeneral} and~\eqref{exprPsiGeneral}. We obtain
\bea\label{solftrho}
f \!\!&=& 1 + \left[\frac{1 + 16 a_2 +4 \cL}{12} + \frac{4\cL}{3}\trho \right]\frac{e^{-2\trho}}{a_0^2} +  \bigg[\frac{1 +8(a_2 +3\cL) - 16(8a_2^2-6a_2\cL +5\cL^2)}{144} \nn\\ [2mm]
&& +\; \frac{1 - 32a_2 + 12\cL}{18}\cL\,\trho - \frac{8\cL^2}{9 }\,\trho^2 \bigg] \frac{e^{-4\trho}}{a_0^4} \; + \; \mathcal O(e^{-5\trho})\,
\eea
and
\bea
\Psi \!\!\!&=&\!\!\! -2a_0^2 e^{2\trho} + \frac{1}{2} + 4 a_2 - 2\cL  + 4\cL\trho +\bigg[ \frac{-1+192 a_4 +16(\cL - 2 a_2) - 8 (44a_2^2 - 20a_2\cL +3\cL^2)}{48}\nn \\ [2mm]
&& \!\!\!+\; \frac 53 \cL(\cL - 12 a_2) \trho -10\cL^2 \trho^2  \bigg] \frac{e^{-2\trho}}{a_0^2}
\,+\, \mathcal O(e^{-3\trho})\,.
\eea
Passing from the coordinates $y,\hat\psi$ to the coordinates $t,\psi$ introduced in~\eqref{changepsi}, namely
\be
\hat{\psi} \ =\ \psi + \chi\,t
\,,\qquad y \ =\ t\,,\qquad {\rm with}\;\; \chi \,=\, -\frac{2}{1 - 4 \cL}\,,
\ee
the five-dimensional metric can be expressed as
\be
\diff s^2 \ = \ g_{\rho\rho} \diff \rho^2 + g_{\theta\theta}\left( \sigma_1^{\,2} + \sigma_2^{\,2} \right) + g_{\psi\psi}\sigma_3^{\,2} + g_{tt}\diff t^2 + 2 g_{t\psi} \,\sigma_3\, \diff t\,,
\ee
with
\bea 
g_{\rho\rho} &=& f^{-1}\,,\qquad\qquad\;\;\; g_{\theta\theta} \ =\ f^{-1}a^2 \,,\qquad\qquad\;\;\; g_{\psi\psi} \ = \ - f^2 \Psi^2 + f^{-1}(2aa')^2\,, \nn\\ [2mm]
g_{tt} & = & - f^2(1 + \chi \Psi)^2 + \chi^2 f^{-1} (2aa')^2 \,,\quad g_{t\psi} \ =\ -f^2 (1+\chi \Psi)\Psi + \chi f^{-1} (2aa')^2\,.\qquad\label{MetricCompsImplicit}
\eea
By substituting the UV solution for $a$, $f$, $\Psi$, these metric components read:
\bea
g_{\rho\rho} &=& 1 - \left[ \frac{1 + 16a_2+4\cL}{12} - \frac{4\cL}{3}\trho\right] \frac{e^{-2\trho}}{a_0^2} \,+\, \mathcal{O}(e^{-3\trho})\,,\nn \\[2mm]
 g_{\theta\theta} &=& a_0^2\, e^{2\trho} + \frac{-1 + 8a_2 - 4 \cL}{12} + \frac{2\cL}{3}\trho \,+\, \mathcal{O}(e^{-\trho})\,,\nn\\ [2mm]
g_{\psi\psi} &=& \left( 1 - 4\cL \right)\left[ a_0^2\,e^{2\trho} + \frac{-1 + 8 a_2 + 20 \cL}{12} + \frac{2\cL}{3}\trho
\right] \,+\, \mathcal{O}(e^{-\trho})\,,\nn \\[2mm]
g_{tt} &=& -\frac{4a_0^2}{1-4\cL} e^{2\trho} + \frac{2+8a_2 + 8\cL(1+\trho)}{12\cL -3} \,+\, \mathcal{O}(e^{-\trho})\,,\nn\\ [2mm]
g_{t\psi} &=& \mathcal{O}(e^{-2\trho})\,.\label{UVmetric}
\eea
In order to see the parameters $a_4$, $a_6$ appear in the metric one needs to go one order further. We will not present this here, as the expressions become cumbersome, but will do it in the next subsection when we will turn to Fefferman--Graham coordinates.

In a gauge ensuring that the five-dimensional supersymmetry parameter $\epsilon$ does not depend on $t$, the gauge potential can be written as
\be\label{formofA}
A \ = \ A_t \,\diff t + A_\psi\, \sigma_3\,,
\ee
with
\bea
A_t &=& \frac{\sqrt{3}}{2} \left(f + \chi f\Psi +\chi \frac{4 a'^2 + 2 a a''}{3} \right) ,\nn \\[2mm]
A_\psi &=& \frac{\sqrt{3}}{2}\,\left(f\Psi +\frac{4 a'^2 + 2 a a'' -1}{3}\right) .\label{AcompImplicit}
\eea
By plugging the UV expansion in, these read
\bea
A_t \!\!&=&\!\! \frac{1}{2\sqrt 3} + \frac{5-256 a_2^2 -384 a_4 +32 \cL -232 \cL^2 + 32 a_2 (2-5\cL)}{48\sqrt3(1-4\cL)}\frac{e^{-2\trho}}{a_0^2} + \mathcal O(e^{-3\trho})\,, \nn \\ [2mm]
A_\psi \!\!\!&=&\!\! -\frac{2}{\sqrt 3}\cL + \left[  \frac{1+256a_2^2 +384a_4 - 32\cL + 136 \cL^2 +32a_2 (1-7\cL)}{96\sqrt 3} + \frac{\cL(1-4\cL)}{\sqrt 3}\trho \right]\frac{e^{-2\trho}}{a_0^2}\nn \\ [1mm]
\!\!&&\!\!+\, \mathcal O(e^{-3\trho})\,.
\eea
The leading order behavior of the metric and of the gauge field determines the background fields for the dual field theory living on the boundary, as displayed in eqs.~\eqref{bdrymetric} and~\eqref{bdryA}.

In the main text, we will also need the leading order expression of $*_5 F$. Starting from~\eqref{formulaFappendix}, this is found to be
\bea
*_5 F &=&  -\, \frac{4\cL}{\sqrt 3} \diff \rho \wedge \diff t \wedge \sigma_3 \nn \\ [2mm]
&&+\, \left[\frac{4\cL}{\sqrt 3}\rho + \frac{1 + 32 a_2 + 256 a_2^2 + 384 a_4 - 80 \cL - 224 a_2 \cL + 328 \cL^2}{24\sqrt 3 (1-4\cL)}  \right]\,\diff t\wedge  \sigma_1\wedge \sigma_2 \nn \\ [2mm]
&& +\, \frac{5 + 64 a_2 - 256 a_2^2 -384 a_4 + 32 \cL - 160 a_2 \cL - 232 \cL^2}{48\sqrt 3}\sigma_1 \wedge \sigma_2 \wedge \sigma_3\,. \label{starF_UV}
\eea

\subsection{Fefferman--Graham coordinates}\label{FGappendix}

In the following, we show that the five-dimensional metric and gauge field can be expressed in a Fefferman--Graham asymptotic expansion, which implies that the UV solution determined above is AlAdS. The general Fefferman--Graham form of the metric is
\be
\diff s^2 \ = \ \frac{\diff \newrho^2}{\newrho^2} +  g_{ij}(x,r) \diff x^i\diff x^j  \,,    
\ee
where 
\be\label{FGexpansion}
g(x,r) \ = \ \newrho^2 \left[g^{(0)} + \frac{g^{(2)}}{r^2} + \frac{g^{(4)} + \tilde g^{(4)}\log r^2}{r^4} + \ldots\right]\,,
\ee
while in the gauge $A_\newrho = 0$ the Maxwell field has the form
\be
A(x,\newrho) \ = \ A^{(0)} + \frac{A^{(2)} + \tilde A^{(2)}\log \newrho^2}{\newrho^2} + \ldots \,.
\ee
Since at leading order in $\rho\to \infty$ the UV solution presented above in this appendix satisfies $g_{\rho\rho}\to 1$, $g_{ij} = \mathcal O(e^{2\rho})$ and $A= \mathcal O(1)$, 
it is immediately compatible with the Fefferman--Graham form upon setting $\rho = \log r$. However, this is not as obvious at subleading orders, as $g_{\rho\rho} = f^{-1}$ is a non-trivial function of $\rho$, and the $g_{ij}$ components of the metric naively seem to have ``logarithmic'' (when $\rho = \log r$) terms too early in the expansion. In order to show that  the solution can be put in Fefferman--Graham form, we need perform a change of radial coordinate, so that
\be
f^{-1/2} \diff \rho \ = \  \frac{\diff \newrho}{\newrho}\,.
\ee
By solving this equation perturbatively at large $\trho$, we find
\bea\label{changeradialcoord}
a_0^2\, \newrho^2 &=& a_0^2\,e^{2\trho} + \frac{1+16 a_2 +12\cL}{24} + \frac{2\cL}{3} \trho  +\, \bigg[ \frac{3+104\cL - 16(48a_2^2-8a_2\cL+ 15 \cL^2)}{2304 } \nn \\ [2mm]
&& + \;\frac{\cL (\cL -12a_2)}{18}\, \trho  - \frac{\cL^2}{3}\trho^2 \bigg] \frac{e^{-2\trho}}{a_0^2} +\, \mathcal{O}(e^{-3\trho}) \,.
\eea
One can now check that the five-dimensional metric is consistent with the Fefferman--Graham expansion. Using the coordinates $(t,\theta,\phi,\psi, r)$, we obtain
\be
\diff s^2 \ = \ \frac{\diff \newrho^2}{\newrho^2} + g_{\theta\theta}\left( \sigma_1^{\,2} + \sigma_2^{\,2} \right) + g_{\psi\psi}\sigma_3^{\,2} + g_{tt}\diff t^2 + 2 g_{t\psi} \,\diff t\, \sigma_3\,,
\ee
where $g_{\theta\theta}$, $g_{\psi\psi}$, $g_{tt}$ and $g_{t\psi}$ depend only on $r$ and have an expansion of the form~\eqref{FGexpansion}. Specifically, we find:
\bea
g^{(0)}_{\theta\theta} \!&=&\! a_0^2\,,\qquad g^{(2)}_{\theta\theta} \,=\, -\frac{3 + 20\cL}{24}\,,\qquad \tilde g^{(4)}_{\theta\theta} \,=\, \frac{\cL(1-4\cL)}{6a_0^2} \,, \nn \\ [2mm]
g^{(4)}_{\theta\theta} &=& \frac{-1 + 1536  a_4 - 120 \cL + 
 16 (64 a_2^2 - 24 a_2 \cL + 
    37 \cL^2)}{768 a_0^2}\,,
\eea
\bea
&& g_{\psi\psi}^{(0)} \,=\, a_0^2(1-4\cL)\,,\qquad g_{\psi\psi}^{(2)} \,=\, \frac{(1-4\cL)(28\cL - 3)}{24}\,,\qquad \tilde g_{\psi\psi}^{(4)} = -\frac{\cL(1-4\cL)^2}{3a_0^2}\,,\\ [2mm]
&&g_{\psi\psi}^{(4)} \,=\, -\frac{1}{{62208 a_0^2}}\big[225 + 2985984 a_6 + 12 (576 a_2 - 2401 \cL) + 
 18432 a_4 (258 a_2 - 7 \cL) \nn \\ [1mm] 
&& + 48 (576 a_2^2 - 3128 a_2 \cL + 3471 \cL^2) + 
 64 (9216 a_2^3 - 216 a_4 + 1248 a_2^2 \cL + 6696 a_2 \cL^2 - 3355 \cL^3) \big],\nn\eea
\be
g_{t\psi}^{(0)} = g_{t\psi}^{(2)} = \tilde g_{t\psi}^{(4)} = 0\,, \qquad g_{t\psi}^{(4)}  = -2g_{\theta\theta}^{(4)} - 2\frac{g_{\theta\theta}^{(0)}}{g_{\psi\psi}^{(0)}}g_{\psi\psi}^{(4)} + \frac{8(\cL - 16a_2 +64a_2\cL + 38 \cL^2)-5}{192a_0^2}\,.
\ee
\bea\label{solgamma}
g_{tt}^{(0)}\! &=&\! -\frac{4 a_0^2}{1-4\cL}\,,\qquad\qquad g_{tt}^{(2)} \,=\, -\frac{4\cL+3}{6(1-4\cL)}\,, \qquad\qquad\tilde g_{tt}^{(4)} \,=\, 0\,, \nn \\ [2mm]
g_{tt}^{(4)} \!&=&\!  8\,\frac{g_{\theta\theta}^{(0)}}{g^{(0)}_{\psi\psi}}g_{\theta\theta}^{(4)} + 4\left(\frac{g_{\theta\theta}^{(0)}}{g_{\psi\psi}^{(0)}}\right)^2 \!g_{\psi\psi}^{(4)} - \frac{1}{48\,g_{\psi\psi}^{(0)}}\left[3+2\left(1-\frac{g_{\psi\psi}^{(0)}}{g_{\theta\theta}^{(0)}}\right) + 11\left( 1- \frac{g_{\psi\psi}^{(0)}}{g_{\theta\theta}^{(0)}} \right)^{\!2} \,\right]\!.\nn\\
\eea
According to the standard AdS/CFT rules, the $g^{(0)}$ coefficients are to be interpreted as source background fields for the dual field theory, while the $g^{(4)}$ are related to the expectation value of the dual energy-momentum tensor.
Let us count how many free parameters we have at this stage. We already discussed that the boundary metric has two independent parameters (plus the trivial one obtained by rescaling $t$). Indeed, $g_{t\psi}^{(0)}$ can always be set to zero by a change of local coordinates, as done here, and $g_{tt}^{(0)}$ can be set to any value by rescaling $t$. We can thus identify the free parameters with $g_{\theta\theta}^{(0)}$ and $g_{\psi\psi}^{(0)}$, which are arbitrary because $a_0$ and $\cL$ are.
As expected from general considerations about solutions on AlAdS backgrounds, we see that the sources fix $g^{(2)}_{ij}$ and $\tilde g_{ij}^{(4)}$.
Regarding the parameters related to vacuum expectation values, we find that three of them are arbitrary. Indeed, $g_{\theta\theta}^{(4)}$, $g_{\psi\psi}^{(4)}$ and $g_{t\psi}^{(4)}$ are arbitrary because they are independent functions of the free parameters $a_2$, $a_4$, $a_6$, while $g_{tt}^{(4)}$ is fixed in terms of the other expectations values and the sources as in~\eqref{solgamma}. The regularity condition imposed in the IR will constrain these vevs so that no freedom is left once the sources are fixed.

The gauge field in the new radial variable $\newrho$ is
\be
A \ = \ \left[A_t^{(0)} + \frac{A_t^{(2)} + \tilde A_t^{(2)} \,\log\newrho^2 }{\newrho^2}\right] \diff t + \bigg[ A_\psi^{(0)} + \frac{A_\psi^{(2)} + \tilde A_\psi^{(2)} \,\log\newrho^2}{\newrho^2} \bigg] \sigma_3 + \mathcal{O}(\newrho^{-3})\,,
\ee
with
\bea
A_t^{(0)} \!&=&\! \frac{1}{2\sqrt 3} \,,\qquad \tilde A_t^{(2)} \,=\, 0\,, \nn\\[2mm] 
A_t^{(2)} \!&=&\! \frac{5 - 256 a_2^2 - 384 a_4 + 32 \cL - 232 \cL^2 + 32 a_2 (2 - 5 \cL)}{48\sqrt 3\, a_0^2 (1-4\cL)} \,,\nn\\ [2mm] 
A_\psi^{(0)} \!\!&=&\!\! -\frac{2}{\sqrt 3}\,\cL \,,\qquad \tilde A_\psi^{(2)} \,=\, \frac{\cL(1-4\cL)}{2\sqrt 3\, a_0^2}\,,\nn \\ [2mm] 
A_\psi^{(2)} &=& \frac{1 + 256 a_2^2 + 384 a_4 - 32 \cL + 136 \cL^2 - 32 a_2 (7 \cL-1)}{96\sqrt 3\, a_0^2}\,.
\eea
Because of supersymmetry, the solution for the gauge field does not contain any additional free parameter with respect to the metric.

We have thus proved that the UV solution can be put in Fefferman--Graham form.
We can also check that our findings agree with general results on AlAdS backgrounds. For instance, an asymptotic analysis of the five-dimensional Einstein equations in the presence of a cosmological constant~\cite{deHaroSoloSkenderis} and a Maxwell field~\cite{Taylor:2000xw} shows that the first subleading component of the metric is determined as
\be\label{g(2)general}
g_{ij}^{(2)} \ = \ -\,\frac{1}{4} \left( R_{ij} - \frac{1}{6} R \, g^{(0)}_{ij} \right),
\ee
while the trace of $g^{(4)}$ is fixed by
\be\label{Traceg(4)general}
{\rm Tr}\, g_{(4)} \ =\ \frac{1}{4} {\rm Tr} \, g_{(2)}^2  + \frac{1}{12} F_{(0)}^2 \,.
\ee
Here, $R_{ij}$ is the Ricci tensor of $g^{(0)}_{ij}$, $R$ its curvature and the indices are raised using $g^{(0)}$.\footnote{In writing eqs.~\eqref{g(2)general}, \eqref{Traceg(4)general} we took into account two differences between the conventions adopted here and those of \cite{Taylor:2000xw}: $R_{\mu\nu\eta\lambda}^{\rm there} =  - R_{\mu\nu\eta\lambda}^{\rm here}$ and $F^{\rm there}=2F^{\rm here}$.\label{comparisonTaylor}} The traceless part of $g^{(4)}$ is instead not constrained by the Einstein equation. We checked that these conditions are indeed satisfied by our solution.

\section{More on renormalisation}

\subsection{Energy-momentum tensor and R-current}\label{EnMomTandRcurrent}

In the following we give the explicit expressions for the holographically renormalised energy-momentum tensor and R-current of our family of solutions. These are computed using~\eqref{HoloEnMomTensor} and~\eqref{ExpressionRcurrent}, and turn out to depend on the UV parameters $a_0$, $a_2$, $a_4$, $a_6$ and $\sq^2 = 1-4\cL$. However, we know that the full solution only has one independent parameter. In the main text we showed how one can eliminate two of them: $a_4$ is determined by the global smoothness of the solution as discussed in section~\ref{sec:globalconstraint}, while $a_6$ is fixed by a Ward identity as explained in section~\ref{sec:HoloCharges}. We find that their explicit expressions are
\bea
a_4 &=& \frac{1}{384}\left( 5 + 64 a_2 - 256 a_2^2 + 32 \cL - 160 a_2 \cL - 104 \cL^2 \right)\,,\nn\\ [2mm]
a_6 &=& \frac{1}{93312}\big[ 72  + 675 c - 6000 c^2 + 8840 c^3 + 3 a_2 (-261 - 5856 c + 15304 c^2)\nn \\ [1mm]
&& \qquad \quad + \, 288 a_2^2 (-90 + 197 c)  + 80640 a_2^3 \big]\label{sola6}\,.
\eea
The relation between the remaining parameters has been determined numerically, see fig.~\ref{fig:otherUVparametersPlots}. For the energy-momentum tensor we find
\be
\langle T_{ij}\rangle \diff x^i \diff x^j \ = \  \langle T_{\theta\theta}\rangle(\sigma_1^{\,2}+\sigma_2^{\,2}) + \langle T_{\psi \psi }\rangle\sigma_3^{\,2} + 2 \langle T_{t\psi }\rangle \diff t\, \sigma_3 + \langle T_{tt}\rangle \diff t^2\,,
\ee
with, using~\eqref{sola6},
\bea
\langle T_{\theta\theta}\rangle &=&  \frac{\ell}{8\pi G a_0^2}\,\frac{32 + 16 (16 a_2 -5) \sq^2 + 67 \sq^4}{384}  \,,   \nn \\ [2mm]
\langle T_{\psi\psi }\rangle &=&  \frac{\ell}{8\pi G a_0^2}\, \frac{-64 + 480 \sq^2 + 24 (192 a_2 -53) \sq^4 + 1117 \sq^6}{3456} \,, \nn \\ [2mm]
\langle T_{t \psi }\rangle &=&  \frac{1}{8\pi G a_0^2}\, \frac{ ( \sq^2 -1 )^3}{27 \sq^2}  \,, \nn \\ [2mm]
\langle T_{tt}\rangle &=& \frac{1}{8\pi G a_0^2\ell} \left( \frac{2}{27 \sq^4} + \frac{1}{9 \sq^2} - \frac{7}{36} + \frac{89}{864} \sq^2\right)\,.
\label{allthestress}
\eea
It is easy to check that $\langle T_{ij}\rangle$ is covariantly conserved and traceless with respect to the boundary metric~\eqref{bdrymetric}. 
The renormalised electric current, corresponding to the dual field theory R-current, reads
\be
\langle j \rangle \ = \  \frac{1}{144 \sqrt 3 \pi G \ell^2 a_0^4} \left[ \ell (\sq^2 -1)^2\frac{\partial}{\partial t}     +  \left( 72 a_2 - 14 + \frac{6}{\sq^2} + \frac{25 \sq^2}{2}  \right)  \frac{\partial}{\partial \psi}\right]
\ee
and is covariantly conserved with respect to the boundary metric.

We can verify that evaluating the associated charges via the definitions in section~\ref{sec:HoloCharges} gives the results presented in the main text.
Recall that $u \, = \,  \frac{\sq}{2a_0} \frac{\partial}{\partial t}\,$, and that the volume form on the three-sphere at the boundary is
\be
{\rm vol}(S^3_{\rm bdry}) \ = \ \ell^3a_0^3\, \sq\, \sigma_1\wedge\sigma_2\wedge\sigma_3 \ = \ -\ell^3a_0^3\, \sq\,  \sin\theta\, \diff\theta\wedge\diff\phi\wedge\diff\psi \,.
\ee 
With our choice of orientation, $\diff^3x = -\diff\theta\wedge\diff\phi\wedge\diff\psi$. So integrating we have $\int {\rm vol}(S^3_{\rm bdry}) = 2\pi^2\sq (2\ell a_0)^3$, which when $\sq=1$ is the volume of a round 3-sphere of radius $2\ell a_0$.

Then for the energy and the angular momentum we find
\bea
\Egrav  &=& \int_{S^3_{\rm bdry}} u^i\langle T_{it}\rangle \, {\rm vol}(S^3_{\rm bdry})\, \ =\ \,  \frac{\pi\ell^2}{G} \left( \frac{2}{27 \sq^2} + \frac{1}{9} - \frac{7}{36} \sq^2 + \frac{89}{864} \sq^4\right) , \nn \\ [2mm]
J &=& \int_{S^3_{\rm bdry}} u^i\langle T_{i\psi}\rangle \, {\rm vol}(S^3_{\rm bdry})\ = \ \, \frac{\pi\ell^3}{27 G}\, ( \sq^2 -1 )^3 \,,
\eea
while the electric charge is
\be
Q \ = \ \int_{S^3_{\rm bdry}} u_i \langle j^i \rangle \, {\rm vol}(S^3_{\rm bdry})\ = \  -\frac{2\pi\ell^2}{9\sqrt 3 G} (\sq^2 -1)^2\,.
\ee
These are indeed the expressions appearing in the main text.


\subsection{Ambiguities in the energy-momentum tensor and R-current}
\label{AmbiguousApp}

Consider the standard finite counterterms~\eqref{ambij} for the on-shell action.
These lead to the following ambiguity in the holographic energy-momentum tensor
\be\label{ambiguityTij}
\Delta T_{ij} \ = \  -\frac{2}{\sqrt{h}}\frac{\delta \Delta S}{\delta h^{ij}} \ = \ \frac{\ell^3}{4\pi G} \left[ 2 \beta\,  B_{ij} + \gamma\, H_{ij} + \frac{\delta}{\ell^2} \left( 2F_{ik}F_j{}^k - \frac{1}{2}h_{ij} F_{kl}F^{kl} \right) \right],
\ee
where we used the fact that the metric variation of the Euler density vanishes identically in four dimensions. The tensor $H_{ij}$ is given by
\be
H_{ij} = -\frac{1}{\sqrt{h}} \frac{\delta}{\delta h^{ij}}\int \diff^4 x \sqrt{h} R^2 \;=\; 2 \nabla_i \nabla_j R - 2 h_{ij}\, \square R + \frac 12 h_{ij} R^2 - 2 R \, R_{ij}\,,
\ee
while $B_{ij}$ is the Bach tensor following from the variation of the Weyl square term:
\bea\label{BachTensor}
B_{ij} &=& -\frac{1}{2\sqrt{h}} \frac{\delta}{\delta h^{ij}}\int \diff^4 x \sqrt{h} \, C_{ijkl}C^{ijkl} \\ [2mm] 
&=& \frac 13\nabla_i\nabla_j R - \square R_{ij} + \frac 16 h_{ij}\, \square R - 2 R_{ ikjl}R^{kl} + \frac 23 R R_{ij} + \frac 12 h_{ij}\left( R_{kl}R^{kl} - \frac 13 R^2 \right).\nn
\eea
These can also be deduced from the expressions given in~\cite{BirrellDavies}, after adapting them to $(-,+,+,+)$ signature and to the curvature tensor conventions in our footnote~\ref{FootnoteConventionsCurvature}. 
Both $H_{ij}$ and $B_{ij}$ are covariantly conserved, and $B_i{}^i=0\,$. 

Including the new counterterm~\eqref{RicciFormTerm} also affects the energy-momentum tensor, as
\be
\frac{1}{\sqrt h}\frac{\delta}{\delta h^{ij}}\int \diff^4 x \sqrt h\, \mathcal R_{kl}\mathcal R^{kl}  \ =\ 2\, \mathcal R_{ik} \mathcal R_j{}^k - \frac{1}{2}h_{ij}\mathcal R_{kl} \mathcal R^{kl} + 2 \nabla_l\nabla_k \left( \mathcal R_i{}^k \complstr_j{}^l\right)  \,,
\ee
where the variation is done at fixed $\complstr_i{}^j$.
Specializing these formulae to our background, one can check that the ambiguities in the energy-momentum tensor affect the energy $\Egrav$ but not the angular momentum $J$. Specifically, $\Egrav$ is shifted by the same amount as the on-shell action $I/\Delta_t$, in such a way that both relations~\eqref{BPSrelation} and~\eqref{RelFromChainRule} continue to hold.

Finally, we note that there is also an ambiguity in the R-current: the variation of the counterterms~\eqref{ambij} with respect to the boundary gauge field $A_i$ yields $\Delta j^i \sim \nabla_j F^{ji}$. However, this does not modify the temporal component of the current and therefore does not affect the charge $Q$ defined in~\eqref{defElCharge}.

\end{document}